\documentclass{article}

\usepackage{arxiv}

\usepackage[utf8]{inputenc} 
\usepackage[T1]{fontenc}    
\usepackage{hyperref}       
\usepackage{url}            
\usepackage{booktabs}       
\usepackage{amsmath}        
\usepackage{amssymb}        
\usepackage{nicefrac}       
\usepackage{microtype}      
\usepackage{graphicx}       
\usepackage{natbib}         
\usepackage{doi}            
\usepackage{xcolor}         
\usepackage{CJK}            
\usepackage{enumerate}      
\usepackage{bm}             
\usepackage{indentfirst} 

\hypersetup{
  colorlinks=true,
  linkcolor=red,
  citecolor=orange,
  urlcolor=blue,
}

\graphicspath{{../figure}}

\title{Resonance phenomena of thermo-bioconvection generated by chemotactic bacteria under unsteady heat condition
}

\author{\href{https://orcid.org/0000-0002-4875-8174}{\includegraphics[scale=0.06]{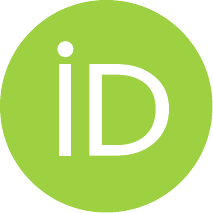}\hspace{1mm}Hideki Yanaoka (柳岡英樹)}
\thanks{Email address for correspondence: yanaoka@iwate-u.ac.jp} \\
    Faculty of Science and Engineering, Iwate University, \\
    4-3-5 Ueda, Morioka, Iwate 020-8551, Japan \\
	\And
	Daichi Koizumi (古泉大地) \\
	Division of Science and Engineering, \\
    Graduate School of Arts and Sciences, Iwate University, \\
    4-3-5 Ueda, Morioka, Iwate 020-8551, Japan \\
	\texttt{} \\
}

\renewcommand{\shorttitle}{Resonance phenomena of thermo-bioconvection generated by chemotactic bacteria under unsteady heat condition}

\makeatletter
\hypersetup{
pdfusetitle,
bookmarksnumbered,
pdftitle={\@title},
pdfsubject={},
pdfauthor={Hideki Yanaoka, Daichi Koizumi},
pdfkeywords={Bio-fluid mechanics, Bioconvection, Thermal convection, Bacteria, Oxygen, Transport phenomena, Resonance phenomena, Numerical simulation}
}
\makeatother

\def\be{\bm{e}}
\def\bu{\bm{u}}
\def\bx{\bm{x}}
\def\bD{\bm{D}}
\def\bI{\bm{I}}
\def\bJ{\bm{J}}
\def\bV{\bm{V}}

\begin{document}

\begin{CJK*}{UTF8}{ipxg} 
\maketitle
\end{CJK*}

\begin{abstract}
Bioconvection is a phenomenon caused by microorganisms with tactic properties. 
To effectively utilize bioconvection for industrial purposes, 
it is necessary to find a way to control it. 
In this study, we performed a three-dimensional numerical analysis of thermo-bioconvection 
generated by a suspension of chemotactic bacteria 
under unsteady heating conditions at the bottom. 
Under unsteady heating conditions, 
thermal convection and bioconvection coexist, 
and unsteady thermo-bioconvection occurs around plumes. 
When the frequency of the temperature fluctuation is low, 
thermo-bioconvection follows the temperature fluctuation. 
However, as the frequency increases, 
the ability of thermo-bioconvection to follow the temperature fluctuation deteriorates. 
A resonance phenomenon occurs at the frequency 
where the instability of the suspension owing to the density difference 
between the bacteria and water is maintained 
and where thermo-bioconvection can follow temperature fluctuations. 
At the resonance frequency, 
the transport characteristics of bacteria and oxygen throughout the entire region 
within the suspension improve significantly. 
As the amplitude of temperature fluctuations and thermal Rayleigh number increase, 
the interference between thermal convection and bioconvection intensifies, 
leading to a noticeable improvement in transport characteristics 
owing to the resonance phenomenon. 
At this time, the amplitude of temperature fluctuations and thermal Rayleigh number 
do not almost affect the resonance frequency. 
This study demonstrated the possibility of thermal control of transport properties in bioconvection.
\end{abstract}


\section{INTRODUCTION}

Microorganisms inhabit various environments on Earth \citep{Omori_et_al_2003}. 
Some of these microorganisms respond to external stimuli 
and move in specific directions. 
These responses are called taxis. 
Microorganisms' responses to gravity, light, and chemicals are denoted as gravitaxis, 
phototaxis, and chemotaxis, respectively. 
In suspension, when a certain quantity of microorganisms accumulates 
near a free surface owing to taxis, the microorganism cells fall, 
resulting in the generation of bioconvection \citep{Platt_1961} 
as the cells are denser than water \citep{Hart&Edwards_1987}.

Many studies have been made on bioconvection 
\citep{Pedley&Kessler_1992, Hillesdon_et_al_1995, Hillesdon&Pedley_1996, Bees&Hill_1997, Metcalfe&Pedley_1998, Czirok_et_al_2000, Ghorai&Hill_2000, Metcalfe&Pedley_2001, Yanaoka_et_al_2007, Yanaoka_et_al_2008, Williams&Bees_2011, Chertock_et_al_2012, Kage_et_al_2013, Karimi&Paul_2013, Yanaoka&Nishimura_2022}. 
Microorganisms have been utilized for environmental cleanup in various fields 
\citep{Omori_et_al_2000, Hirooka&Nagase_2003, Omori_et_al_2003}. 
Bioconvection can be applied in other applications, 
including driving micromechanical systems 
\citep{Itoh_et_al_2001, Itoh_et_al_2006}, 
mixing chemicals \citep{Geng&Kuznetsov_2005}, 
detecting toxicity \citep{Noever&Matsos_1991a, Noever&Matsos_1991b, Noever_et_al_1992}, 
and controlling microorganisms in biochips. 
\citet{Bees&Croze_2014} reported that it may be possible to utilize microbial taxis 
in the aeration process within bioreactors, 
the recovery of microorganisms, and the prevention of biological adhesion to the walls of the container. 
Additionally, from the perspective of reducing the environmental load, 
the production of biofuels using microorganisms has attracted attention. 
In research using microorganisms to produce bioethanol, 
a biofuel, contact between microorganisms and biomass was promoted, 
achieving high-speed bioethanol production \citep{Tanaka_et_al_2010}. 
Bioconvection caused by the chemotactic \textit{Bacillus subtilis} bacteria enhances 
the mixing of substances in a suspension \citep{Tuval_et_al_2005}. 
Thus, it is conceivable that bioconvection could be applied to bioethanol production. 
To realize a recycling-based society, it is necessary to create innovative technologies, 
and from the perspective of utilizing microorganisms in various fields, 
it is significant to elucidate the behavior of microorganisms with taxis in fluids.

Studies on the control of bioconvection have been performed to utilize microorganisms 
for engineering purposes. 
\citet{Itoh_et_al_2001, Itoh_et_al_2006} proposed a method to control the location of bioconvection 
by utilizing the negative electrotaxis of \textit{Tetrahymena}. 
They reported that \textit{Tetrahymena} concentrated near the electrode through electrotaxis, 
generating bioconvection there. 
\citet{Kuznetsov_2005-10} numerically analyzed the stability of bioconvection 
generated by microorganisms with negative geotaxis under the application of vertical vibrations. 
The study suggested that high-frequency vibrations could potentially be employed to regulate bioconvection. 
Various methods can be considered to control bioconvection. 
A simple method for changing the flow field is to heat the bottom wall of the container. 
The impact of temperature gradients on bioconvection has been investigated. 
\citet{Kuznetsov_2005-3, Kuznetsov_2005-8} explored the stability of a suspension of chemotactic bacteria 
when heated from the bottom wall and found that heating destabilized the suspension 
and promoted the development of bioconvection. 
\citet{Alloui_et_al_2006} examined the stability of a suspension of geotactic bacteria 
when subjected to heating and cooling from the bottom wall 
and demonstrated that these temperature variations affected the convection cell pattern. 
Therefore, as heating the bottom wall affects the stability of a suspension and the convection pattern, 
it may be possible to control bioconvection by heat. 
Furthermore, various studies have been conducted on nano-bioconvection 
in suspensions containing nanoparticles and bacteria 
\citep{Kuznetsov_2011, Geng&Kuznetsov_2005, Uddin_et_al_2016, Zadeha_et_al_2020}. 
Recently, bioconvection in suspensions containing microorganisms 
and nanoparticles under an applied magnetic field has been investigated 
\citep{Naseem_et_al_2017, Khan_et_al_2020, Shi_et_al_2021}. 
New applicative research on bioconvection is underway. 
However, previous studies failed to capture the phenomenon of three-dimensional bioconvection 
because they performed a stability analysis 
or solved the fundamental equations using similarity transformation. 
\citet{Biswas_et_al_2022} investigated mixed thermal bioconvection 
in a W-shaped container filled with a suspension containing copper nanoparticles 
under an applied magnetic field. 
This study reported results that will be useful for designing devices 
that operate in equipment such as microbial fuel cells, 
using the technique to control nanoparticles with a magnetic field. 
However, it is worth noting that three-dimensional bioconvection has not yet been investigated. 
Bioconvection with multiple microbial plumes is intricate, 
and the details of the transport characteristics in three-dimensional bioconvection have not been clarified. 
Furthermore, the changes in convection patterns and transport properties 
with the control of three-dimensional bioconvection have not been thoroughly investigated.

In studies on natural convection, 
researchers have investigated the impact of unsteady heating on thermal convection. 
Both numerical \citep{Kazmierczak&Chinoda_1992} and experimental \citep{Mantle_et_al_1994} evidence have shown 
that unsteady heating enhances heat transfer compared to steady heating. 
Furthermore, previous studies on natural convection in a container heated from the side 
investigated the effect of the frequency of temperature fluctuations on heat transfer characteristics 
\citep{Paolucci&Chenoweth_1989, Kwak&Hyun_1996, Kwak_et_al_1998} 
and revealed the existence of a resonance phenomenon 
in which the time-dependent changes in heat transfer characteristics and convection velocity become significantly large 
when the frequency of internal gravity waves, i.e., waves generated in a density-stratified fluid with buoyancy as the restoring force, 
coincides with the frequency of temperature fluctuations. 
Additionally, it has been reported that the time-averaged heat transfer characteristics 
and convection velocity are significantly improved at the resonance frequency. 
However, an existing study on natural convection in a bottom-heated container 
did not observe the resonance phenomenon \citep{Mantle_et_al_1994}.

As with the characteristics of thermal convection, 
unsteady heating can affect the transport properties of substances in bioconvection. 
However, previous studies on thermo-bioconvection generated by heating a suspension 
have not explored the impact of temperature fluctuations on the container walls 
on bioconvection and transport properties. 
Therefore, for establishing a method for controlling bioconvection, 
it is significant to clarify the effects of unsteady heating on thermo-bioconvection.

From this perspective, in this study, 
we perform a three-dimensional numerical analysis of the thermo-bioconvection 
generated by a suspension of chemotactic bacteria under conditions of unsteady bottom heating 
and clarify the effects of the amplitude and frequency of temperature fluctuations 
on the thermo-bioconvection pattern and the transport characteristics of bacteria and oxygen.

The remainder of the paper is summarized as follows: 
Section \ref{fundamental_equation} describes the computational model, 
microbial modeling, fundamental equations, and computational methods. 
Section \ref{calculation_condition} describes the computational conditions and parameters. 
Additionally, we explain the method for evaluating the transport properties. 
Section \ref{discussion} presents the behavior of unsteady thermo-bioconvection 
obtained through this analysis method and clarifies the material transport properties. 
Finally, Section \ref{summary} summarizes the results.

\section{NUMERICAL PROCEDURES}
\label{fundamental_equation}

\begin{figure}[t]
\centering
\includegraphics[width=110mm]{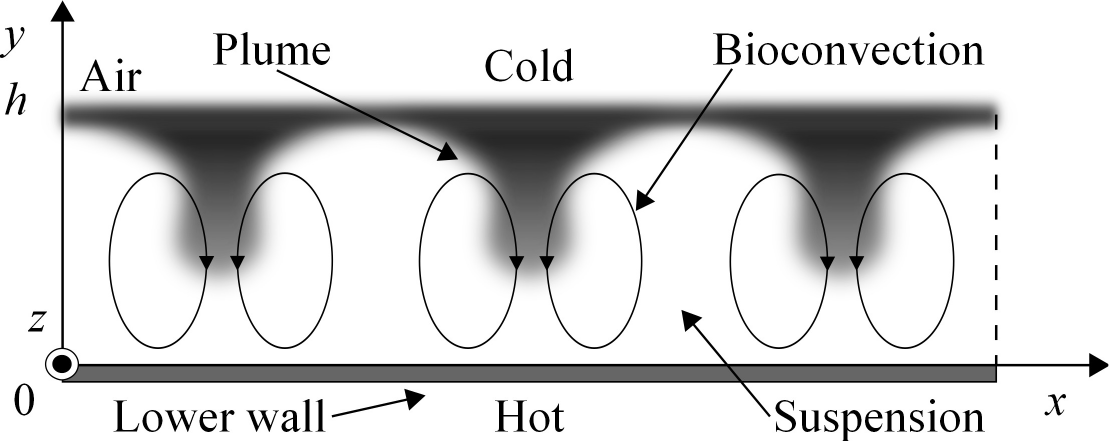} \\
\centering
\caption{Flow configuration and coordinate system.}
\label{flow_model}
\centering
\end{figure}

Figure \ref{flow_model} shows the flow configuration and coordinate system. 
A suspension in a chamber contains bacterial cells, 
and the depth of the suspension is denoted as $h$. 
The origin is located at the bottom wall of the chamber. 
The $x$- and $y$-axes represent the horizontal and vertical direction, respectively, 
while the $z$-axis indicates the direction perpendicular to the page. 
We assume that the suspension is sufficiently dilute 
for hydrodynamic cell--cell interactions to be negligible 
and consider an incompressible viscous fluid. 
The bottom wall of the container containing the suspension is heated, while the temperature at the water surface is kept constant and lower than that of the heated surface. A low heating temperature is set, and the heating process is assumed not to exterminate the microorganisms.

The fundamental equations consist of the continuity equation, 
the momentum equation under the Boussinesq approximation, 
and the transport equations for cells and oxygen 
\citep{Hillesdon_et_al_1995, Hillesdon&Pedley_1996}. 
The nondimensionalized fundamental equation is given as follows:
\begin{equation}
  \nabla \cdot \bu = 0,
  \label{continuity}
\end{equation}
\begin{equation}
  \frac{Wo^2}{Re} \frac{\partial \bu}{\partial t} 
  + \nabla \cdot (\bu \otimes \bu) 
  = - Eu \nabla p + \frac{1}{Re} \nabla^{2} \bu 
  + \frac{\Gamma}{Re^2 Sc} n \be 
  - \frac{Ra}{Re^2 Pr} \theta \be,
  \label{navier-stokes}
\end{equation}
\begin{equation}
  \frac{Wo^2}{Re} \frac{\partial \theta}{\partial t} 
  + \nabla \cdot (\bu \theta) = \frac{1}{Re Pr} \nabla^{2} \theta,
  \label{energy}
\end{equation}
\begin{equation}
  \frac{Wo^2}{Re} \frac{\partial n}{\partial t} 
  + \nabla \cdot \left[ \bu n + \frac{1}{Re Sc} H(c) 
  \left( \gamma n \nabla c - \nabla n \right) \right] = 0,
  \label{cell_concentration}
\end{equation}
\begin{equation}
  \frac{Wo^2}{Re} \frac{\partial c}{\partial t} 
  + \nabla \cdot \left( \bu c - \frac{\delta}{Re Sc} \nabla c \right) 
  = - \frac{\beta \delta}{Re Sc} H(c) n,
  \label{oxygen_concentration}
\end{equation}
where $t$, $\bu$, $p$, ${\bf e}=-\hat{\bf y}$, $\theta$ $n$, and $c$ represent 
time, flow velocity at the coordinate $\bx$, pressure, unit vector in the direction of gravity, 
temperature, cell concentration, and oxygen concentration, respectively. 
Regarding the reference values used for nondimensionalization, 
the length is $L_\mathrm{ref}$, velocity is $U_\mathrm{ref}$, 
pressure is $P_\mathrm{ref}$, and time is $T_\mathrm{ref}$. 
The temperature difference is defined as $\Delta \theta = \theta_{H \mathrm{s}}-\theta_C$, 
using the temperature $\theta_{H \mathrm{s}}$ on the high-temperature side for steady-state heating 
and the temperature $\theta_C$ on the low-temperature side. 
The initial concentrations of bacteria and oxygen are denoted as $n_0$ and $c_0$, respectively, 
and the minimum oxygen concentration at which microorganisms can survive is denoted as $c_\mathrm{min}$. 
Using these reference values, 
the variables in the fundamental equations are nondimensionalized as follows:
\begin{equation}
   \bx^{*} = \frac{\bx}{L_\mathrm{ref}}, \quad
   \bu^{*} = \frac{\bu}{U_\mathrm{ref}}, \quad
   p^{*} = \frac{p}{P_\mathrm{ref}}, \quad
   t^{*} = \frac{t}{T_\mathrm{ref}}, \quad
   \theta^{*} = \frac{\theta-\theta_C}{\Delta \theta}, \quad
   n^{*} = \frac{n}{n_0}, \quad
  c^{*} = \frac{c - c_\mathrm{min}}{c_{0}-c_\mathrm{min}},
\end{equation}
where $*$ represents the nondimensional variable 
and is omitted in the fundamental equations. 
The dimensionless parameters in these fundamental equations are defined as follows: 
\[
   \beta = \frac{K_0 n_0 L_\mathrm{ref}^2}{D_c (c_0 - c_\mathrm{min})}, \quad 
   \gamma = \frac{b V_s}{D_{n0}}, \quad 
   \delta = \frac{D_c}{D_{n0}},
\]
\[
   Re = \frac{U_\mathrm{ref} L_\mathrm{ref}}{\nu}, \quad 
   Wo = L_\mathrm{ref} \sqrt{\frac{1}{\nu_f T_\mathrm{ref}}}, \quad 
   Eu = \frac{P_\mathrm{ref}}{\rho U_\mathrm{ref}^2}, \quad 
   Pr = \frac{\nu}{\alpha} =\frac{Sc}{Le}
\]
\begin{equation}
   Le = \frac{\alpha}{D_{n0}}, \quad 
   Sc = \frac{\nu}{D_{n0}}, \quad 
   \Gamma = \frac{V n_0 g L_\mathrm{ref}^3 (\rho_n - \rho)}{\nu D_{n0} \rho}, \quad 
   Ra = \frac{g \beta_T \Delta \theta L_\mathrm{ref}^3}{\nu \alpha}.
\label{parameter}
\end{equation}
where $\rho$ and $\nu$ represent the density and kinematic viscosity of the fluid, respectively. 
$g$, $\beta_T$, and $\alpha$ represent the gravity acceleration, 
volume expansion coefficient, and thermal diffusion coefficient, respectively. 
$\rho_n$ and $V$ represent the density and volume of the microorganism, respectively. 
$D_{n0}$ and $D_c$ represent the diffusion coefficients of microorganisms and oxygen, respectively. 
$V_s$, $b$, and $K_0$ represent the mean swimming speed, a constant, 
and the rate of oxygen consumption by microorganisms, respectively. 
The parameter $\beta$ represents the strength of oxygen consumption 
relative to its diffusion, 
$\gamma$ represents a measure of the relative strengths of directional 
and random swimming, 
and $\delta$ represents the ratio of oxygen diffusivity to cell diffusivity. 
$Re$, $Wo$, $Eu$, $Pr$, $Le$, and $Sc$ represent the Reynolds, Womersley, 
Euler, Prandtl, Lewis, and Schmidt numbers, respectively. 
$\Gamma$ and $Ra$ represent the Rayleigh numbers for bioconvection and thermal convection, respectively.

\citet{Berg&Brown_1972} found that the swimming velocity of microorganisms 
has both directional and random components. 
In this study, similarly to \citet{Hillesdon_et_al_1995}, random and directional swimming of microorganisms are modeled as cell diffusion and average swimming velocity, respectively. 
The cell diffusion is assumed to be isotropic, 
and the cell diffusivity tensor $\bD_n$ is modeled as $\bD_n = D_{n0} H(c^*) \bI$. 
Here, $H(c^*)$ and $\bI$ represent a step function and the identity tensor, respectively. 
Subsequently, the directional swimming of the cell is modeled as an average swimming velocity. 
The average cell swimming velocity vector $\bV$ is modeled as being proportional 
to the oxygen concentration gradient and is defined as 
$\bV = b V_s H(c^*) \nabla c^*$. 
The oxygen consumption rate $K$ by cells is modeled as $K = K_0 H(c^*)$ 
like that of \citet{Hillesdon_et_al_1995}. 
In this study, $H(c^*)$ is modeled with an approximate equation, 
$H(c^*)=1-\exp(-c^*/c^*_1)$, to suppress discontinuous changes 
due to the step function \citep{Hillesdon_et_al_1995}. 
Here, $c^*_1$ is 0.01 \citep{Hillesdon_et_al_1995, Yanaoka_et_al_2007, Yanaoka_et_al_2008}.
\citet{Hillesdon_et_al_1995} investigated the effect of $c_1^*$ 
on the concentration distribution in a stationary field, 
and the qualitative trend did not change. 
In the present study, $c_1^*$ is set to a small value. 
Therefore, in a shallow chamber treated in the present study, 
as oxygen is transported to the bottom, 
the step function $H(c^*)$ is approximately 1.0, 
and the modeling of $H(c^*)$ does not affect the calculation results.

The governing equations are solved using the simplified marker and cell (SMAC) method \citep{Amsden&Harlow_1970}. 
This study uses the Euler implicit method and implicit midpoint rule for the time differentials 
for analyzing steady and unsteady flows, respectively, 
and the second-order central difference scheme for the space differentials. 
Similarly to existing studies \cite{Yanaoka&Inafune_2023,Yanaoka_2023}, 
we used a simultaneous relaxation method for velocity and pressure.

\section{CALCULATION CONDITIONS}
\label{calculation_condition}

\subsection{Calculation conditions}

This study examines the behavior of chemotactic bacteria in response to oxygen 
as microorganisms with taxis, 
with a specific focus on investigating the impact of temperature fluctuations 
on three-dimensional thermos-bioconvection. 
Initially, we obtain the steady-state field under the condition of isothermal heating 
without temperature fluctuation. 
As the initial condition for steady fields, 
the suspension is stationary and isothermal, 
with constant concentrations of bacteria and oxygen. 
To model the initial state in an experiment of bioconvection, 
we introduce low initial disturbances to the bacterial concentration in the suspension, 
drawing from a previous study \citep{Ghorai&Hill_2002}. 
The initial concentration is defined as
\begin{equation}
  n = n_0 \Bigl[ 1 + \varepsilon E(x,y) \Bigr],
\end{equation}
where $\varepsilon=10^{-2}$ and 
$A(x,y)$ is a random number generated within the range of $-1$ to 1. 
The random number is generated using the Mersenne twister method \citep{Matsumoto&Nishimura_1998}, 
with the initial value provided by the linear congruent method.

A non-slip boundary condition is set at the bottom wall. 
Furthermore, the gradients perpendicular to the wall are to be assumed zero 
for the bacterial and oxygen concentrations, 
and the lower wall surface is heated at a uniform temperature. 
At the free surface, a slip boundary condition is imposed for the velocity field. 
The flux of bacterial concentration is zero, and the oxygen concentration is constant. 
The water surface is cooled at a uniform temperature. 
Periodic boundary conditions are imposed in the $x$- and $z$-directions 
for the velocity, concentration, and temperature fields. 
Using the steady-state field obtained under the above conditions as the initial condition, 
we analyze unsteady thermal bioconvection when temperature fluctuations are applied to the bottom wall of the container. 
To model unsteady heating, the temperature of the hot wall, $\theta_H$, 
fluctuates sinusoidally around the steady isothermal heating temperature, 
$\theta_{H \mathrm{s}}$, as follows:
\begin{equation}
 \theta_H = \theta_{H \mathrm{s}} + \theta_\mathrm{A} \sin (2 \pi f t),
\end{equation}
where $\theta_\mathrm{A}$ and $f$ are the amplitude and frequency of the temperature fluctuation, respectively.

The authors \citep{Yanaoka&Nishimura_2022} investigated the influence of the computational domain 
on the computational results. 
Their findings indicated that the wavelengths at $L = 10h$ and $L = 20h$ are nearly identical. 
Furthermore, the wavelength of the bioconvection pattern observed in a previous experiment 
\citep{Czirok_et_al_2000} was determined to be approximately $\lambda/h = 1.0$. 
Based on these results \citep{Czirok_et_al_2000, Yanaoka&Nishimura_2022}, 
we decided to establish the calculation domain in the $x$ and $z$ directions to $L = 10h$, 
which corresponds to approximately 10 times the wavelength. 
Addionally. the reference values used for nondimensionalization are 
$L_\mathrm{ref} = h$, $U_\mathrm{ref} = \sqrt{V n_0 g (\rho_n - \rho) h/\rho}$, 
$T_\mathrm{ref} = h^2/D_{n0}$, and $p_\mathrm{ref} = \rho U_\mathrm{ref}^2$. 

Nondimensional parameters are determined using the properties of 
\textit{Bacillus subtilis}, which responds to oxygen gradients. 
The depth of the suspension is set to $h = 0.24$ cm. 
The diffusion coefficient of bacteria $D_{n0}$ varies in different studies, 
ranging from $10^{-6}$ to $10^{-5}$ cm$^2$/s \citep{Tuval_et_al_2005}. 
This study takes $D_{n0}=10^{-5}$ cm$^2$/s 
and adopts values from \citet{Hillesdon&Pedley_1996} for the other properties. 
Therefore, the nondimensional base parameters are 
$\beta = 1$, $\gamma = 10$, $\delta = 2$, and $Sc = 1000$. 
In this study, we vary the Rayleigh numbers $\Gamma$ and $Ra$ for bioconvection and thermal convection, respectively.

In this calculation, 
we investigate the frequency response of thermo-bioconvection 
when the frequency $f$ of the temperature fluctuation is changed. 
The nondimensional amplitude and frequency are defined as 
$\Theta_\mathrm{A} = \theta_\mathrm{A}/\Delta \theta$ 
and $F = {f/(D_{n0}/h^2)}$, respectively. 
The nondimensional amplitude $\Theta_\mathrm{A}$ of the temperature fluctuation 
is restricted to $\Theta_\mathrm{A} < 1$ to ensure that the temperature of the hot wall does not drop below that of the cold wall. 
In a previous study \citep{Mantle_et_al_1994}, 
an electric circuit comprising a transformer, variable resistor, timer, 
and heater was employed to experimentally explore the impact of heating with temperature fluctuations on thermal convection. 
It was confirmed that the frequency of the temperature fluctuations could be controlled 
in the range of $f = 0 - 0.5$ Hz. 
The time scale based on thermal diffusion, $\tau_{\theta} = h^2/\alpha$, is $\tau_{\theta} = 40.3$ s. 
Converting this to a dimensionless frequency gives $F = 143$. 
If the temperature fluctuates at a frequency lower than this frequency, 
there is enough time for the heating to affect the area near the top surface. 
Hence, as it is believed that the characteristics of the thermo-bioconvection change around this frequency, 
the frequency of the temperature fluctuation is set to $F = 0 - 2000$. 
The range of this frequency corresponds to $f = 0 - 0.347$ Hz. 
When $F = 0$, thermal convection is a steady field without temperature fluctuations, 
and when $F > 0$, it is an unsteady field affected by temperature fluctuations.

This study used four uniform grids with dimensions of $76 \times 46 \times 76$ (grid1), 
$101 \times 61 \times 101$ (grid2), $126 \times 76 \times 126$ (grid3), 
and $151 \times 91 \times 151$ (grid4) 
to examine the grid dependency of the numerical results. 
We confirmed that the grid resolution of grid2 was suitable 
for obtaining reliable results; 
hence, the results of grid2 are presented below. 
The time intervals $\Delta t$ used in the calculations are 
$\Delta t/(h^2/D_{n0}) = 1 \times 10^{-5}$, $5 \times 10^{-6}$, 
$4 \times 10^{-6}$, and $2 \times 10^{-6}$ for grid1, grid2, grid3, and grid4, respectively.

After the flow, temperature, and concentration fields have reached a statistically stationary state, 
we perform data sampling. 
The sampling time for all frequencies is three periods. 
To confirm the validity of the sampling time, 
we compared the results using sampling times of five and three periods 
and confirmed no differences. 
In the results below, the dimensionless time of the period $2 \pi$ is represented as $T$, 
and the start time for data sampling is $T = 0$.

\subsection{Evaluation of calculation results}

To quantitatively evaluate the calculation results, 
we define the surface friction coefficient and Nusselt number. 
The surface friction coefficient is given by
\begin{equation}
  C_f = 2 \frac{1}{Re} \frac{\partial V^*}{\partial y^*},
\end{equation}
where $V^*$ represents the velocity along the lower wall where $y^* = 0$, defined as $V^* = \sqrt {u^{*2} + w^{*2}}$.

The Nusselt number at the lower wall is determined by
\begin{equation}
  Nu = - \frac{\partial \theta^*}{\partial y^*}.
\end{equation}
The average Nusselt number at the lower surface (area $A$) can be calculated 
through surface integration, as shown below:
\begin{equation}
  Nu_\mathrm{ave} = \frac{1}{A} \int_A Nu dA.
\end{equation}
Similarly, we define the average surface friction coefficient $C_{f, \mathrm{ave}}$.
We define total fluxes to evaluate the transport of heat, bacteria, and oxygen.
The total flux of temperature $\bJ_{\theta}^mathrm{total}$ is determined as follows:
\begin{equation}
  \bJ_{\theta}^\mathrm{total} = \bu \theta - \frac{1}{Re Pr} \nabla \theta.
\end{equation}
The fluxes arising from convection and diffusion are expressed by the following equations, respectively:
\begin{equation}
  \bJ_{\theta}^\mathrm{conv} = \bu \theta,
\end{equation}
\begin{equation}
  \bJ_{\theta}^\mathrm{diff} = - \frac{1}{Re Pr} \nabla \theta,
\end{equation}
where the superscripts conv and diff denote convection and diffusion, respectively.
Additionally, the magnitude of the total flux vector $J_{\theta}^\mathrm{total}$ is given 
using the components $\bJ_{\theta}^\mathrm{total} = (J_{\theta,x}, J_{\theta,y}, J_{\theta,z})$ as follows:
\begin{equation}
  J_{\theta}^\mathrm{total} 
  = \sqrt{J_{\theta,x}^{2} + J_{\theta,y}^{2} + J_{\theta,z}^{2}}.
\end{equation}

Subsequently, the total flux of bacterial concentration $\bJ_n^\mathrm{total}$ is defined as follows:
\begin{equation}
  \bJ_n^\mathrm{total} = \bu n - \frac{1}{Re Sc} H(c) \nabla n 
  + \frac{1}{Re Sc} H(c) \gamma n \nabla c.
\end{equation}
The fluxes arising from convection, diffusion, and swimming are expressed by the following equations, respectively.
\begin{equation}
  \bJ_n^\mathrm{conv} = \bu n,
\end{equation}
\begin{equation}
  \bJ_n^\mathrm{diff} = - \frac{1}{Re Sc} H(c) \nabla n,
\end{equation}
\begin{equation}
  \bJ_n^\mathrm{swim} = \frac{1}{Re Sc} H(c) \gamma n \nabla c,
\end{equation}
where the superscripts conv, diff, and swim denote convection, diffusion, and swimming, respectively. 
Furthermore, the magnitude of the total flux vector $J_n^\mathrm{total}$ is given 
using the components $\bJ_n^\mathrm{total} = (J_{n,x}, J_{n,y}, J_{n,z})$  as follows:
\begin{equation}
  J_n^\mathrm{total} = \sqrt{J_{n,x}^{2} + J_{n,y}^{2} + J_{n,z}^{2}}.
\end{equation}

The total oxygen flux $\bJ_c^\mathrm{total}$ is defined as follows:
\begin{equation}
  \bJ_c^\mathrm{total} = \bu c - \frac{\delta}{Re Sc} \nabla c.
\end{equation}
The fluxes arising from convection and diffusion are expressed by the following equations, respectively.
\begin{equation}
  \bJ_c^\mathrm{conv} = \bu c,
\end{equation}
\begin{equation}
  \bJ_c^\mathrm{diff} = - \frac{\delta}{Re Sc} \nabla c,
\end{equation}
where the superscripts conv and diff indicate convection and diffusion, respectively. 
Furthermore, the magnitude of the total flux vector $J_c^\mathrm{total}$ is given 
using the components $\bJ_c^\mathrm{total} = (J_{c,x}, J_{c,y}, J_{c,z})$  as follows:
\begin{equation}
  J_c^\mathrm{total} = \sqrt{J_{c,x}^{2} + J_{c,y}^{2} + J_{c,z}^{2}}.
\end{equation}

To evaluate the transport characteristics of heat, bacteria, and oxygen in the entire domain $V$, we define the integral of the total flux as follows:
\begin{align}
  J_{\theta}^\mathrm{int} &= \int_V |J_{\theta}^\mathrm{total}| dV, \\
  J_{n}^\mathrm{int} &= \int_V |J_{n}^\mathrm{total}| dV, \\
  J_{c}^\mathrm{int} &= \int_V |J_{c}^\mathrm{total}| dV.
\end{align}

In this study, to evaluate the convective characteristics during unsteady heating, 
we use the nondimensionalized downward flow integral $|v_\mathrm{down}^\mathrm{int}|$, 
the total flux integral $J^\mathrm{int}$, 
the kinetic energy integral $KE^\mathrm{int}$, 
the average Nusselt number $Nu_\mathrm{ave}$, 
and the average surface friction coefficient $C_{f,\mathrm{ave}}$ 
as characteristic values $\phi$. 
The increase rate of the convective characteristics $\phi^{\star}$ is defined 
by the ration of the time-averaged characteristic value $\phi$ 
at an arbitrary frequency to the characteristic value $\phi_s$ in steady heating, 
as follows:
\begin{equation}
  \phi^{\star} = \frac{\phi}{\phi_s}.
\end{equation}

\section{RESULTS AND DISCUSSION}
\label{discussion}

\subsection{Comparison with theoretical values}

\begin{figure}[!t]
\centering
\begin{minipage}{0.48\linewidth}
\centering
\includegraphics[trim=0mm 0mm 0mm 0mm, clip, width=75mm]{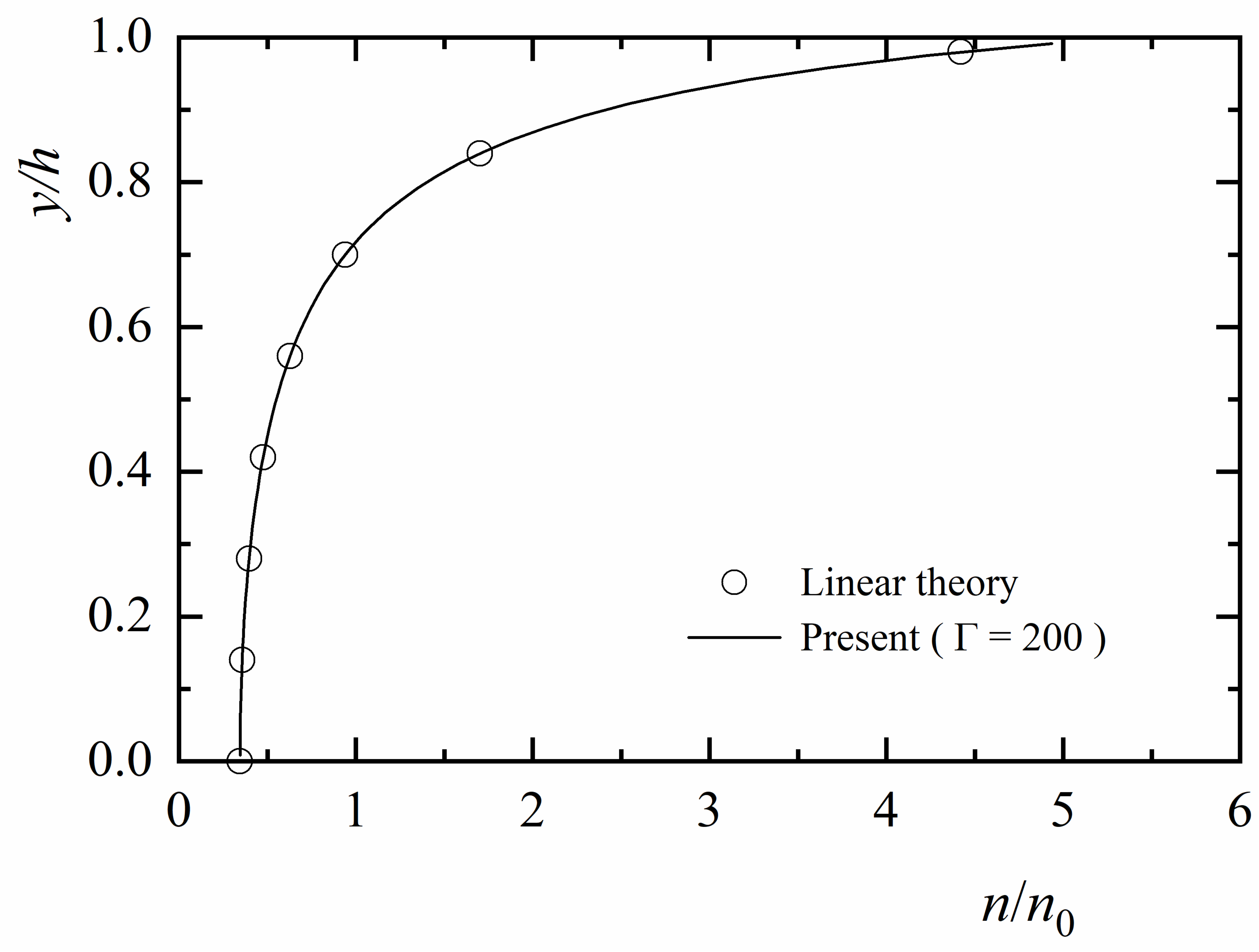}\\
(a) Bacteria \\
\end{minipage}
\begin{minipage}{0.48\linewidth}
\centering
\includegraphics[trim=0mm 0mm 0mm 0mm, clip, width=75mm]{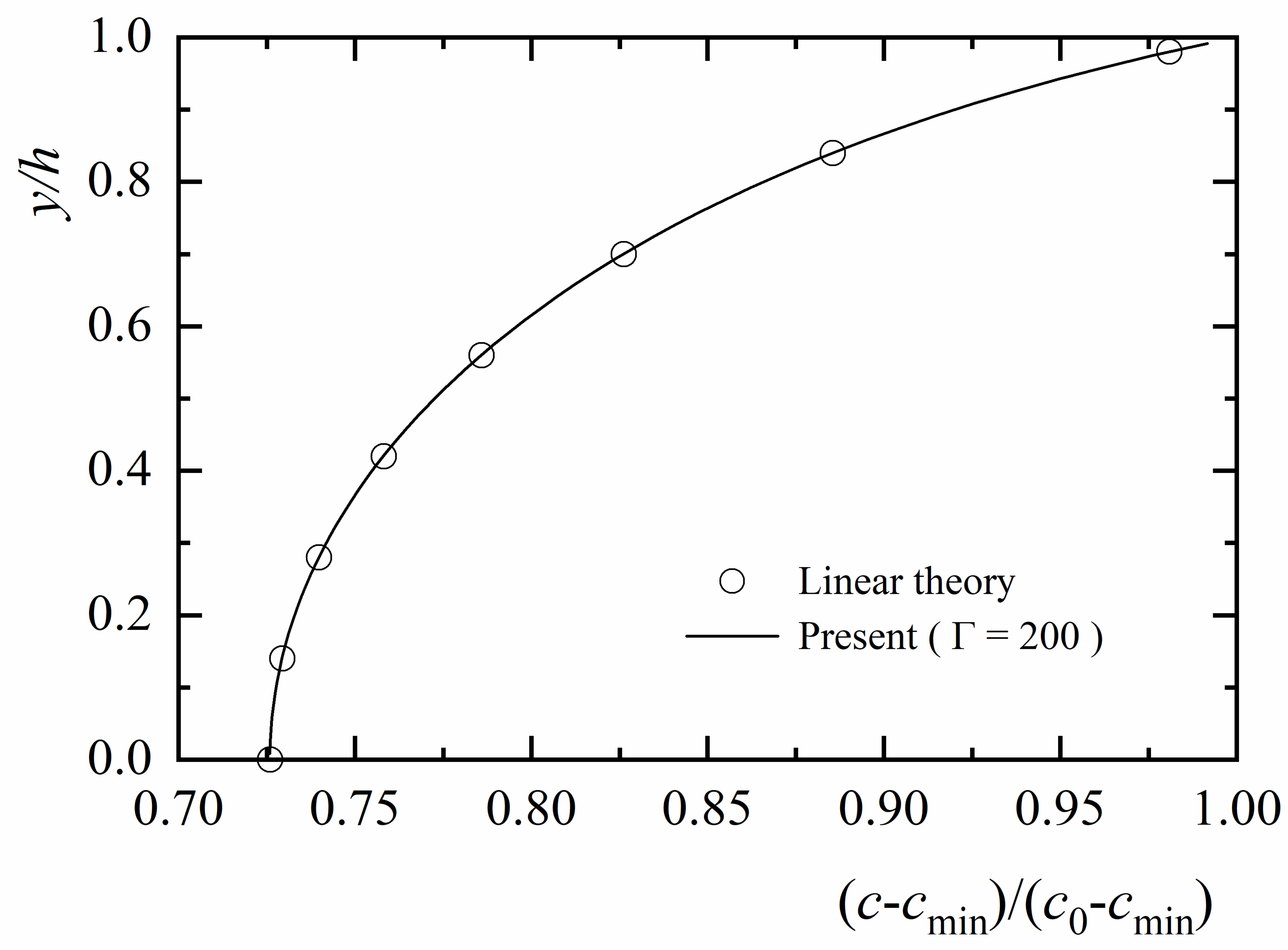}\\
(b) Oxygen \\
\end{minipage}
\caption{Comparison of concentration distributions with linear theoretical solution.}
\label{linear_theory}
\end{figure}

We compared the calculation results with the linear theoretical value 
\citep{Metcalfe&Pedley_1998} for a steady-state field 
and demonstrated the validity of the numerical method. 
In this field, no bioconvection occurs when the bottom wall is not heated. 
The Rayleigh numbers are $\Gamma = 200$ and $Ra = 0$, 
with $\Gamma$ set to be lower than the critical Rayleigh number at which convection occurs. 
The condition $Ra = 0$ indicates that the bottom wall is not heated 
and no temperature gradient occurs in the suspension. 
Figure \ref{linear_theory} shows the bacterial and oxygen concentration distributions 
in the $y$-direction at $x/h = 5.0$ and $z/h = 5.0$. 
The calculation result supports the linear theoretical value.

\subsection{Convection structure}

\subsubsection{Onset of bioconvection}

\begin{figure}[!t]
\centering
\begin{minipage}{0.48\linewidth}
\centering
\includegraphics[trim=0mm 0mm 0mm 0mm, clip, width=70mm]{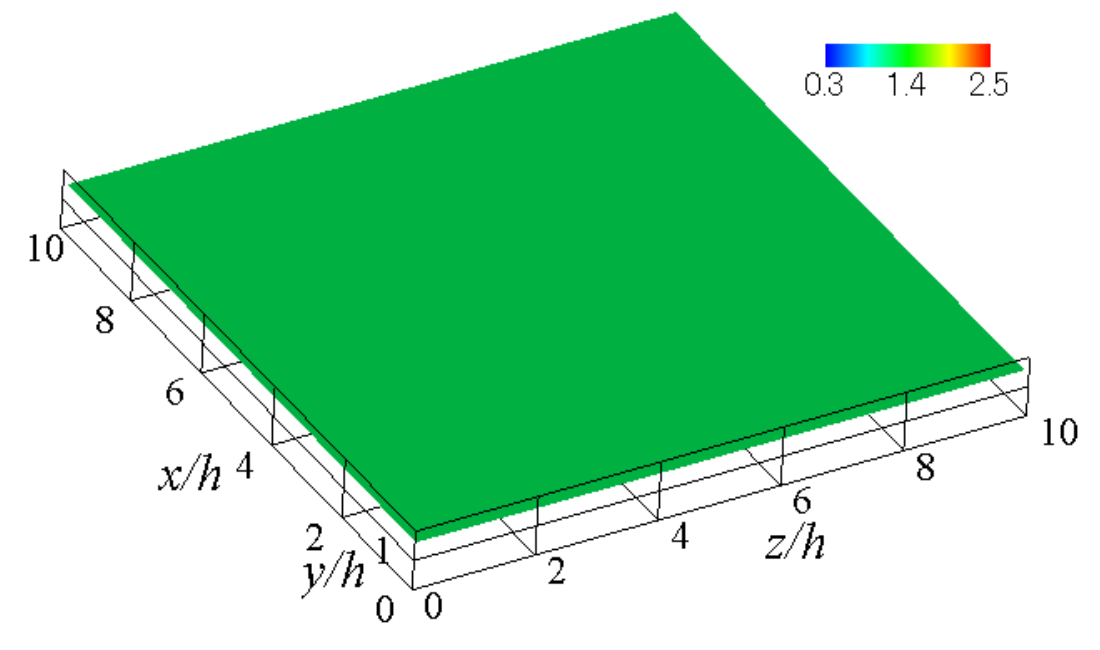} \\
(a) Isosurface of bacterial concentration \\
\end{minipage}
\begin{minipage}{0.48\linewidth}
\centering
\includegraphics[trim=0mm 0mm 0mm 0mm, clip, width=70mm]{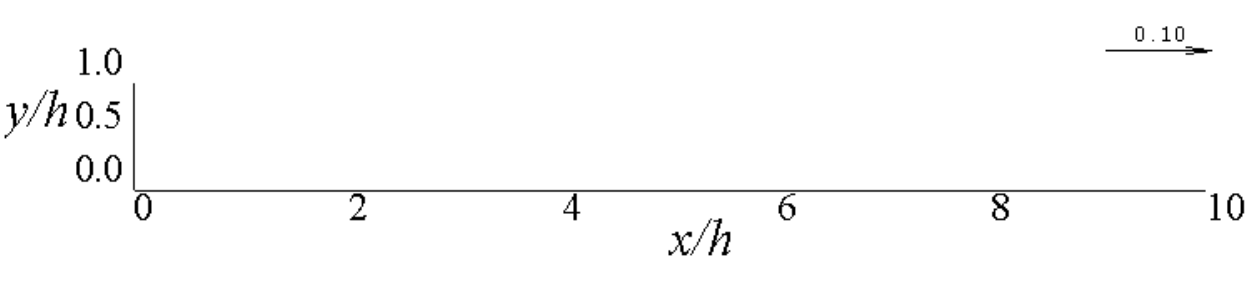} \\
(b) Velocity vectors \\
\includegraphics[trim=0mm 0mm 0mm 0mm, clip, width=70mm]{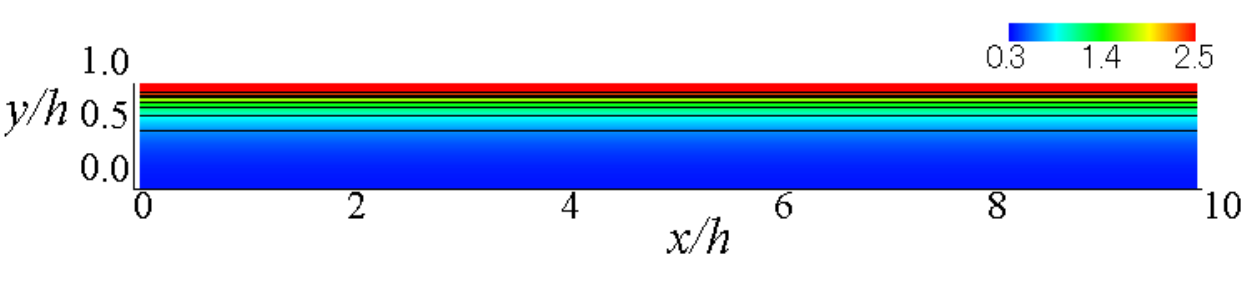} \\
(c) Bacteria \\
\includegraphics[trim=0mm 0mm 0mm 0mm, clip, width=70mm]{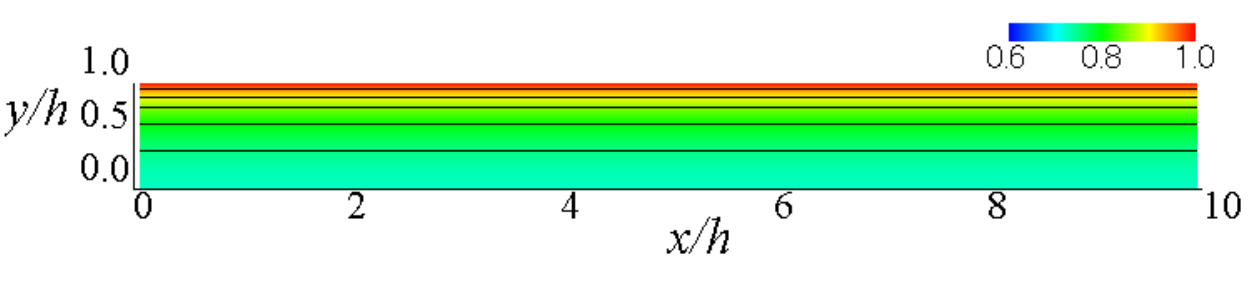} \\
(d) Oxygen \\
\end{minipage}
\caption{Isosurface of bacterial concentration, 
velocity vectors, and contours of bacterial and oxygen concentrations 
at $z/h = 4.95$ for $\Gamma = 200$ and $Ra = 0$: 
Isosurface value is 1.0. 
Contour intervals are 0.275 from 0.3 to 2.5 for bacteria and 0.04 from 0.6 to 1.0 for oxygen.}
\label{g2r200rt0}
\centering
\end{figure}

\begin{figure}[!t]
\centering
\begin{minipage}{0.48\linewidth}
\centering
\includegraphics[trim=0mm 0mm 0mm 0mm, clip, width=70mm]{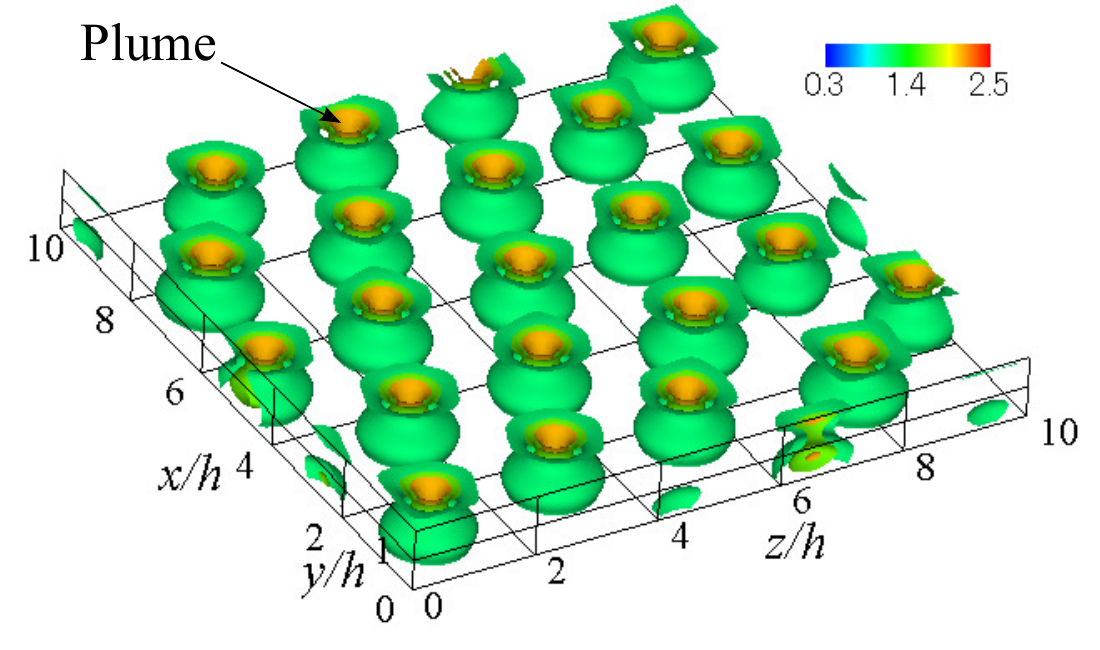} \\
(a) Isosurface of bacterial concentration \\
\end{minipage}
\begin{minipage}{0.48\linewidth}
\centering
\includegraphics[trim=0mm 0mm 0mm 0mm, clip, width=70mm]{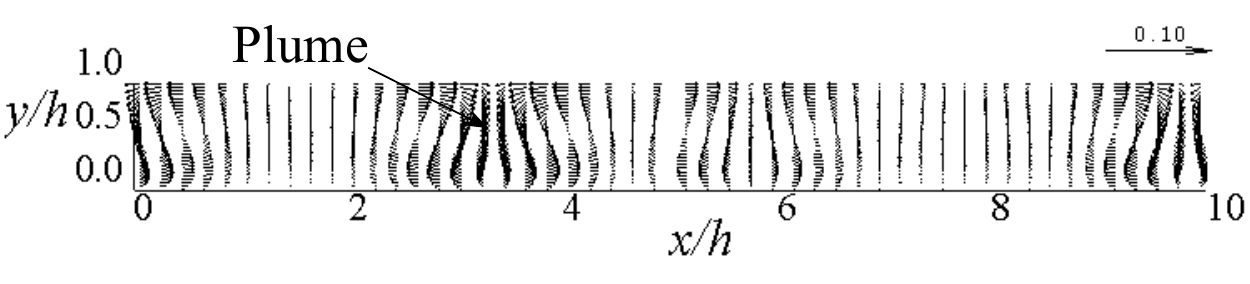} \\
(b) Velocity vectors \\
\includegraphics[trim=0mm 0mm 0mm 0mm, clip, width=70mm]{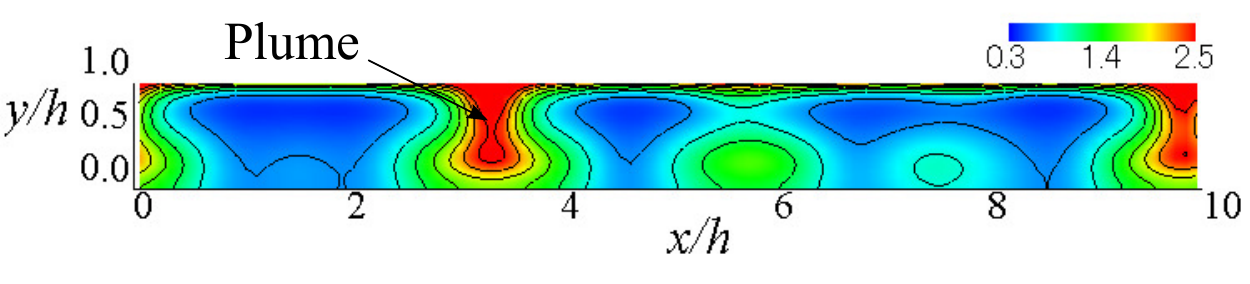} \\
(c) Bacteria \\
\includegraphics[trim=0mm 0mm 0mm 0mm, clip, width=70mm]{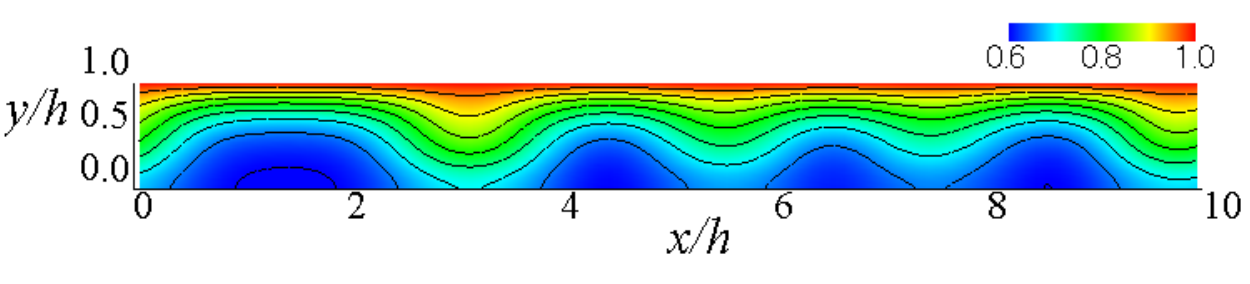} \\
(d) Oxygen \\
\end{minipage}
\caption{Isosurface of bacterial concentration, 
velocity vectors, and contours of bacterial and oxygen concentrations 
at $z/h = 6.45$ for $\Gamma = 1000$ and $Ra = 0$: 
Isosurface values are 0.8, 1.05, and 1.3. 
Contour intervals are 0.275 from 0.3 to 2.5 for bacteria and 0.04 from 0.6 to 1.0 for oxygen.}
\label{g2r1000rt0}
\end{figure}

First, we clarify the characteristics of three-dimensional bioconvection. 
Figures \ref{g2r200rt0} and \ref{g2r1000rt0} display the isosurfaces of bacterial concentration, 
as well as the velocity and concentration fields at the $x$--$y$ cross sections 
at $z/h = 4.95$ and $6.45$ while keeping $Ra = 0$ fixed and varying $\Gamma = 200$ and $1000$. 
The velocity distribution in Fig. \ref{g2r200rt0}(b) shows no convection. 
The oxygen concentration distribution in Fig. \ref{g2r200rt0}(d) indicates 
a decrease in oxygen concentration from the water surface to the bottom wall 
owing to oxygen diffusion from the water surface. 
Both the bacterial concentration isosurface in Fig. \ref{g2r200rt0}(a) 
and the bacterial concentration distribution in Fig. \ref{g2r200rt0}(c) are two-dimensional. 
The bacteria swim owing to chemotaxis toward the water surface, 
where the oxygen concentration is high, 
resulting in a high bacterial concentration near the water surface.

\begin{figure}[!t]
\centering
\begin{minipage}{0.48\linewidth}
\centering
\includegraphics[trim=0mm 0mm 0mm 0mm, clip, width=70mm]{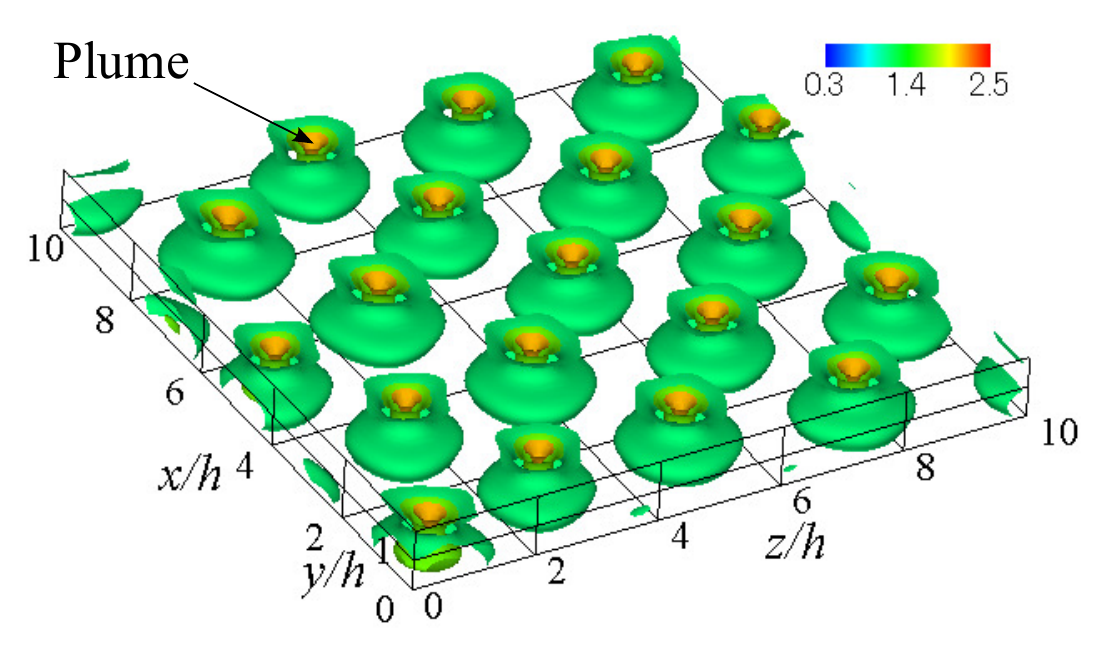} \\
(a) Isosurface of bacterial concentration \\
\includegraphics[trim=0mm 0mm 0mm 0mm, clip, width=70mm]{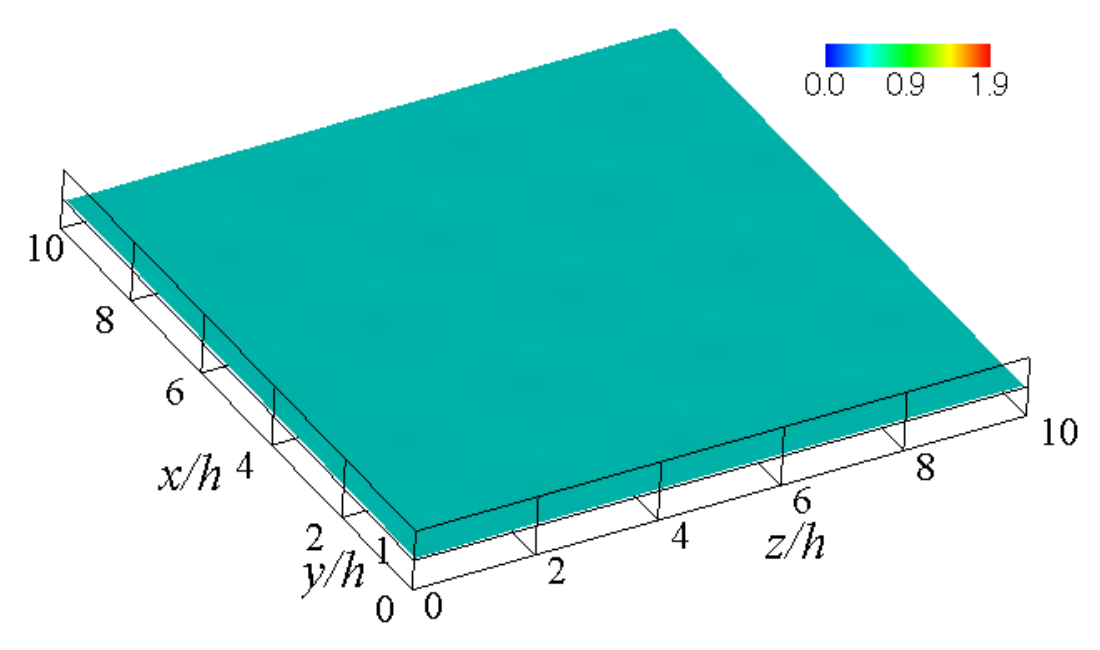} \\
(b) Isosurface of temperature \\
\end{minipage}
\begin{minipage}{0.48\linewidth}
\centering
\includegraphics[trim=0mm 0mm 0mm 0mm, clip, width=70mm]{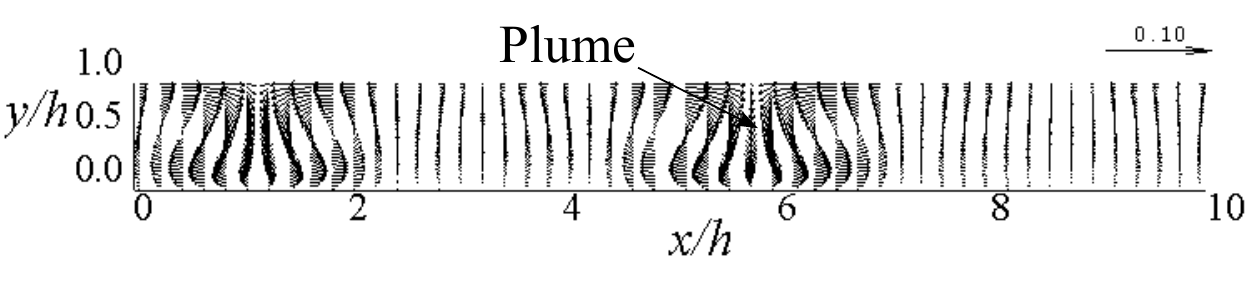} \\
(c) Velocity vectors \\
\includegraphics[trim=0mm 0mm 0mm 0mm, clip, width=70mm]{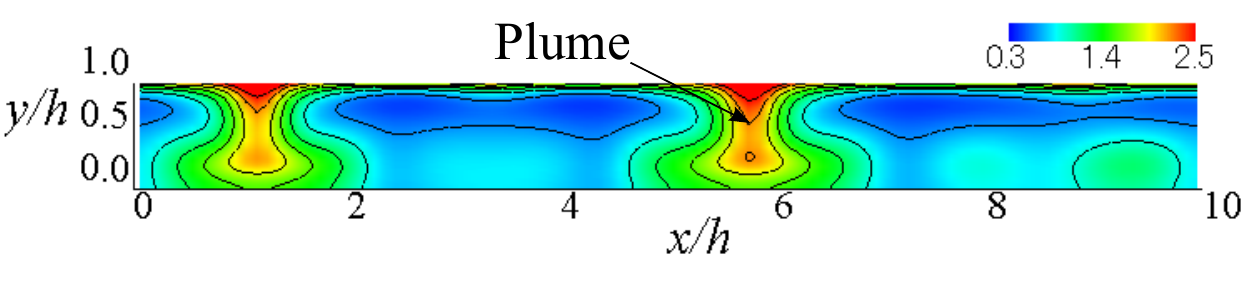} \\
(d) Bacteria \\
\includegraphics[trim=0mm 0mm 0mm 0mm, clip, width=70mm]{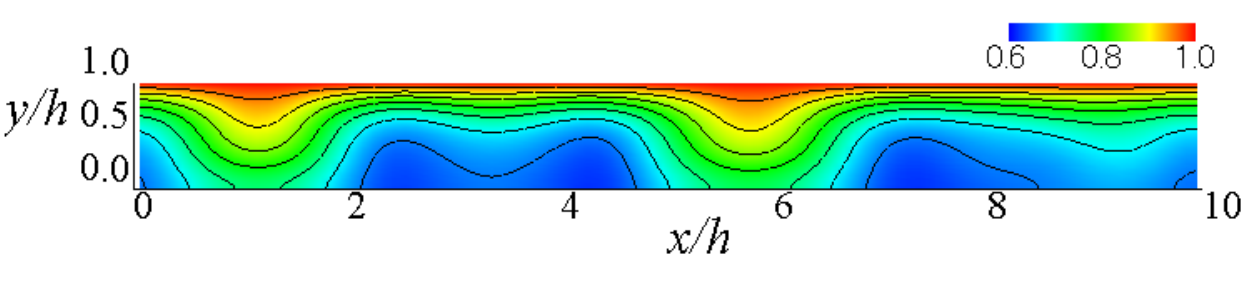} \\
(e) Oxygen \\
\includegraphics[trim=0mm 0mm 0mm 0mm, clip, width=70mm]{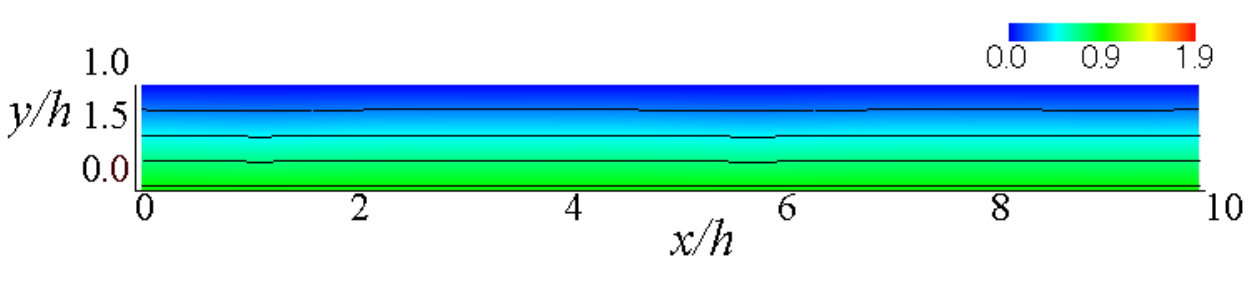} \\
(f) Temperature \\
\end{minipage}
\caption{Isosurface of bacterial concentration, 
isosurface of temperature, 
velocity vectors, and contours of 
bacterial and oxygen concentrations and temperature 
at $z/h = 2.75$ for $\Gamma = 1000$ and $Ra = 500$: 
Isosurface values are 0.8, 1.05, and 1.3 for bacteria 
and 0.5 for temperature. 
Contour intervals are 0.275 from 0.3 to 2.5 for bacteria, 
0.04 from 0.6 to 1.0 for oxygen, and 0.19 from 0.0 to 1.9 for temperature.}
\label{g2r1000rt500}
\end{figure}

Subsequently, we discuss the results for $\Gamma = 1000$. 
In Fig. \ref{g2r1000rt0}(a), 
the isosurface of bacterial concentration reveals the presence of columnar regions 
with high bacterial concentration, indicating the formation of plumes. 
Multiple plumes are observed in this scenario. 
This plume formation phenomenon has also been confirmed 
in previous experiments on bioconvection 
\citep{Bees&Hill_1997, Janosi_et_al_1998, Czirok_et_al_2000} 
and numerical analyses \citep{Ghorai&Hill_2000, Chertock_et_al_2012, Karimi&Paul_2013}. 
Examining the velocity distribution in Fig. \ref{g2r1000rt0}(b), 
we can observe a downward flow toward the lower wall at the plume center, 
while an upward flow toward the water surface occurs between the plumes. 
This result indicates that bioconvection occurs around the plumes. 
In Fig. \ref{g2r1000rt0}(c), 
the bacterial concentration distribution shows that bacteria concentrated 
near the water surface at around $x/h=3.35$ and $9.85$ sink toward the lower wall. 
The oxygen concentration distribution in Fig. \ref{g2r1000rt0}(d) is convex 
toward the lower wall in the plume areas, 
confirming that oxygen transport owing to convection is active. 
By comparing Figs. \ref{g2r200rt0} and \ref{g2r1000rt0}, 
we can conclude that an increase in the Rayleigh number for bioconvection 
strengthens the instability of the suspension and generates bioconvection.

\subsubsection{Changes in the structure of thermo-bioconvection}

To clarify the change in the bioconvection structure resulting from heating from the bottom wall, 
we investigated the impact of increasing the Rayleigh number $Ra$ 
on the velocity, concentration, and temperature fields. 
Figure \ref{g2r1000rt500} shows the isosurfaces of bacterial concentration 
and temperature for $\Gamma = 1000$ and $Ra = 500$, 
as well as the velocity, concentration, and temperature fields 
at the $x$--$y$ cross-section at $z/h = 2.75$. 
The isosurface of bacterial concentration in Fig. \ref{g2r1000rt500}(a) 
indicates the occurrence of multiple plumes. 
As $Ra$ increases, the instability of the suspension increases, 
leading to more intensive interference between bioconvection and thermal convection. 
Consequently, the downward flow at the plume center becomes faster 
in Fig. \ref{g2r1000rt500}(c) compared to the results shown in Fig. \ref{g2r1000rt0}(b). 
\citet{Kuznetsov_2005-8} clarified the phenomenon 
in which heating the bottom wall destabilizes the suspension. 
The distributions of bacterial and oxygen concentrations in Figs. \ref{g2r1000rt500}(d) and (e) 
suggest that an increase in the convection velocity expands the area of high oxygen concentration 
at the plume center compared to the results in Figs. \ref{g2r1000rt0}(c) and (d). 
Additionally, the active transport of materials owing to convection between the plumes 
increases the concentration of bacteria and oxygen around the plume. 
From the temperature distribution in Fig. \ref{g2r1000rt500}(f), 
it is evident that there is almost no heat transport owing to convection, 
and heat conduction is dominant as the Lewis number in this study is $Le=143$.

\begin{figure}[!t]
\centering
\begin{minipage}{0.48\linewidth}
\centering
\includegraphics[trim=0mm 0mm 0mm 0mm, clip, width=70mm]{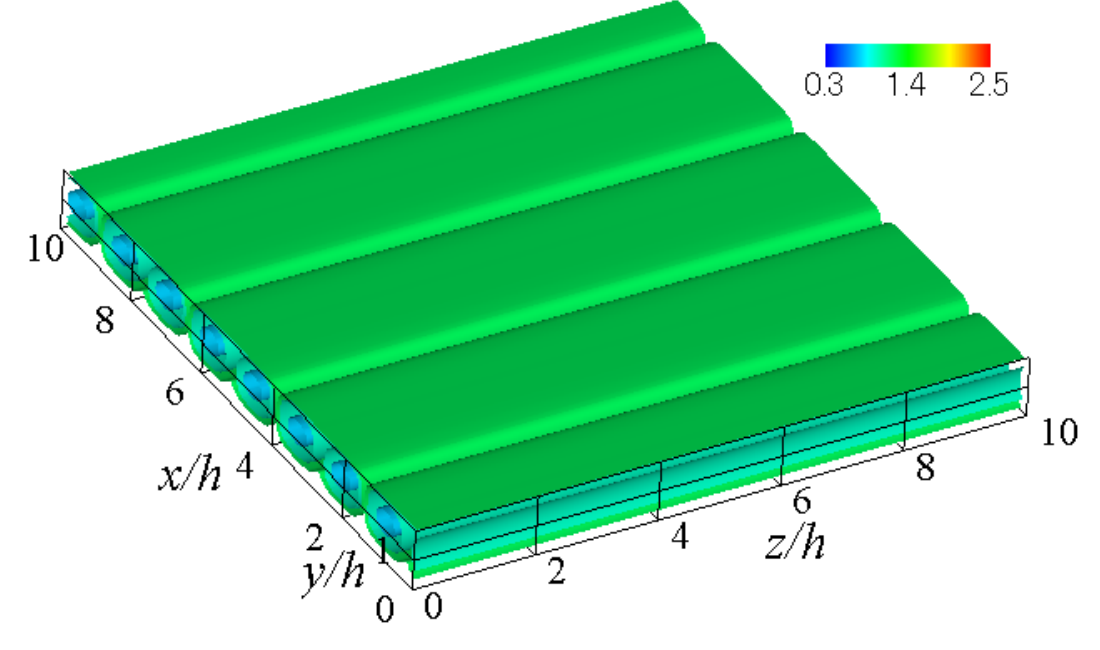} \\
(a) Isosurface of bacterial concentration \\
\includegraphics[trim=0mm 0mm 0mm 0mm, clip, width=70mm]{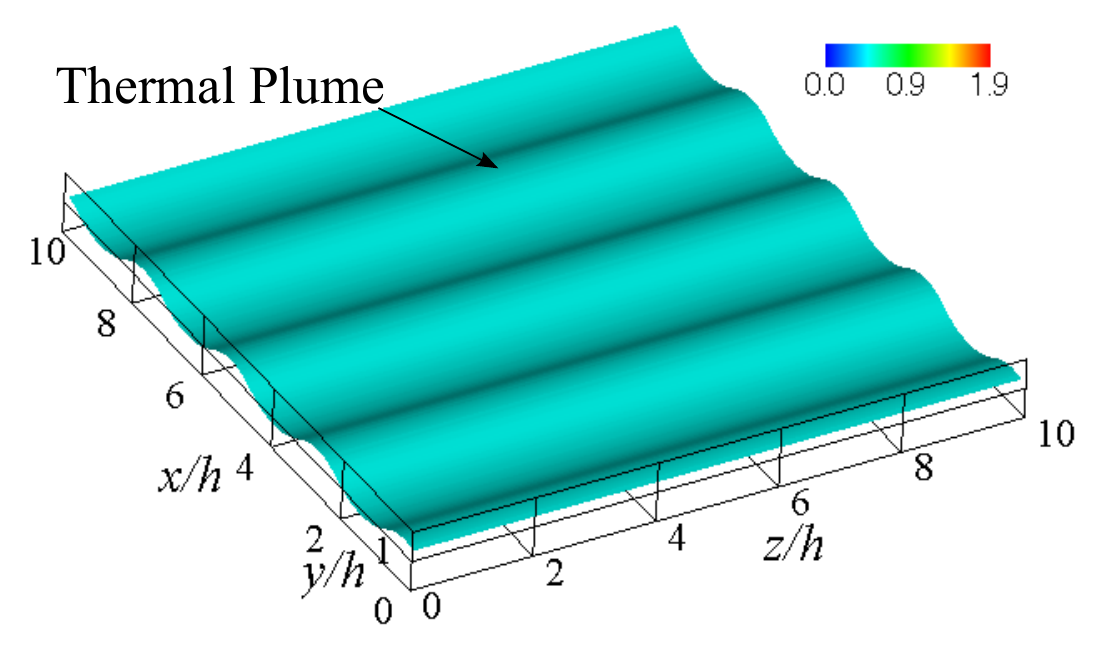} \\
(b) Isosurface of temperature
\end{minipage}
\begin{minipage}{0.48\linewidth}
\centering
\includegraphics[trim=0mm 0mm 0mm 0mm, clip, width=70mm]{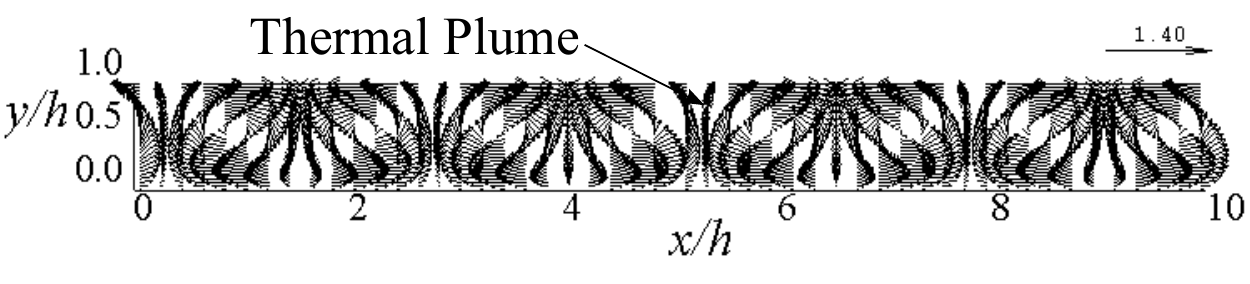} \\
(c) Velocity vectors \\
\includegraphics[trim=0mm 0mm 0mm 0mm, clip, width=70mm]{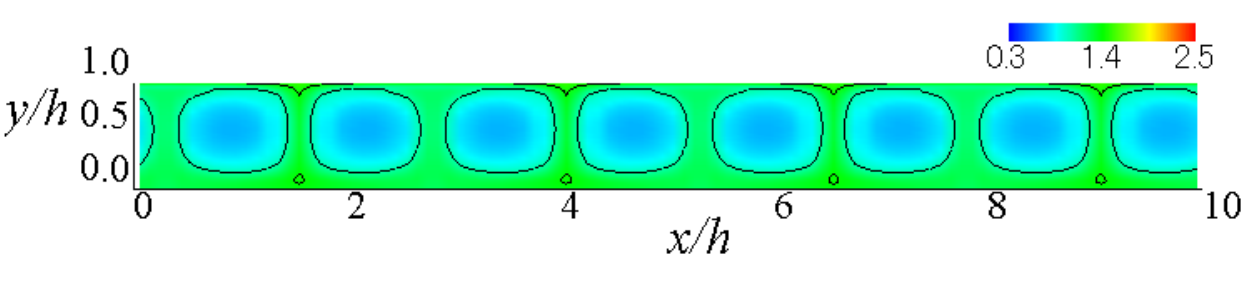} \\
(d) Bacteria \\
\includegraphics[trim=0mm 0mm 0mm 0mm, clip, width=70mm]{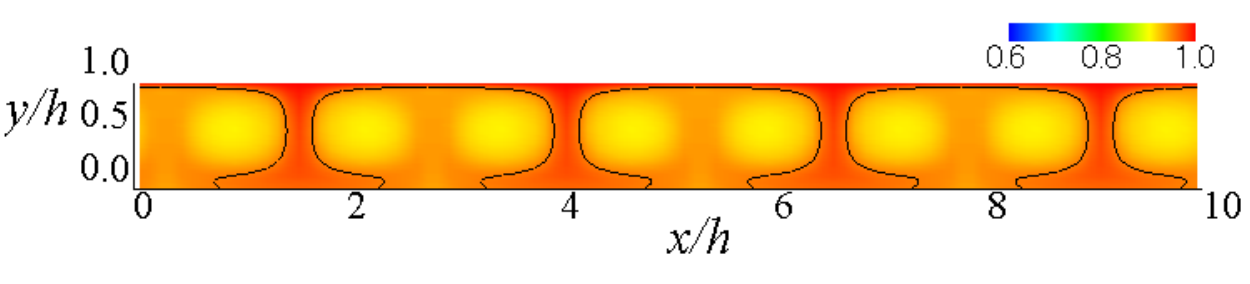} \\
(e) Oxygen \\
\includegraphics[trim=0mm 0mm 0mm 0mm, clip, width=70mm]{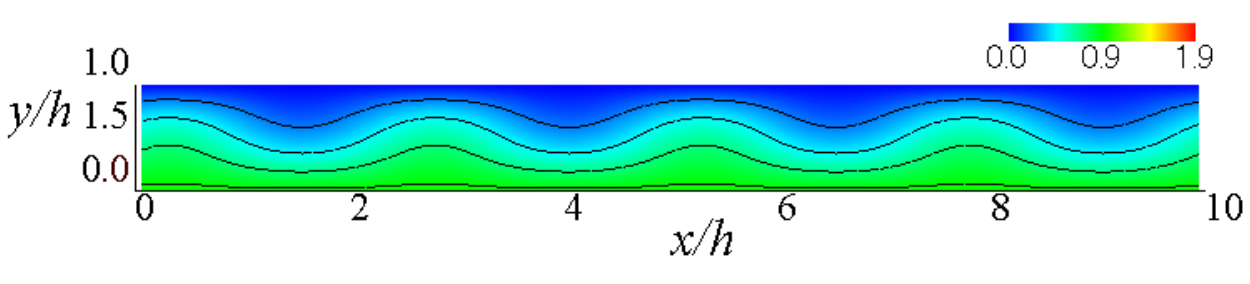} \\
(f) Temperature \\
\end{minipage}
\caption{Isosurface of bacterial concentration, 
isosurface of temperature, 
velocity vectors, and contours of 
bacterial and oxygen concentrations and temperature 
at $z/h = 4.95$ for $\Gamma = 1000$ and $Ra = 1200$: 
Isosurface values are 0.75, 0.975, 1.2 for bacteria and 0.5 for temperature. 
Contour intervals are 0.275 from 0.3 to 2.5 for bacteria, 
0.04 from 0.6 to 1.0 for oxygen, and 0.19 from 0.0 to 1.9 for temperature.}
\label{g2r1000rt1200}
\end{figure}

Figure \ref{g2r1000rt1200} shows isosurfaces of bacterial concentration 
and temperature for $\Gamma = 1000$ and $Ra = 1200$, 
as well as the velocity, concentration, and temperature fields 
at the $x$--$y$ cross-section at $z/h = 4.95$. 
As can be seen from Figs. \ref{g2r1000rt1200}(a) and (d), 
no plume is formed, and the bacterial concentration is distributed in a roll shape. 
In Fig. \ref{g2r1000rt1200}(f), unlike the results in Fig. \ref{g2r1000rt500}(f), 
thermal plumes are formed near $x/h = 0.35$, $2.85$, $5.35$, and $7.85$. 
Thus, it is clear that as $Ra$ increases, the suspension changes 
from a field dominated by bioconvection to a field dominated by thermal convection.

\subsection{Thermo-bioconvection under steady heating conditions}

\subsubsection{Comparison of experiments and calculations regarding wavelengths of bioconvection patterns}

To verify the validity of the present calculation results, we compare the experiment and calculation results concerning the wavelength of the bioconvection pattern. 
\citet{Bees&Hill_1997} conducted experiments on bioconvection formed by 
single-celled alga \textit{Chlamydomonas nivalis} 
and investigated the wavelengths of the bioconvection patterns 
at the onset of bioconvection and in a long-term pattern. 
They found that the wavelength decreases with increasing cell concentration 
and that the cell concentration significantly affects the wavelength of the pattern. 
Similarly, \citet{Czirok_et_al_2000} experimentally investigated the bioconvection 
formed by \textit{Bacillus subtilis} 
and measured the wavelength of the bioconvection pattern 
at the onset of bioconvection. 
They clarified that the wavelength decreases with increased cell concentration 
measured by optical density measurements. 
We calculated the Rayleigh number of the experimental data. 
to compare our numerical results with the previous experimental results. 
The volume of a cell $V$, density ratio of a cell to water 
$(\rho_n-\rho)/\rho$, water kinematic viscosity $\nu$, 
and cell diffusivity $D_{n0}$ were not described in Bees and Hill's paper. 
Hence, we used values from \citet{Ghorai&Hill_2002}. 
Likewise, $V$, $(\rho_n-\rho)/\rho$, $\nu$, and $D_{n0}$ 
were not mentioned in \citet{Czirok_et_al_2000}. 
Hence, we took $\nu$ from \citet{Hillesdon&Pedley_1996}, 
and other values from \citet{Janosi_et_al_1998}. 
The gravity acceleration $g$ was not reported in these papers 
\citep{Hillesdon&Pedley_1996, Bees&Hill_1997, Janosi_et_al_1998, Czirok_et_al_2000, 
Ghorai&Hill_2002}, and thus we took $g$ from \citet{Pedley_et_al_1988}. 
The numerator of the Rayleigh number equation (\ref{parameter}) 
defined in this paper includes the initial cell concentration $n_0$. 
Thus, it can be considered that increasing the Rayleigh number increases 
the initial cell concentration.

\begin{figure}[!t]
\centering
\includegraphics[trim=0mm 0mm 0mm 0mm, clip, width=100mm]{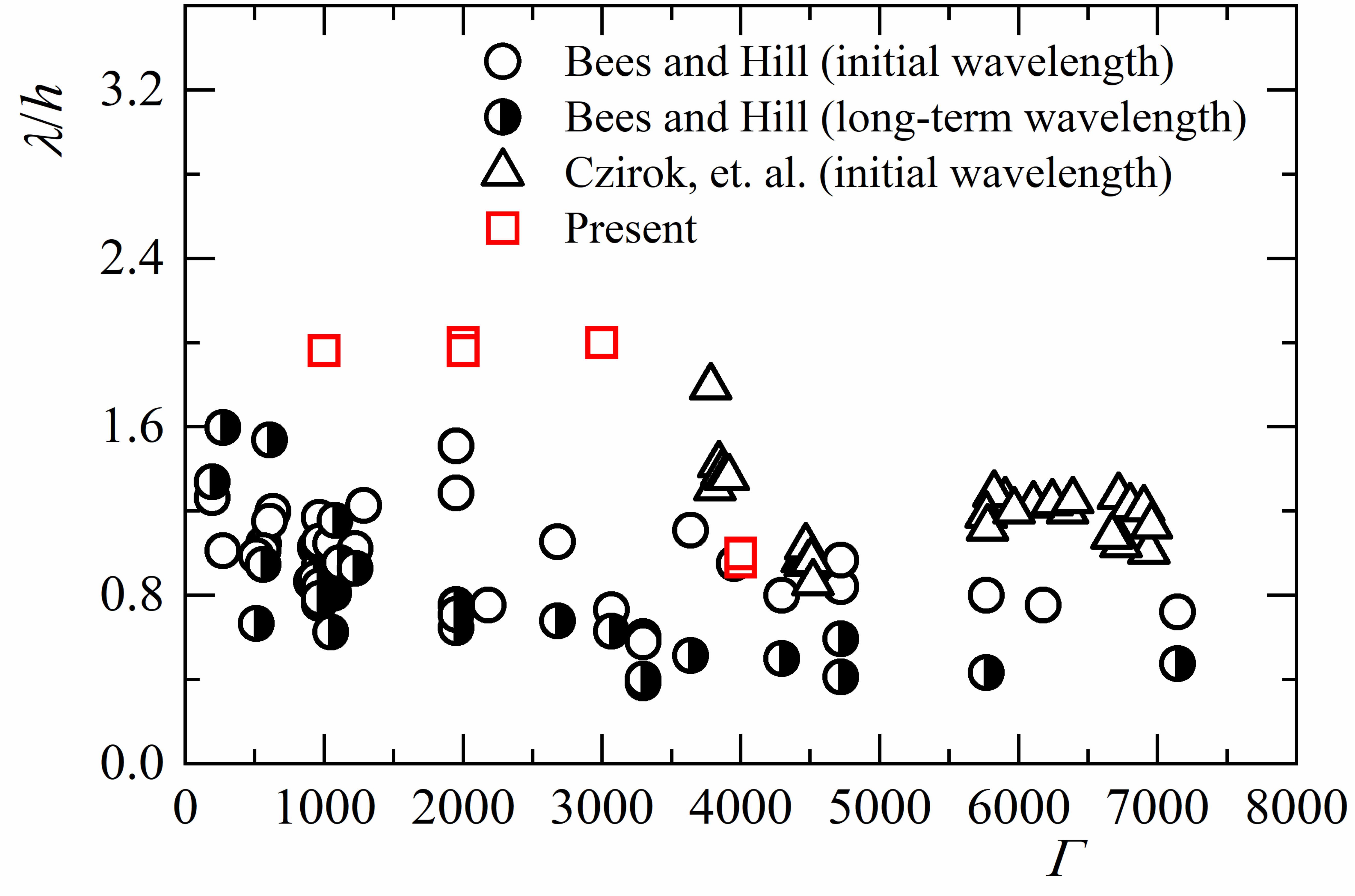} \\
\caption{Comparison of pattern wavelength with previous results.}
\label{comparison_experiment}
\end{figure}

Figure \ref{comparison_experiment} shows a comparison 
between our calculation and experimental values 
\citep{Bees&Hill_1997, Czirok_et_al_2000} for the wavelength $\lambda$ of the bioconvection pattern. 
An existing experimental study \citep{Bees&Hill_1997} reported 
that it was hard to achieve a uniform distribution of cells 
as an ideal initial state of the suspension as fluid motion remained 
after mixing the suspension. 
Consequently, it was shown that the bioconvection patterns changed 
and had different wavelengths in the experiments of \citet{Bees&Hill_1997} 
despite using the same bacterial concentration and suspension depth. 
To simulate the initial state in the existing experiment \citep{Bees&Hill_1997}, 
this study added an initial disturbance to the initial condition of the bacterial concentration. 
It is found from Fig. \ref{comparison_experiment} that the trend of the wavelength of the pattern to decrease 
with increasing Rayleigh number is qualitatively consistent with the experimental results. 
However, there is a quantitative difference between the wavelengths 
in this calculation and the experiments. 
The difference from the experimental results of \citet{Bees&Hill_1997} is due to the use of different microorganisms in both studies. 
This study utilized chemotactic bacteria responding to oxygen, 
whereas \citet{Bees&Hill_1997} used {\it Chlamydomonas nivalis}, 
which exhibits gravitaxis, gyrotaxis, and phototaxis. 
We believe that the the discrepancy from the experimental results of \citet{Czirok_et_al_2000} 
using \textit{Bacillus subtilis} is due to differences in the observation fields. 
Specifically, we observed a steady field of the fully developed bioconvection pattern, 
while \citet{Czirok_et_al_2000} observed the field at the onset of bioconvection.

\subsubsection{Transport properties}

\begin{figure}[!t]
\centering
\begin{minipage}{0.48\linewidth}
\centering
\includegraphics[trim=0mm 0mm 0mm 0mm, clip, width=70mm]{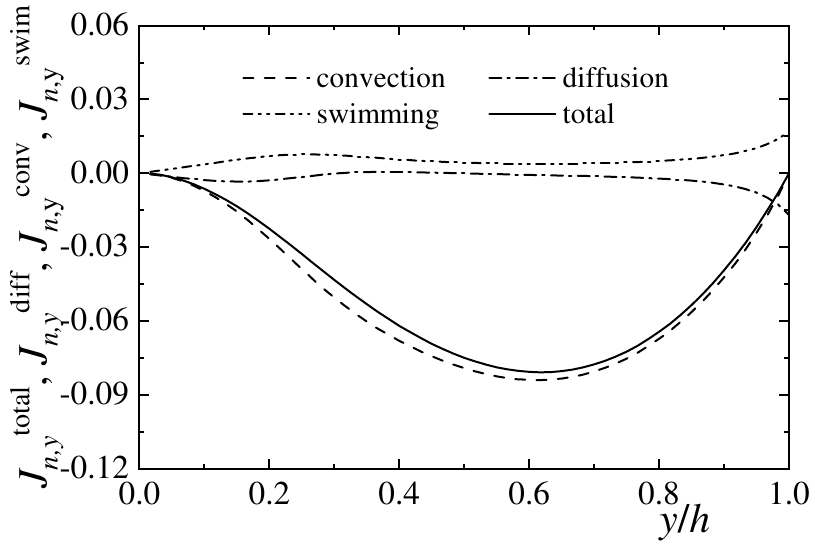} \\
\vspace*{-0.3\baselineskip}
(a) Center of plume \\
\end{minipage}
\begin{minipage}{0.48\linewidth}
\centering
\includegraphics[trim=0mm 0mm 0mm 0mm, clip, width=70mm]{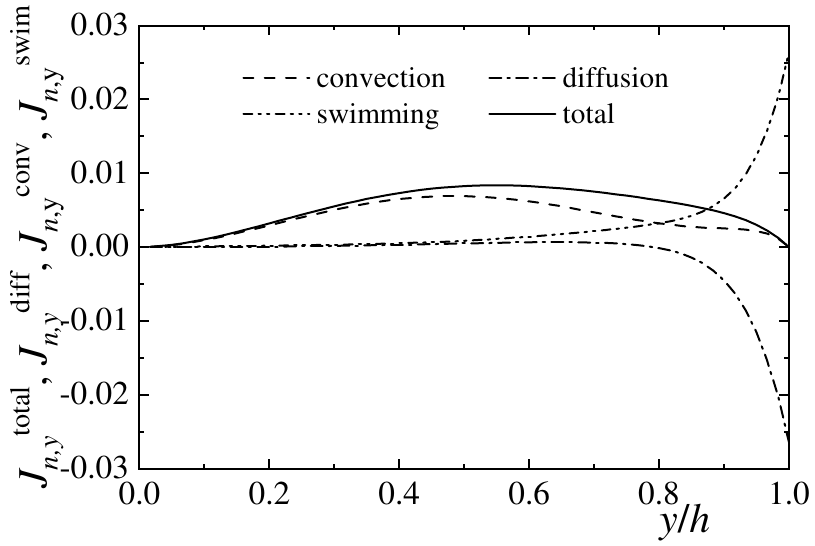} \\
\vspace*{-0.3\baselineskip}
(b) Between plumes
\end{minipage}
\caption{Distributions of bacteria flux 
in $y$--direction at the center of plume and between plumes 
for $\Gamma = 1000$ and $Ra = 500$.}
\label{flux_n_dist_r1000rt500}
\end{figure}

\begin{figure}[!t]
\centering
\begin{minipage}{0.48\linewidth}
\centering
\includegraphics[trim=0mm 0mm 0mm 0mm, clip, width=70mm]{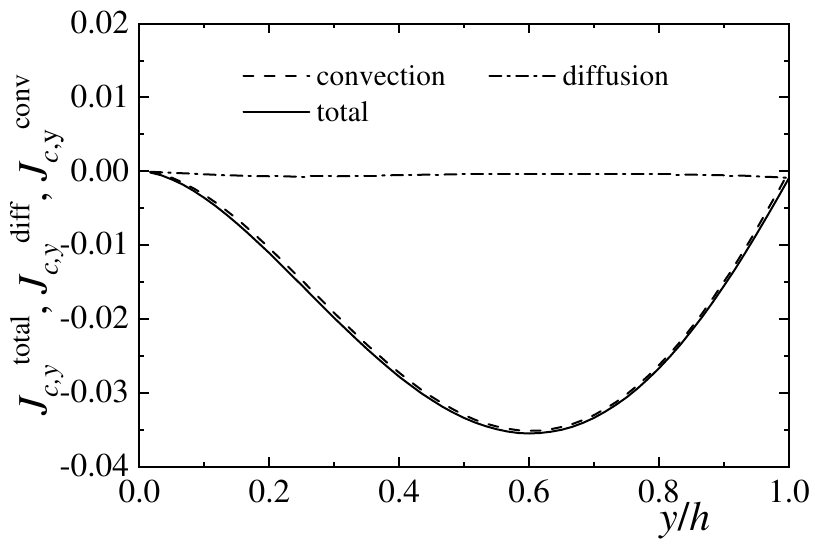} \\
\vspace*{-0.3\baselineskip}
(a) Center of plume \\
\end{minipage}
\begin{minipage}{0.48\linewidth}
\centering
\includegraphics[trim=0mm 0mm 0mm 0mm, clip, width=70mm]{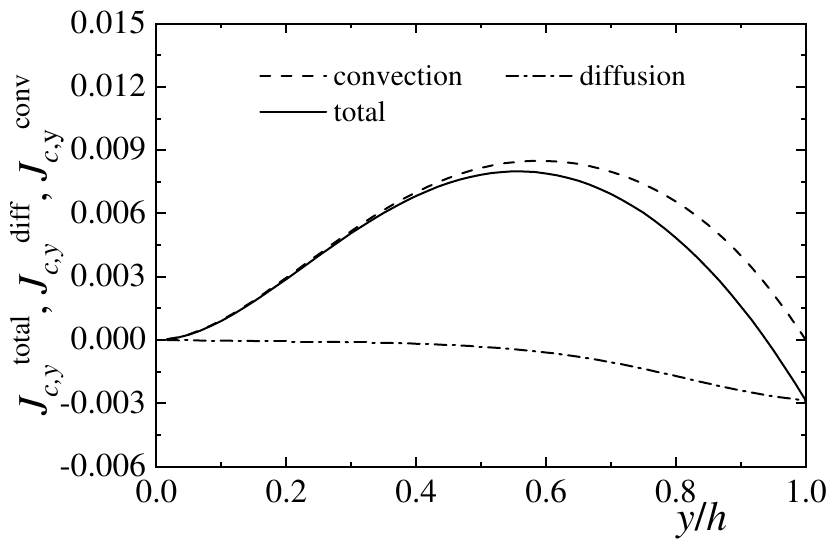} \\
\vspace*{-0.3\baselineskip}
(b) Between plumes \\
\end{minipage}
\caption{Distributions of oxygen flux 
in $y$--direction at the center of plume and between plumes 
for $\Gamma = 1000$ and $Ra = 500$.}
\label{flux_c_dist_r1000rt500}
\end{figure}

\begin{figure}[!t]
\centering
\begin{minipage}{0.48\linewidth}
\centering
\includegraphics[trim=0mm 0mm 0mm 0mm, clip, width=70mm]{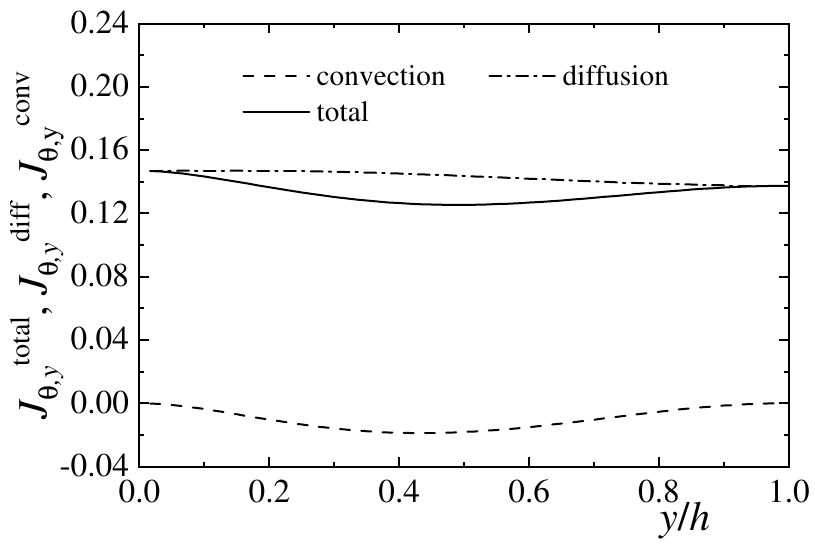} \\
\vspace*{-0.3\baselineskip}
(a) Center of plume \\
\end{minipage}
\begin{minipage}{0.48\linewidth}
\centering
\includegraphics[trim=0mm 0mm 0mm 0mm, clip, width=70mm]{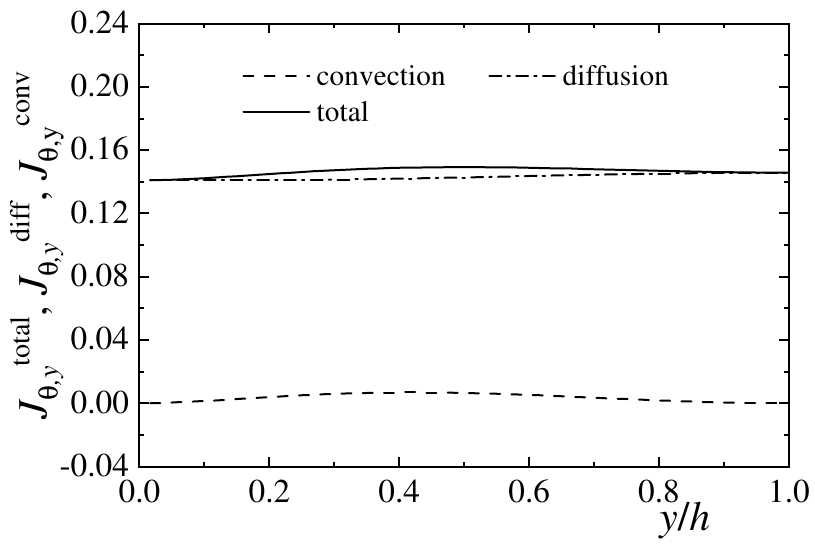} \\
\vspace*{-0.3\baselineskip}
(b) Between plumes \\
\end{minipage}
\caption{Distributions of temperature flux 
in $y$--direction at the center of plume and between plumes 
for $\Gamma = 1000$ and $Ra = 500$.}
\label{flux_t_dist_r1000rt500}
\end{figure}

In a field dominated by bioconvection, 
we clarify the transport characteristics of bacteria, oxygen, and heat 
at the plume center and between plumes. 
Figures \ref{flux_n_dist_r1000rt500}, \ref{flux_c_dist_r1000rt500}, 
and \ref{flux_t_dist_r1000rt500} show the flux distributions of bacteria, oxygen, and heat 
in the $y$-direction at $Ra = 500$, respectively. 
The plume center and the position between the plumes are defined 
as the coordinates $(x/h, z/h) = (5.75, 2.75)$ and $(6.45, 1.25)$ 
where the bacterial concentrations at $y/h = 0.5$ reach maximum and minimum, respectively. 
At the plume center and between plumes, 
convection transport for bacteria and oxygen is dominant. 
For both bacteria and oxygen, 
convection transport between plumes is lower than the value at the plume center. 
Between plumes, transport due to bacterial diffusion and swimming increases near the water surface, 
indicating that the bacteria are active. 
As oxygen is supplied from the water surface, 
transport due to oxygen diffusion is significant near the water surface. 
As for heat transport, owing to the high Lewis number, 
there is almost no convection transport of heat at the plume center and between plumes, 
and diffusive transport is dominant.

\subsubsection{Relationship between thermal Rayleigh number and transport properties}

\begin{figure}[!t]
\centering
\begin{minipage}{0.48\linewidth}
\centering
\includegraphics[trim=0mm 0mm 0mm 0mm, clip, width=75mm]{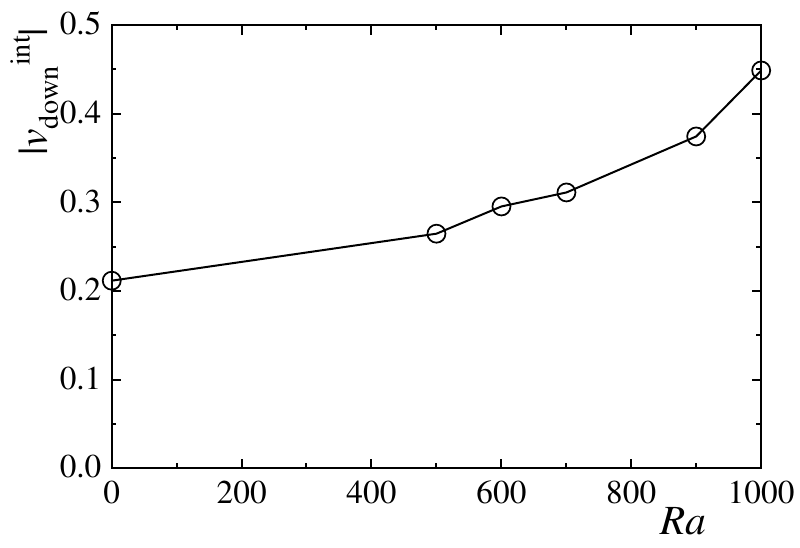} \\
\vspace*{-0.3\baselineskip}
(a) Downward velocity \\
\end{minipage}
\begin{minipage}{0.48\linewidth}
\centering
\includegraphics[trim=0mm 0mm 0mm 0mm, clip, width=79mm]{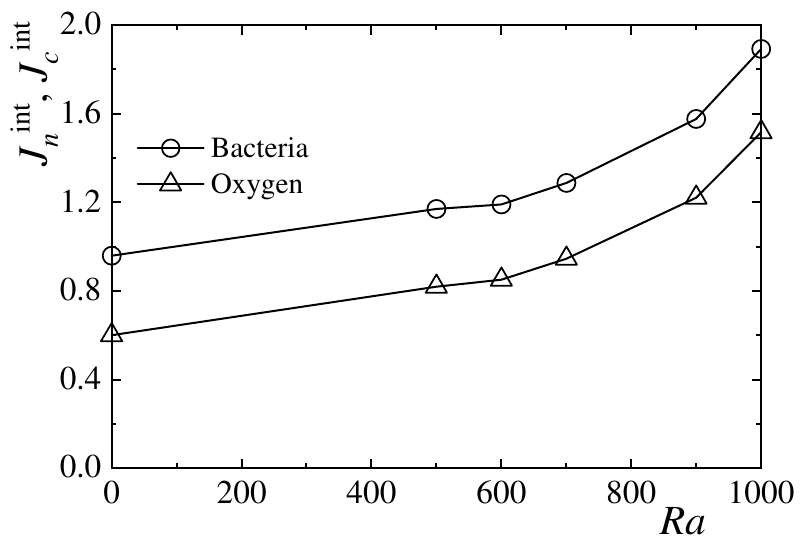} \\
\vspace*{-0.3\baselineskip}
(b) Flux \\
\end{minipage}
\caption{Integral values of downward velocity, 
and total fluxes of bacteria and oxygen for $\Gamma = 1000$.}
\label{int_vdwon_flux_rt0_1000}
\end{figure}

We clarify the changes in the transport characteristics of bacteria and oxygen 
with the heating at the lower wall. 
Figure \ref{int_vdwon_flux_rt0_1000} shows the integral value $|v_\mathrm{down}^\mathrm{int}|$ 
of the downward flow velocity, 
as well as the integral values $J_{n}^\mathrm{int}$ and $J_{c}^\mathrm{int}$ 
of the magnitude of the total flux vector of bacteria and oxygen 
when $Ra$ is varied at $\Gamma = 1000$. 
As $\Gamma = 1000$ exceeds the critical value of the Rayleigh number for bioconvection, 
bioconvection occurs regardless of $Ra$. 
As shown in Figs. \ref{g2r1000rt0}(b) and \ref{g2r1000rt500}(c), 
heating from the bottom wall strengthens the instability of the suspension, 
and the interference between bioconvection and thermal convection increases. 
At this time, the velocities of the downward and upward flows 
at the plume center and between plumes increase, 
causing an increase in $|v_\mathrm{down}^\mathrm{int}|$ with an increase in $Ra$. 
Consequently, the convective transport of substances becomes more active, 
and $J_{n}^\mathrm{int}$ and $J_{c}^\mathrm{int}$ also increase. 
Additionally, when $Ra < 700$, 
$|v_\mathrm{down}^\mathrm{int}|$, $J_{n}^\mathrm{int}$, and $J_{c}^\mathrm{int}$ increase linearly with increasing $Ra$. 
Meanwhile, when $Ra > 700$, 
the effect of heating the bottom wall on the suspension becomes significant, 
resulting in an exponential increase in $|v_\mathrm{down}^\mathrm{int}|$, 
$J_{n}^\mathrm{int}$, and $J_{c}^\mathrm{int}$ with increasing $Ra$ 
due to the increased interference between bioconvection and thermal convection.

\subsubsection{Grid dependence of calculation results}

We confirmed the grid dependency of the calculation results 
for thermo-bioconvection under steady heating. 
The calculation condition was $\Gamma$ = 1000 and $Ra = 1200$, 
representing the highest Rayleigh number combination in steady flow calculations. 
We calculated by varying the initial disturbance of the bacterial concentration 
to compare the results in which the same convection pattern occurred in each grid.

Table \ref{stfluxav_tbio} shows the integral values of the magnitude 
of the total bacterial and oxygen flux vectors, 
$J_{n}^\mathrm{int}$ and $J_{c}^\mathrm{int}$. 
The relative differences in the results obtained using grid1, grid2, and grid3 compared to grid4 
are approximately $-95.458\%$, $0.177\%$, and $0.054\%$ 
for $J_{n}^\mathrm{int}$, respectively, 
and approximately $-96.322\%$, $0.266\%$, and $0.096\%$ 
for $J_{c}^\mathrm{int}$, respectively. 
Based on these results, 
it is considered that the calculation results obtained using grid2 in a steady-state field are valid.

\begin{table}[!t]
\begin{center}
\vspace{1.5mm}
\caption{Integral values of total flux of bacteria and oxygen using different computational grids: 
$\Gamma = 1000$, $Ra = 1200$, $\Theta_\mathrm{A} = 0$, and $F = 0$.}
\label{stfluxav_tbio}
\scalebox{0.9}{
\centering
\begin{tabular}{|c|c|c|}
\hline
        & Bacteria: $J_{n}^\mathrm{int}$ & Oxygen: $J_{c}^\mathrm{int}$ \\ \hline
  grid1 & 1.4937                         & 1.1572  \\ \hline
  grid2 & 32.9412                        & 31.5457 \\ \hline
  grid3 & 32.9009                        & 31.4921 \\ \hline
  grid4 & 32.8830                        & 31.4620 \\ \hline
\end{tabular}}
\end{center}
\end{table}

\subsection{Thermo-bioconvection under unsteady heating conditions}

\subsubsection{Effect of temperature fluctuations on transport properties}

\begin{figure}[!t]
\centering
\begin{minipage}{0.48\linewidth}
\centering
\includegraphics[trim=0mm 0mm 0mm 0mm, clip, width=80mm]{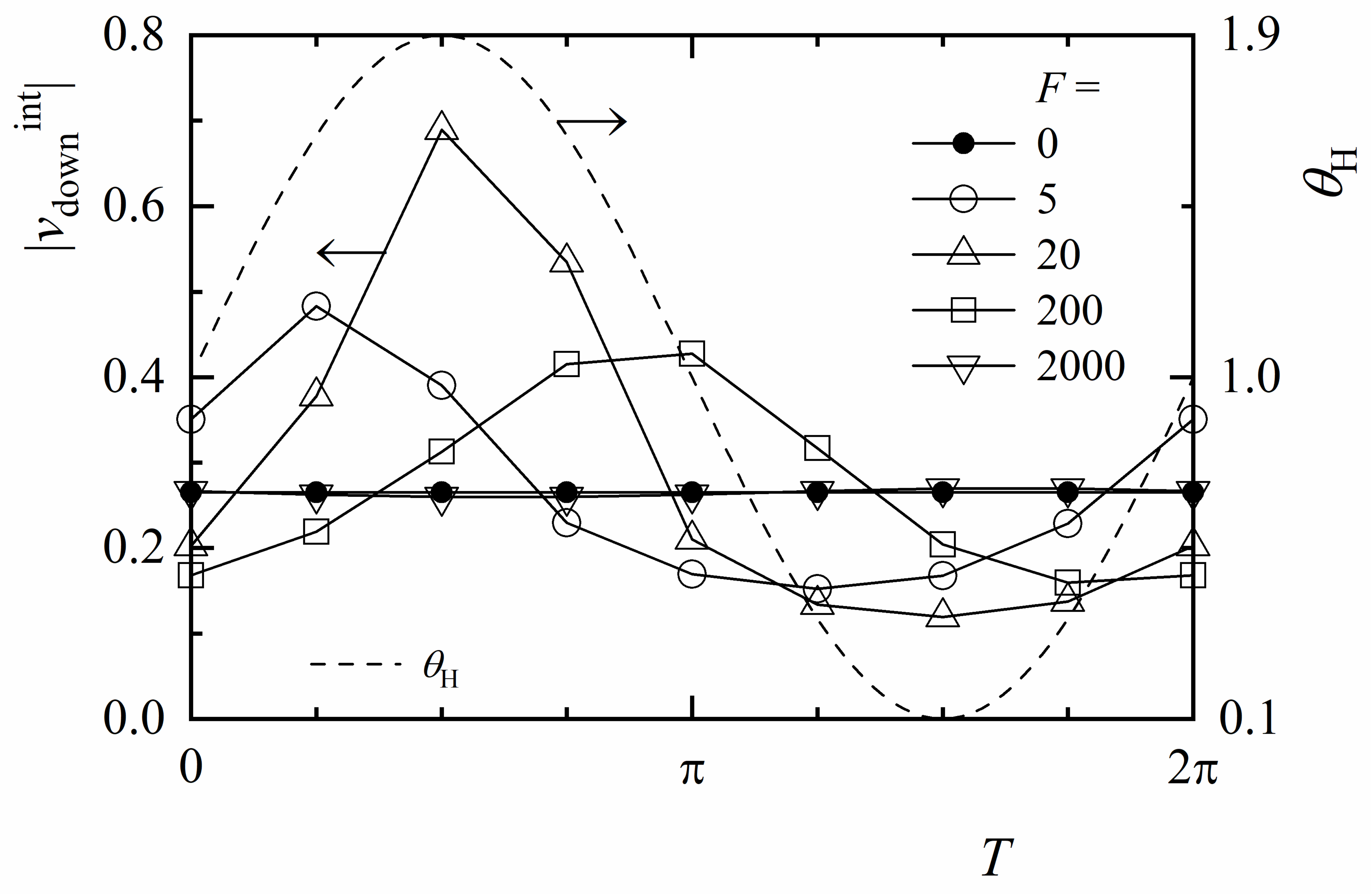} \\
\vspace*{-0.2\baselineskip}
(a) Downward velocity \\
\end{minipage}
\begin{minipage}{0.48\linewidth}
\centering
\includegraphics[trim=0mm 0mm 0mm 0mm, clip, width=80mm]{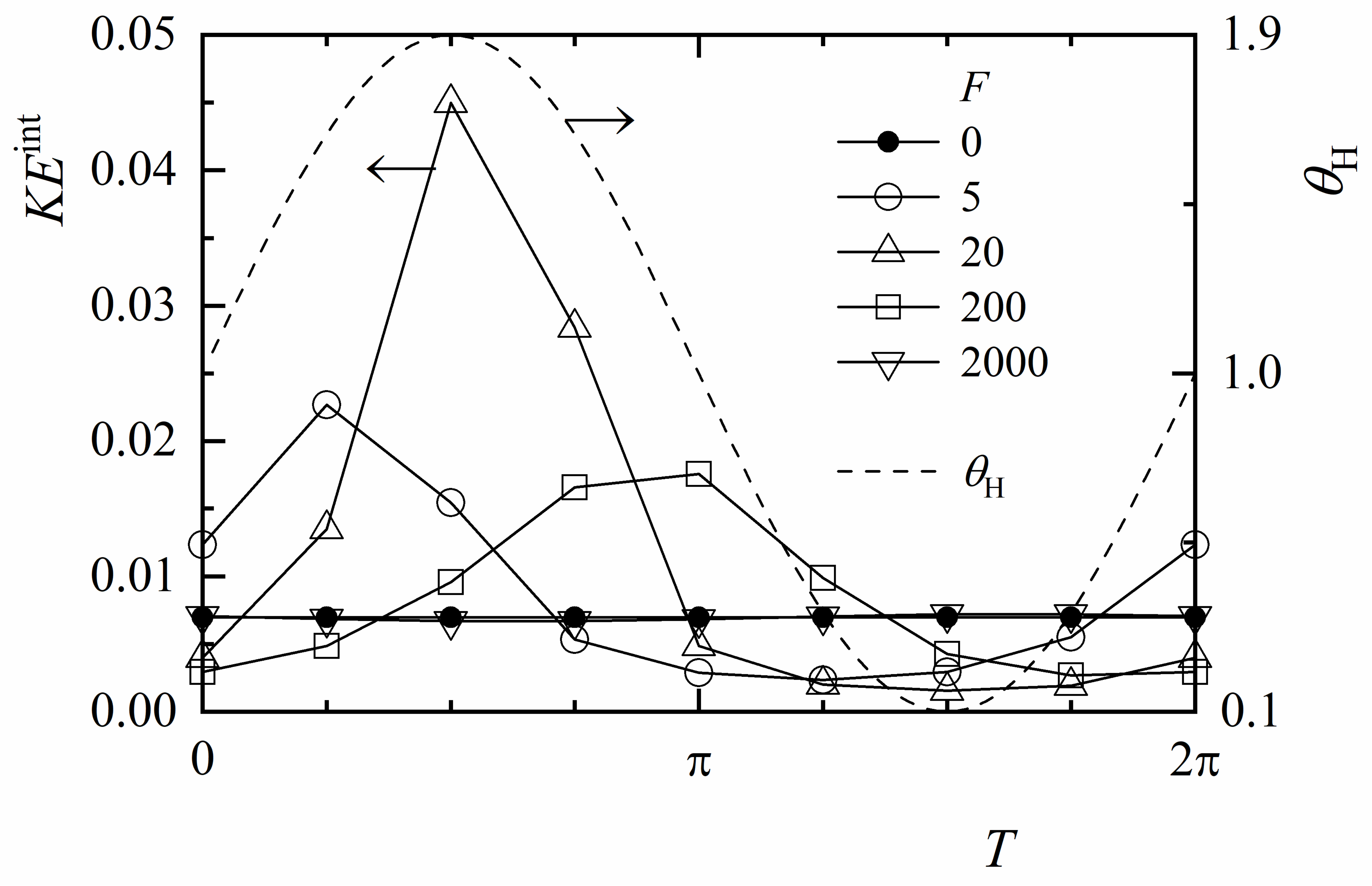}\\
\vspace*{-0.2\baselineskip}
(b) Kinetic energy
\end{minipage}
\caption{Time variations of integral values of downward velocity and kinetic energy for $\Gamma = 1000$, $Ra = 500$, and $\Theta_\mathrm{A} = 0.9$.}
\label{time_vari_vdown_KE_f5_2000}
\end{figure}

\begin{figure}[!t]
\centering
\begin{minipage}{0.48\linewidth}
\centering
\includegraphics[trim=0mm 0mm 0mm 0mm, clip, width=80mm]{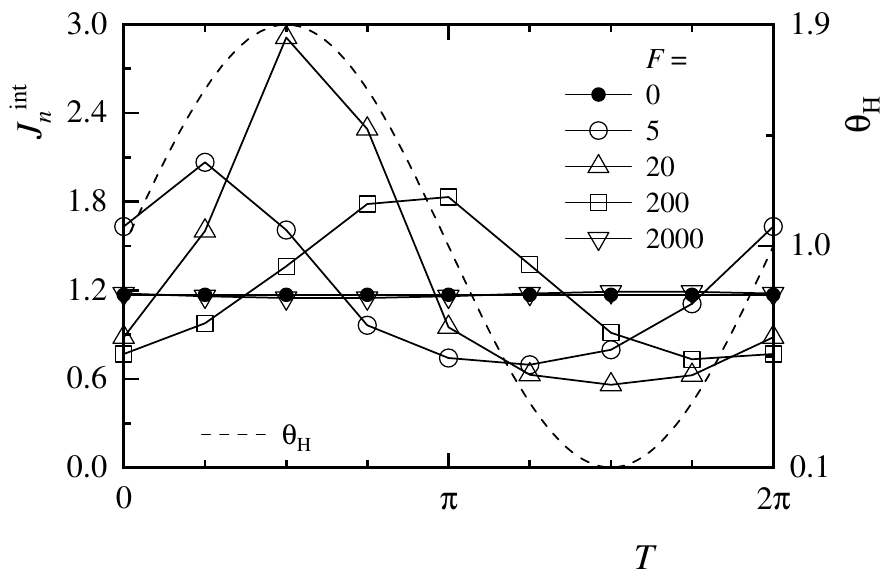} \\
\vspace*{-0.2\baselineskip}
(a) Bacteria \\
\end{minipage}
\begin{minipage}{0.48\linewidth}
\centering
\includegraphics[trim=0mm 0mm 0mm 0mm, clip, width=80mm]{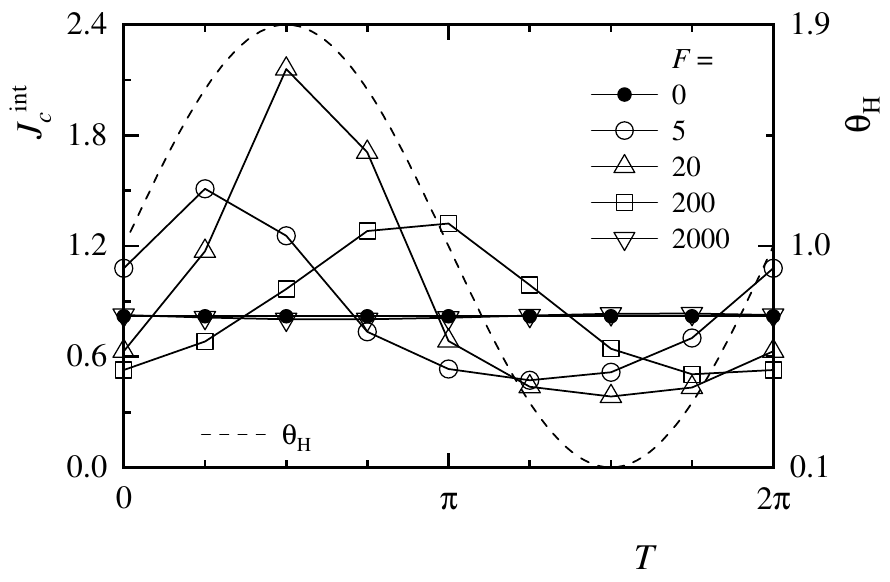}\\
\vspace*{-0.2\baselineskip}
(b) Oxygen
\end{minipage}
\caption{Time variations of integral values of total flux of bacteria and oxygen for $\Gamma = 1000$, $Ra = 500$, and $\Theta_\mathrm{A} = 0.9$.}
\label{time_vari_fluxint_f5_2000}
\end{figure}

We investigate the changes in convective characteristics over time at different frequencies $F$. 
For $\Gamma = 1000$, $Ra = 500$, $\Theta_\mathrm{A} = 0.9$, and $F = 0$, $5$, $20$, $200$, and $2000$, 
the time variations in the integrals of the downward flow velocity and kinetic energy 
$|v_\mathrm{down}^\mathrm{int}|$ and $KE^\mathrm{int}$ are shown in Fig. \ref{time_vari_vdown_KE_f5_2000}, 
and the integrals of the magnitude of the total bacterial and oxygen flux vectors 
$J_{n}^\mathrm{int}$ and $J_{c}^\mathrm{int}$ are shown in Fig. \ref{time_vari_fluxint_f5_2000}. 
As we impose the temperature fluctuation using a sine function with a period of $2 \pi$, 
the wall temperature reaches a maximum at time $T = 2 \pi/4$ and a minimum at $T = 6 \pi/4$. 
For $F = 20$, the thermo-bioconvection follows the temperature fluctuation well. 
At $T = 2 \pi/4$, the convection is strengthened, 
resulting in maximum values for $|v_\mathrm{down}^\mathrm{int}|$, 
$KE^\mathrm{int}$, $J_{n}^\mathrm{int}$, and $J_{c}^\mathrm{int}$. 
Conversely, at $T = 6 \pi/4$, convection weakens, 
leading to minimum valus for $|v_\mathrm{down}^\mathrm{int}|$, 
$KE^\mathrm{int}$, $J_{n}^\mathrm{int}$, and $J_{c}^\mathrm{int}$. 
Compared with the steady heating result of $F = 0$, 
the values for $|v_\mathrm{down}^\mathrm{int}|$, $KE^\mathrm{int}$, 
$J_{n}^\mathrm{int}$, and $J_{c}^\mathrm{int}$ increase by up to $160\%$, 
$546\%$, $149\%$, and $162\%$, respectively, at $T = 2 \pi/4$, 
and decrease by up to $54\%$, $78\%$, $52\%$, and $51\%$, respectively, at $T = 6 \pi/4$. 
Thus, it is found that the enhancement of convection resulting from temperature fluctuations 
is significantly higher for transport properties than the attenuation of convection. 
In the steady thermo-bioconvection shown in Fig. \ref{int_vdwon_flux_rt0_1000}, 
the interaction between bioconvection and thermal convection was stronger at high $Ra$ than at low $Ra$. 
Under unsteady heating, 
as the local thermal Rayleigh number momentarily increases at $T = 2 \pi/4$, 
the interaction is strengthened, 
leading to stronger convection compared to when $F = 0$. 
When the frequency increases to $F = 200$, 
the ability of thermo-bioconvection to follow the temperature fluctuation deteriorates. 
Thus, a phase difference of $2 \pi/4$ occurs 
between the time $T = 2 \pi/4$ when the temperature reaches its maximum 
and the times when $|v_\mathrm{down}^\mathrm{int}|$, $KE^\mathrm{int}$, $J_{n}^\mathrm{int}$, and $J_{c}^\mathrm{int}$ reach their maximums. 
When the frequency reaches $F = 2000$, 
the time variations in $|v_\mathrm{down}^\mathrm{int}|$, 
$KE^\mathrm{int}$, $J_{n}^\mathrm{int}$, and $J_{c}^\mathrm{int}$ become minimal, 
and are almost consistent with the results for $F = 0$. 
At $F = 2000$, the ability of thermo-bioconvection to follow the temperature fluctuation becomes even worse. 
This weakens the impact of temperature fluctuations on the suspension near the water surface 
and significantly decreases the time change in the strength of the interaction 
between bioconvection and thermal convection. 
At low frequency $F = 5$, 
$|v_\mathrm{down}^\mathrm{int}|$, $KE^\mathrm{int}$, 
$J_{n}^\mathrm{int}$, and $J_{c}^\mathrm{int}$ reach a maximum at $T = \pi/4$ 
and decrease at $T = 2 \pi/4$. 
To further understand this cause, 
we investigate the time change of the integral value of the bacterial concentration.

\begin{figure}[!t]
\centering
\begin{minipage}{0.48\linewidth}
\centering
\includegraphics[trim=0mm 0mm 0mm 0mm, clip, width=80mm]{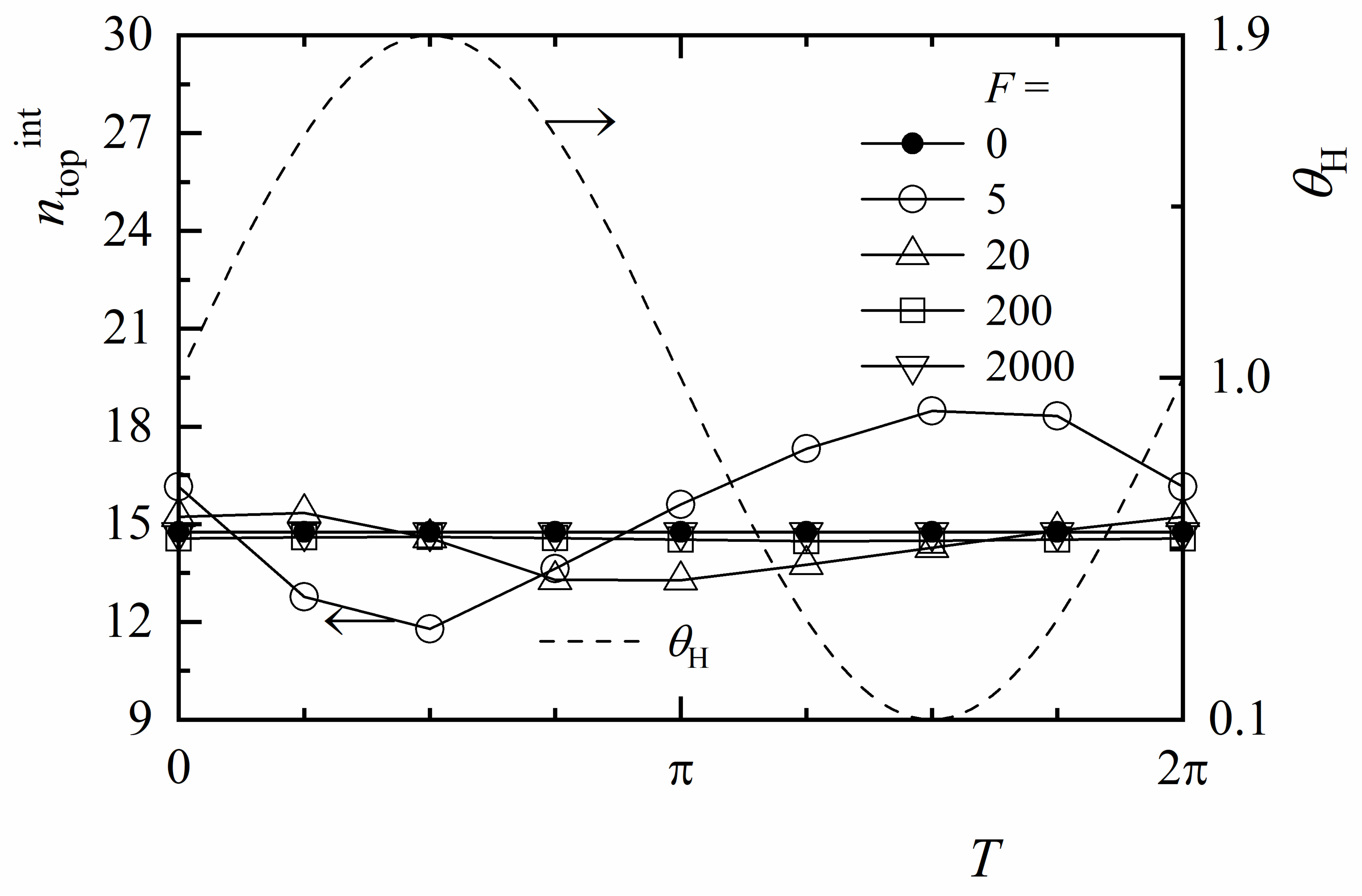}\\
\vspace*{-0.2\baselineskip}
(a) Top \\
\end{minipage}
\begin{minipage}{0.48\linewidth}
\centering
\includegraphics[trim=0mm 0mm 0mm 0mm, clip, width=80mm]{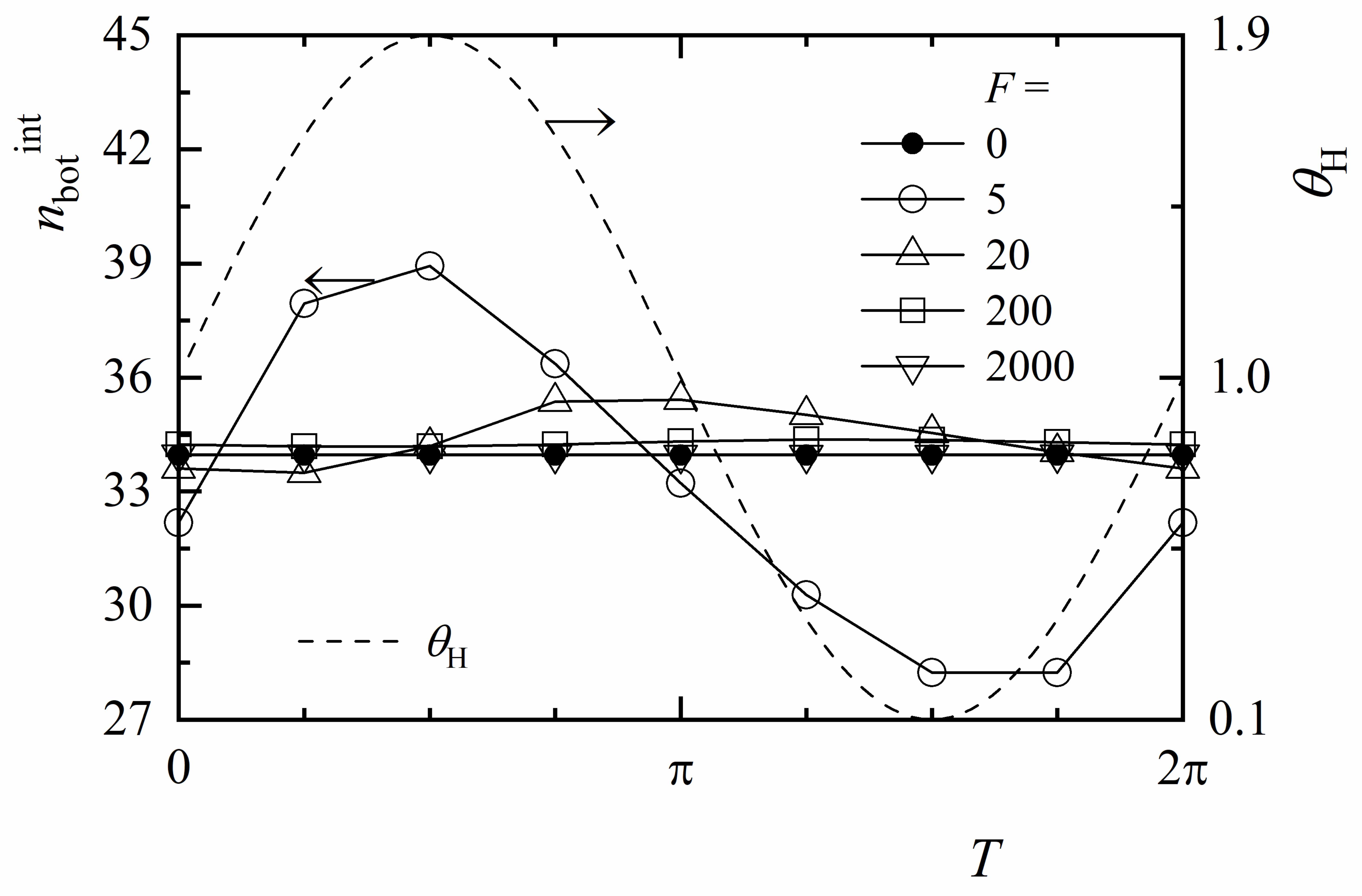}\\
\vspace*{-0.2\baselineskip}
(b) Bottom
\end{minipage}
\caption{Time variations of integral values of bacterial concentrations near surface and bottom 
for $\Gamma = 1000$, $Ra = 500$, and $\Theta_\mathrm{A} = 0.9$.}
\label{time_vari_cint_top_f5_2000}
\centering
\end{figure}

Figure \ref{time_vari_cint_top_f5_2000} shows the time variation 
in the integral value of the bacterial concentration 
under the same conditions as above. 
Figures \ref{time_vari_cint_top_f5_2000}(a) and (b) display 
the integral values near the water surface $y/h = 0.9 - 1.0$ 
and near the bottom $y/h = 0.0 - 0.3$, respectively. 
At the low frequency of $F = 5$, 
there is enough time for the bacteria to be transported during one period 
so that at the time $T = \pi/4$ when the total bacterial flux is at its maximum, 
the bacteria at the upper part are transported by convection, 
decreasing the concentration at the upper part 
and increasing the concentration at the lower part. 
Thus, the instability of the suspension due to density difference weakens, 
and $|v_\mathrm{down}^\mathrm{int}|$ decreases at later times, 
despite the increase in the wall temperature. 
After that, the bacterial concentration at the top side increases over time, 
and the instability of the suspension strengthens; 
thus, at $T = 5 \pi/4 - 6 \pi/4$, $|v_\mathrm{down}^\mathrm{int}|$ increases, 
despite the decrease in the wall temperature. 
We have also confirmed that the integral value of the upward flow, 
$|v_\mathrm{up}^\mathrm{int}|$, shows a similar trend to the integral value of the downward flow, $|v_\mathrm{down}^\mathrm{int}|$.

These results indicate the existence of an optimal frequency of temperature fluctuation 
that momentarily and significantly improves the convective characteristics of thermo-bioconvection. 
In other words, it has become clear that even in thermo-bioconvection, 
there is a resonance phenomenon in which the time change in convective characteristics 
significantly increases at a specific frequency band. 
Similarly to existing research on thermal convection \citep{Paolucci&Chenoweth_1989, Kwak&Hyun_1996, Kwak_et_al_1998}, 
it is considered that a resonance phenomenon occurs 
when the frequencies of internal gravity waves and temperature fluctuations match.

\begin{figure}[!t]
\centering
\begin{minipage}{0.48\linewidth}
\centering
\includegraphics[trim=0mm 0mm 0mm 0mm, clip, width=80mm]{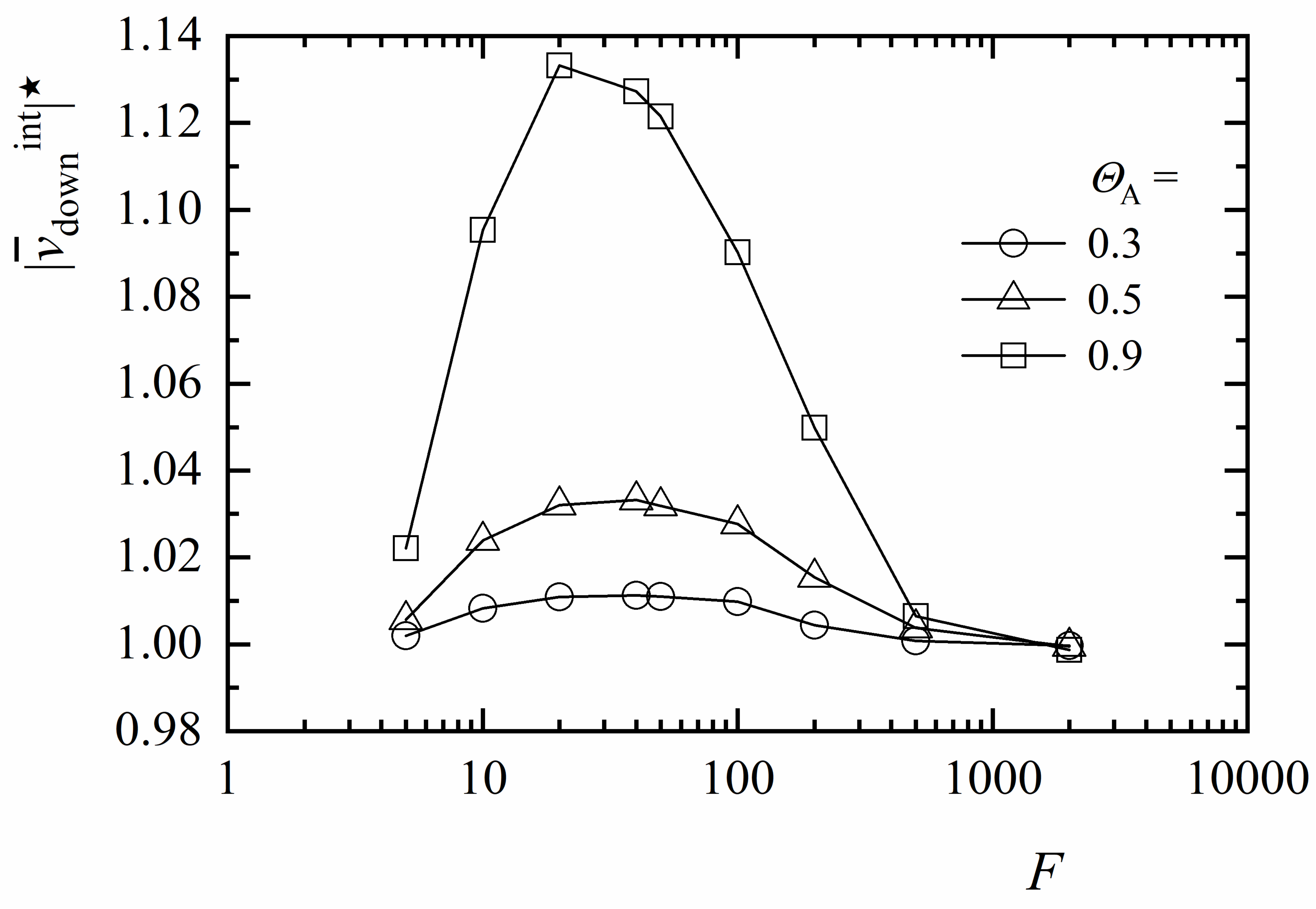} \\
\vspace*{-0.2\baselineskip}
(a) Downward velocity \\
\end{minipage}
\begin{minipage}{0.48\linewidth}
\centering
\includegraphics[trim=0mm 0mm 0mm 0mm, clip, width=80mm]{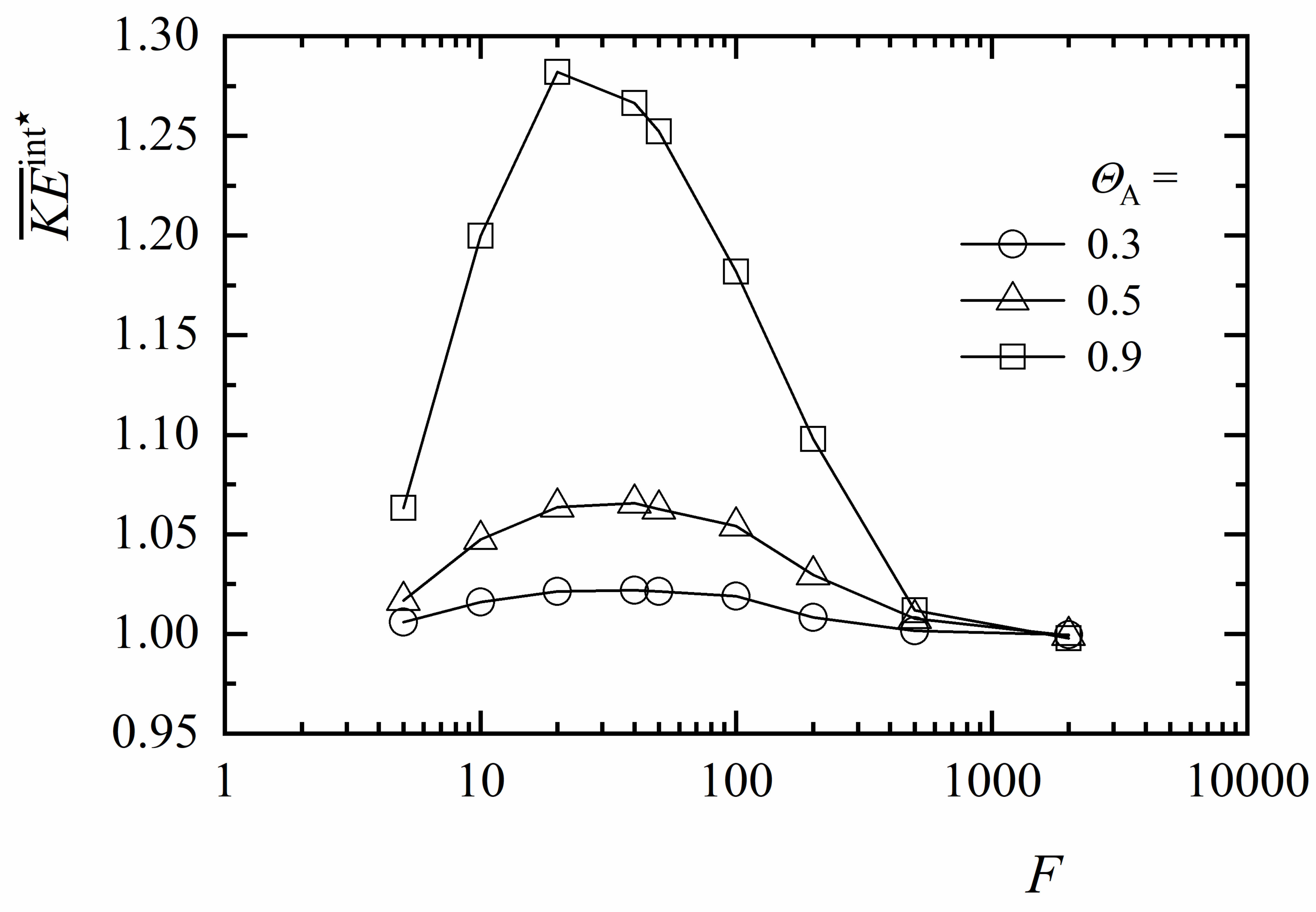}\\
\vspace*{-0.2\baselineskip}
(b) Kinetic energy
\end{minipage}
\caption{Growth rates of time--averaged integral values of downward velocity and kinetic energy for $\Gamma = 1000$ and $Ra = 500$.}
\label{time_ave_vdown_KE_A}
\end{figure}

\begin{figure}[!t]
\centering
\begin{minipage}{0.48\linewidth}
\centering
\includegraphics[trim=0mm 0mm 0mm 0mm, clip, width=80mm]{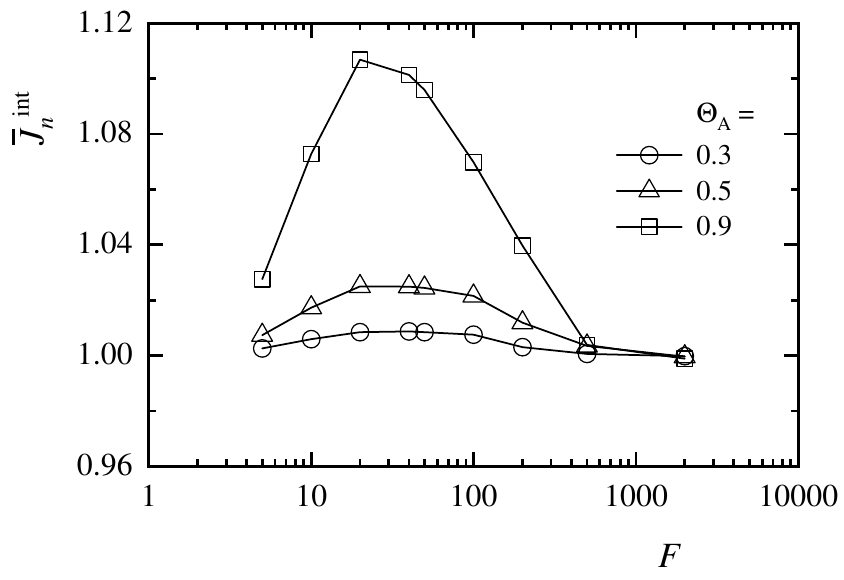} \\
\vspace*{-0.2\baselineskip}
(a) Bacteria \\
\end{minipage}
\begin{minipage}{0.48\linewidth}
\centering
\includegraphics[trim=0mm 0mm 0mm 0mm, clip, width=80mm]{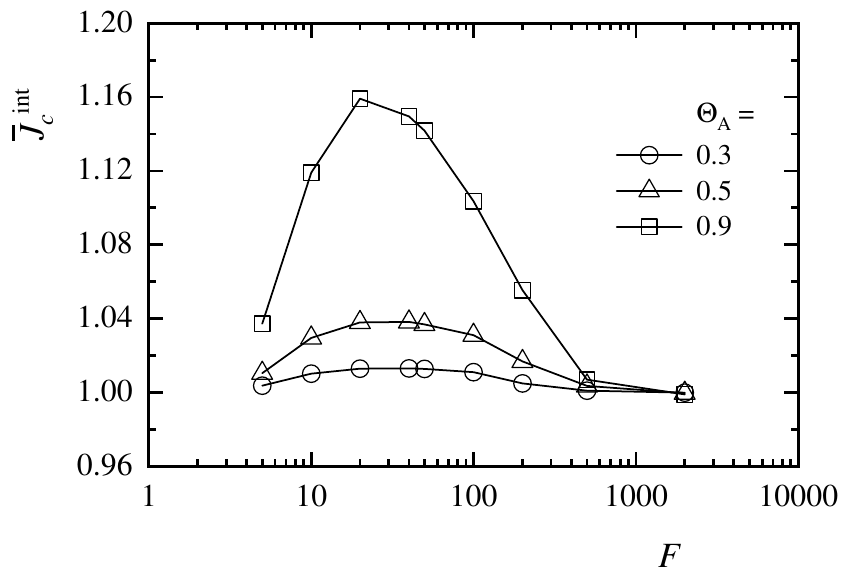}\\
\vspace*{-0.2\baselineskip}
(b) Oxygen
\end{minipage}
\caption{Growth rates of time--averaged integral values of total fluxes of bacteria and oxygen for $\Gamma = 1000$ and $Ra = 500$.}
\label{time_ave_fluxint_A}
\end{figure}

We clarify the changes in time-averaged convective characteristics 
at different frequencies $F$ for the amplitudes $\Theta_\mathrm{A} = 0.3$, $0.5$, and $0.9$. 
We plotted the increase rates $|\overline{v}_\mathrm{down}^\mathrm{int}|^{\star}$ 
and $\overline{KE}^{\mathrm{int}^{\star}}$ of the integrals of the time-averaged downward velocity 
and kinetic energy at $\Gamma = 1000$ and $Ra = 500$ in Fig. \ref{time_ave_vdown_KE_A}, 
and the increase rates $\overline{J}_{n}^{\mathrm{int}^{\star}}$ and $\overline{J}_{c}^{\mathrm{int}^{\star}}$ 
of the integrals of the magnitude of the time-averaged total bacterial and oxygen flux vectors 
in Fig. \ref{time_ave_fluxint_A}. 
Bsed on the result shown in Fig. \ref{time_ave_vdown_KE_A}, 
we observe that within the frequency range of $F = 20 - 40$, 
the values of $|\overline{v}_\mathrm{down}^\mathrm{int}|^{\star}$, 
$\overline{KE}^{\mathrm{int}^{\star}}$, $\overline{J}_{n}^{\mathrm{int}^{\star}}$, 
and $\overline{J}_{c}^{\mathrm{int}^{\star}}$ reach their maximum, 
regardless of the value of $\Theta_\mathrm{A}$. 
At $\Theta_\mathrm{A} = 0.9$ and $F = 20$, 
$|\overline{v}_\mathrm{down}^\mathrm{int}|$, $\overline{KE}^\mathrm{int}$, 
$\overline{J}_{n}^\mathrm{int}$, and $\overline{J}_{c}^\mathrm{int}$ 
significantly increase to the values of $13.3\%$, $28.2\%$, $10.1\%$, and $15.7\%$, respectively, 
compared to the results for steady heating. 
In this way, it is found that there exists the resonant frequency 
at which $|\overline{v}_\mathrm{down}^\mathrm{int}|$, $\overline{KE}^\mathrm{int}$, 
$\overline{J}_{n}^{\mathrm{int}^{\star}}$, and $\overline{J}_{c}^{\mathrm{int}^{\star}}$ are maximum 
and that $\Theta_\mathrm{A}$ does not significantly affect this resonant frequency. 
Furthermore, as $\Theta_\mathrm{A}$ increases, 
the increases in $|\overline{v}_\mathrm{down}^{\mathrm{int}}|^{\star}$, 
$\overline{KE}^{\mathrm{int}^{\star}}$, $\overline{J}_{n}^{\mathrm{int}^{\star}}$, 
and $\overline{J}_{c}^{\mathrm{int}^{\star}}$ at the resonant frequency 
become more noticeable. 
This is because, as $\Theta_\mathrm{A}$ increases, the temperature gradient becomes steeper 
when the temperature reaches its maximum, and the buoyancy force generated becomes larger, 
resulting in stronger interference between bioconvection and thermal convection. 
As the value of $F$ increases from 5 to 20, 
the magnitudes of $|\overline{v}_\mathrm{down}^{\mathrm{int}}|^{\star}$, 
$\overline{KE}^{\mathrm{int}^{\star}}$, $J_{n}^\mathrm{int}$, 
and $J_{c}^\mathrm{int}$ also increase with increasing frequency. 
This is because, as $F$ increases, one period of each temperature fluctuation becomes shorter, 
and at a time when the local thermal Rayleigh number momentarily increases, 
the transport of bacteria from the upper to lower side decreases, 
maintaining the instability of the suspension due to density differences 
and strengthening convection. 
Conversely, in the frequency band of $F = 40 - 2000$, 
the response of thermo-bioconvection to each temperature fluctuation worsens, 
and the enhancement of convection resulting from temperature fluctuations is suppressed. 
Hence, $|\overline{v}_\mathrm{down}^{\mathrm{int}}|^{\star}$, 
$\overline{KE}^{\mathrm{int}^{\star}}$, $\overline{J}_{n}^{\mathrm{int}^{\star}}$, 
and $\overline{J}_{c}^{\mathrm{int}^{\star}}$ decrease with increasing $F$. 
At the high frequency of $F = 2000$, especially, 
the ability of thermo-bioconvection to follow the temperature fluctuation is at its worst, 
and the time variations in convection velocity and convective transport decrease significantly, 
so that $|\overline{v}_\mathrm{down}^{\mathrm{int}}|^{\star}$, 
$\overline{KE}^{\mathrm{int}^{\star}}$, $\overline{J}_{n}^{\mathrm{int}^{\star}}$, 
and $\overline{J}_{c}^{\mathrm{int}^{\star}}$ are approximately $1$.

\begin{figure}[!t]
\centering
\begin{minipage}{0.48\linewidth}
\centering
\includegraphics[trim=0mm 0mm 0mm 0mm, clip, width=80mm]{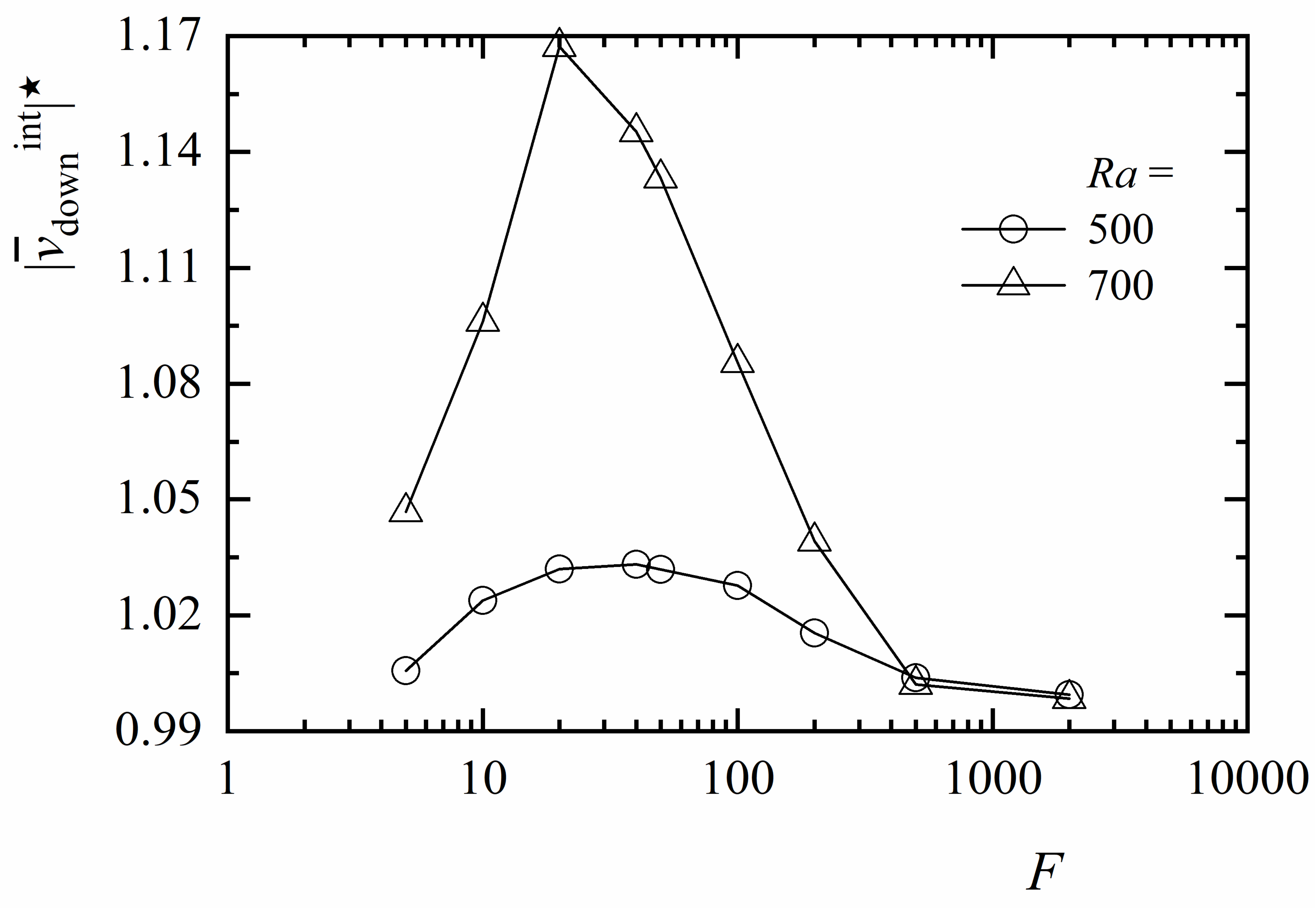} \\
\vspace*{-0.2\baselineskip}
(a) Downward velocity \\
\end{minipage}
\begin{minipage}{0.48\linewidth}
\centering
\includegraphics[trim=0mm 0mm 0mm 0mm, clip, width=80mm]{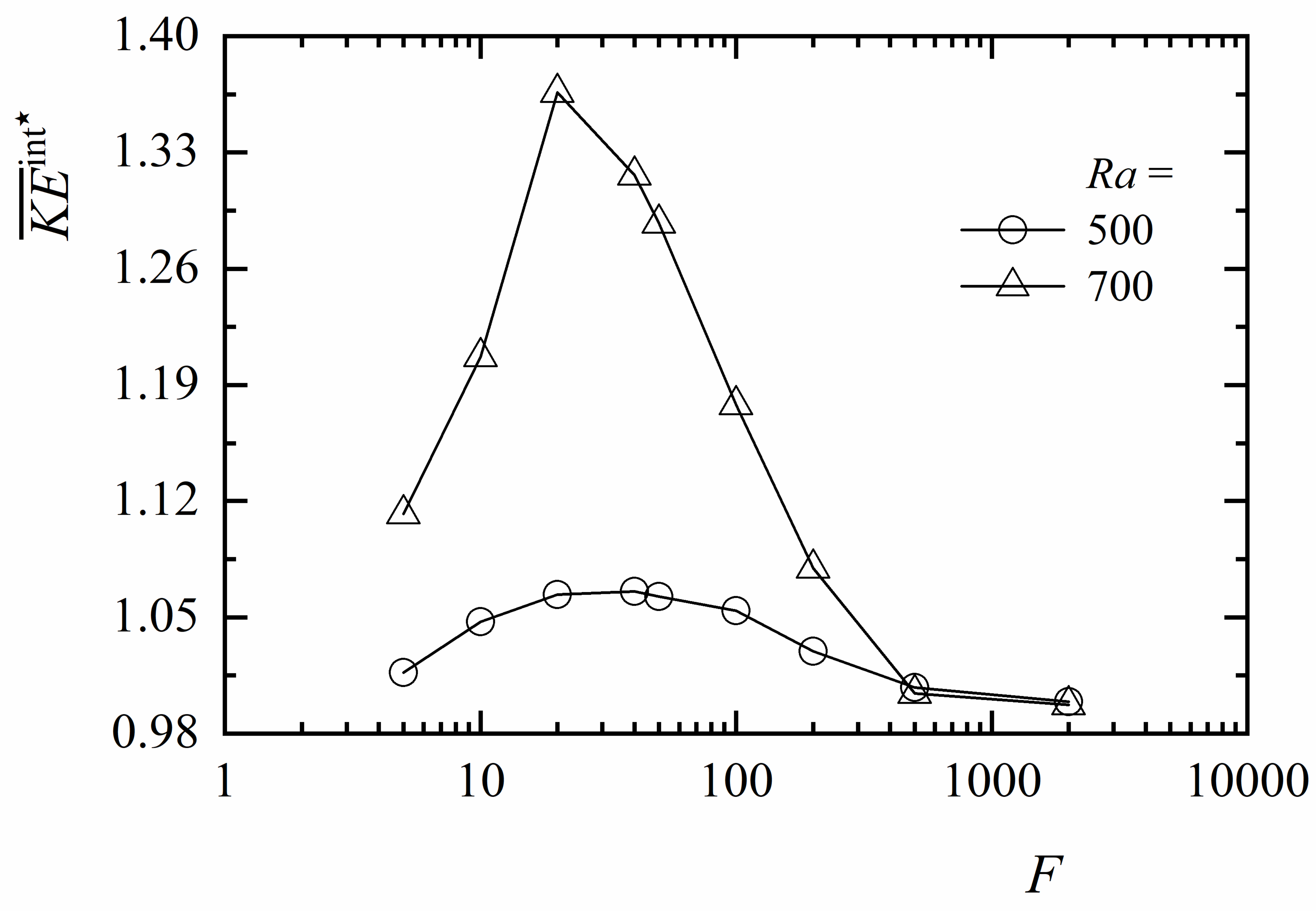}\\
\vspace*{-0.2\baselineskip}
(b) Kinetic energy
\end{minipage}
\caption{Growth rates of time--averaged integral values of downward velocity and kinetic energy for $\Gamma = 1000$ and $Ra = 500$.}
\label{time_ave_vdown_KE_Ra}
\end{figure}

\begin{figure}[!t]
\centering
\begin{minipage}{0.48\linewidth}
\centering
\includegraphics[trim=0mm 0mm 0mm 0mm, clip, width=80mm]{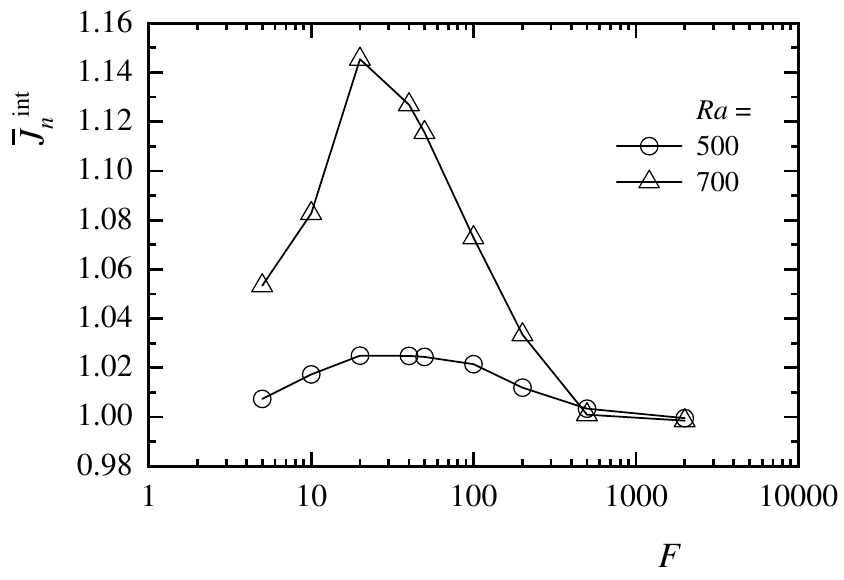} \\
\vspace*{-0.2\baselineskip}
(a) Bacteria \\
\end{minipage}
\begin{minipage}{0.48\linewidth}
\centering
\includegraphics[trim=0mm 0mm 0mm 0mm, clip, width=80mm]{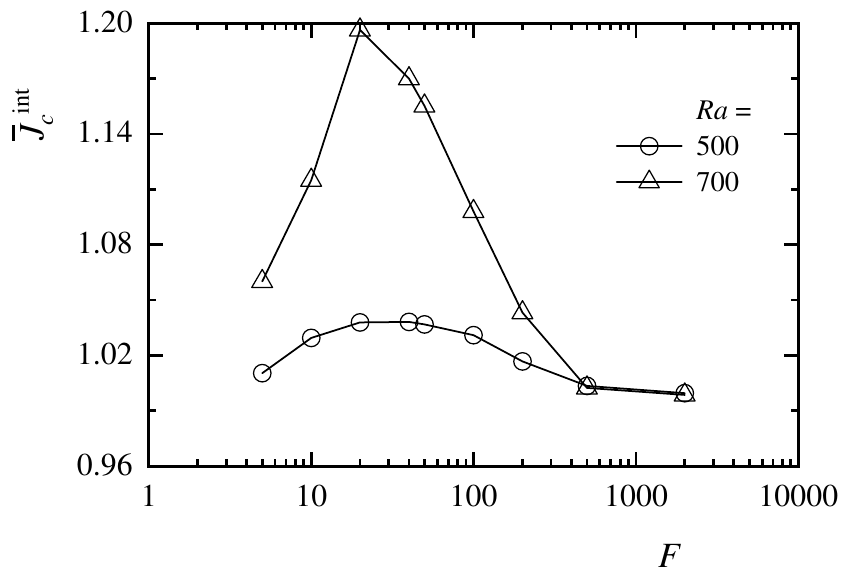}\\
\vspace*{-0.2\baselineskip}
(b) Oxygen
\end{minipage}
\caption{Growth rates of time--averaged integral values of total fluxes of bacteria and oxygen for $\Gamma = 1000$, and $\Theta_\mathrm{A} = 0.5$.}
\label{time_ave_fluxint_Ra}
\end{figure}

Subsequently, we clarify the variations in the time-averaged convective characteristics 
with the difference in $F$ for $Ra = 500$ and $700$. 
We plotted the increase rates $|\overline{v}_\mathrm{down}^\mathrm{int}|^{\star}$ 
and $\overline{KE}^{\mathrm{int}^{\star}}$ 
of the integrals of the time-averaged downward velocity and kinetic energy 
at $\Gamma = 1000$ and $\Theta_\mathrm{A} = 0.5$ in Fig. \ref{time_ave_vdown_KE_Ra}, 
and the increase rates $\overline{J}_{n}^{\mathrm{int}^{\star}}$ and $\overline{J}_{c}^{\mathrm{int}^{\star}}$ 
of the integrals of the magnitude of the time-averaged total bacterial and oxygen flux vectors 
in Fig. \ref{time_ave_fluxint_Ra}. 
The impact of $Ra$ on the time-averaged characteristics of convection and transport 
resembles the trend in Figs. \ref{time_ave_vdown_KE_A} and \ref{time_ave_fluxint_A}, 
and with increasing $Ra$, the increases in $|\overline{v}_\mathrm{down}^{\mathrm{int}}|^{\star}$, 
$\overline{KE}^{\mathrm{int}^{\star}}$, $\overline{J}_{n}^{\mathrm{int}^{\star}}$, 
and $\overline{J}_{c}^{\mathrm{int}^{\star}}$ are more pronounced 
at the resonant frequency. 
Specifically, for $Ra$ = 700, 
$|\overline{v}_\mathrm{down}^\mathrm{int}|$, $\overline{KE}^\mathrm{int}$, 
$\overline{J}_{n}^\mathrm{int}$, and $\overline{J}_{c}^\mathrm{int}$ at $F = 20$ 
increase to $16.7\%$, $36.6\%$, $14.5\%$, and $19.4\%$, respectively, 
compared to the results for steady heating. 
Therefore, it is clear that convection and transport characteristics are significantly enhanced 
at the resonant frequency by the wall temperature fluctuation.

\subsubsection{The effect of temperature fluctuations on the structure of thermo-bioconvection}

\begin{figure}[!t]
\centering
\begin{minipage}{0.48\linewidth}
\centering
\includegraphics[trim=0mm 0mm 0mm 0mm, clip, width=80mm]{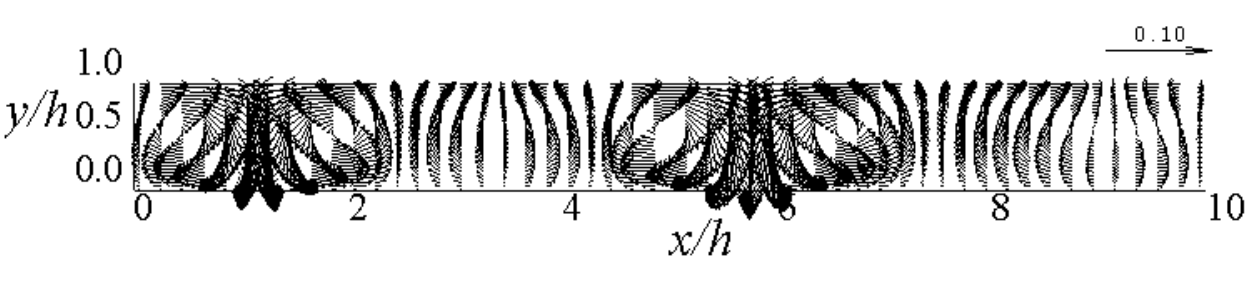} \\
\end{minipage}
\begin{minipage}{0.48\linewidth}
\centering
\includegraphics[trim=0mm 0mm 0mm 0mm, clip, width=80mm]{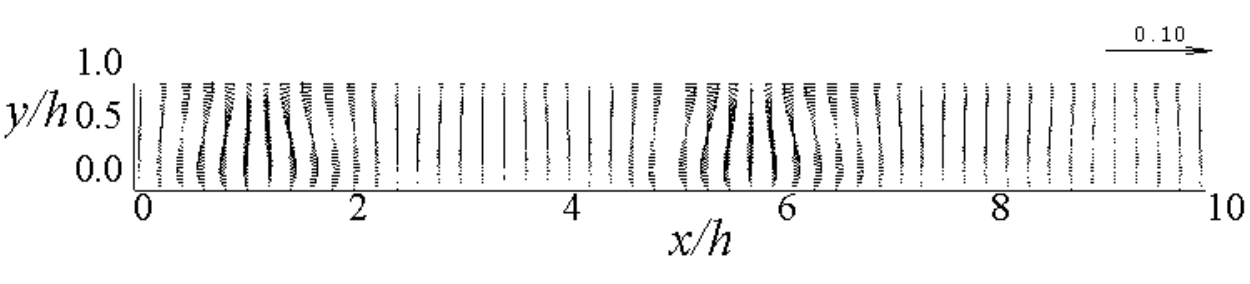} \\
\end{minipage}
\\ (a) Velocity vectors \\
\begin{minipage}{0.48\linewidth}
\centering
\includegraphics[trim=0mm 0mm 0mm 0mm, clip, width=80mm]{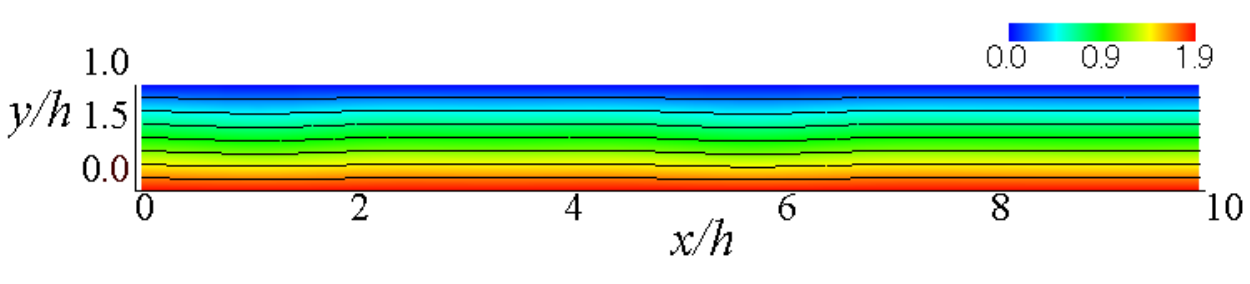} \\
\end{minipage}
\begin{minipage}{0.48\linewidth}
\centering
\includegraphics[trim=0mm 0mm 0mm 0mm, clip, width=80mm]{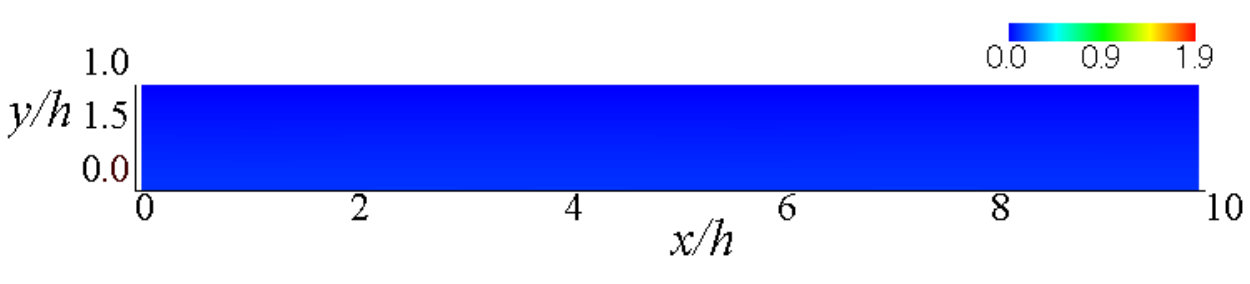} \\
\end{minipage}
\\ (b) Temperature \\
\begin{minipage}{0.48\linewidth}
\centering
\includegraphics[trim=0mm 0mm 0mm 0mm, clip, width=80mm]{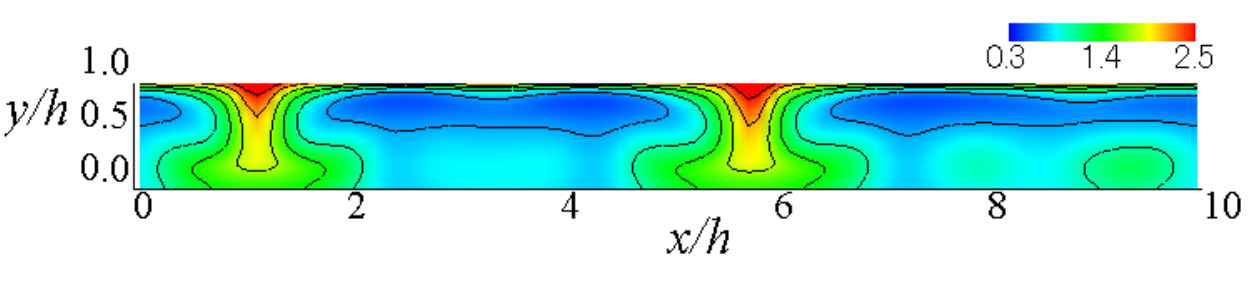} \\
\end{minipage}
\begin{minipage}{0.48\linewidth}
\centering
\includegraphics[trim=0mm 0mm 0mm 0mm, clip, width=80mm]{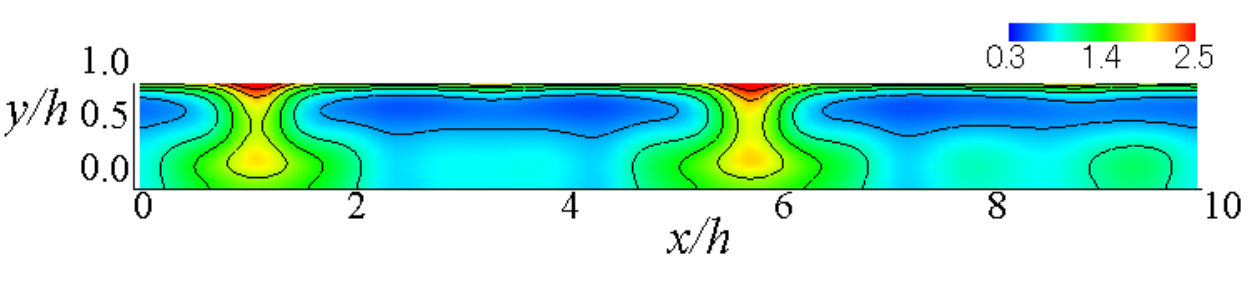} \\
\end{minipage}
\\ (c) Baceria \\
\begin{minipage}{0.48\linewidth}
\centering
\includegraphics[trim=0mm 0mm 0mm 0mm, clip, width=80mm]{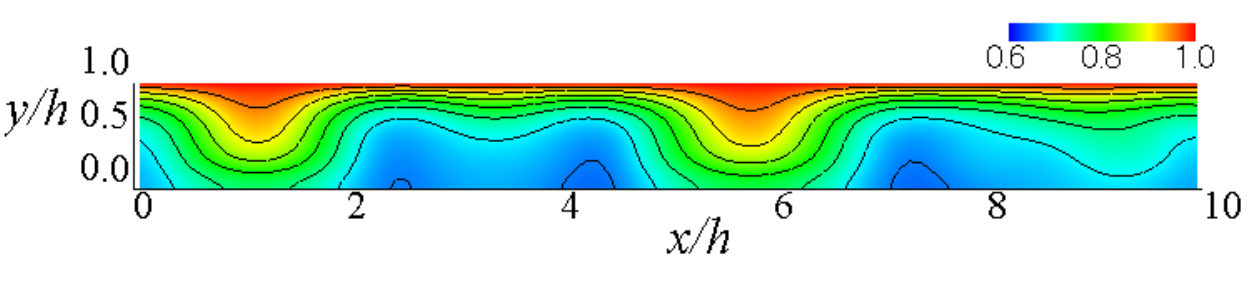} \\
\end{minipage}
\begin{minipage}{0.48\linewidth}
\centering
\includegraphics[trim=0mm 0mm 0mm 0mm, clip, width=80mm]{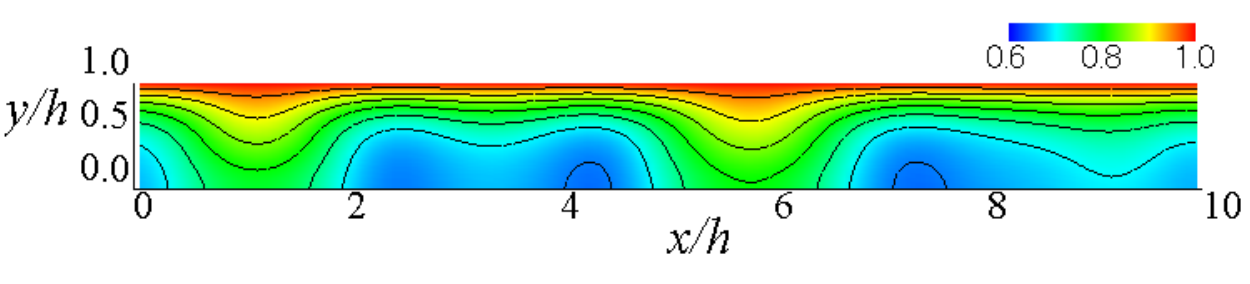} \\
\end{minipage}
\\ (d) Oxygen \\
\caption{Velocity vectors, and contours of bacterial concentration, oxygen and temperature
at $z/h = 2.75$ for $\Gamma = 1000$, $Ra = 500$, $\Theta_\mathrm{A} = 0.9$, and $F = 20$ at $T = 2\pi/4$(left) and $T = 6\pi/4$ (right): 
Contour intervals are 0.275 from 0.3 to 2.5 for bacteria, 0.04 from 0.6 to 1.0 for oxygen, and 0.19 from 0 to 1.9 for temperature.}
\label{time_vari_r1000rt500a90f20}
\end{figure}

\begin{figure}[!t]
\centering
\begin{minipage}{0.48\linewidth}
\centering
\includegraphics[trim=0mm 0mm 0mm 0mm, clip, width=80mm]{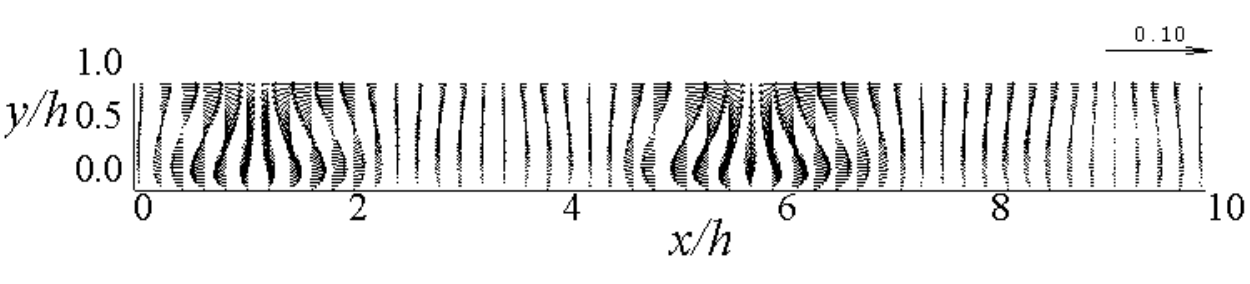} \\
\end{minipage}
\begin{minipage}{0.48\linewidth}
\centering
\includegraphics[trim=0mm 0mm 0mm 0mm, clip, width=80mm]{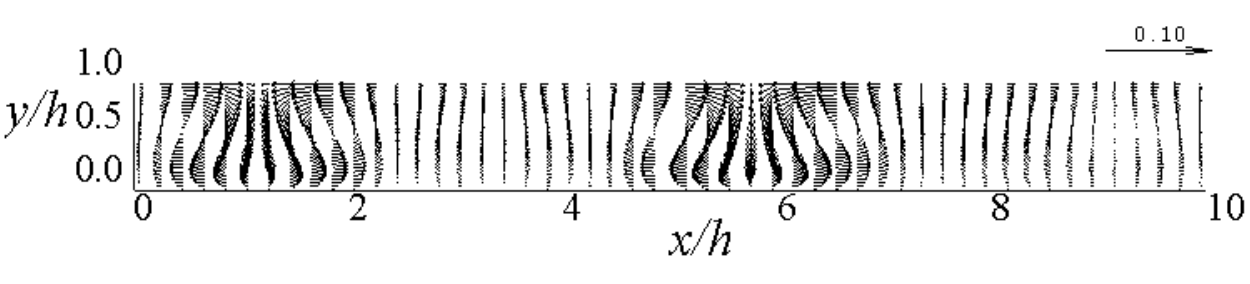} \\
\end{minipage}
\\ (a) Velocity vectors \\
\begin{minipage}{0.48\linewidth}
\centering
\includegraphics[trim=0mm 0mm 0mm 0mm, clip, width=80mm]{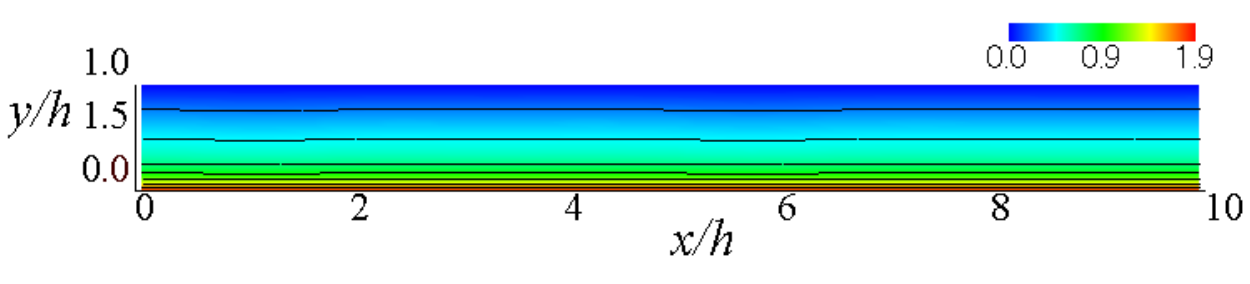} \\
\end{minipage}
\begin{minipage}{0.48\linewidth}
\centering
\includegraphics[trim=0mm 0mm 0mm 0mm, clip, width=80mm]{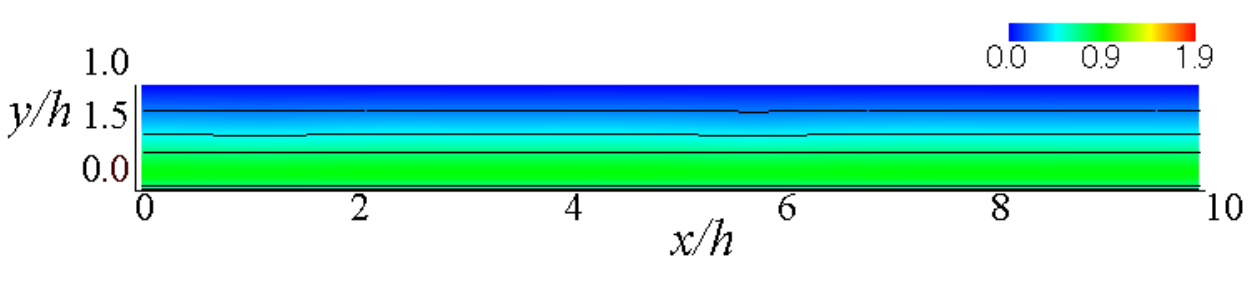} \\
\end{minipage}
\centering
\\ (b) Temperature \\
\begin{minipage}{0.48\linewidth}
\centering
\includegraphics[trim=0mm 0mm 0mm 0mm, clip, width=80mm]{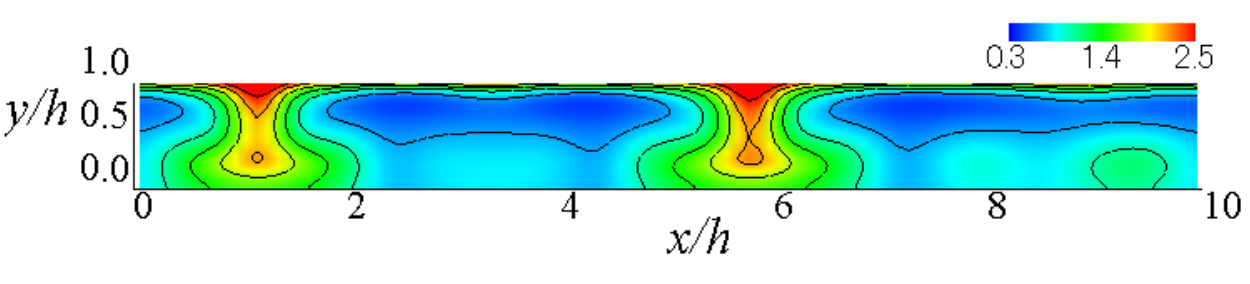} \\
\end{minipage}
\begin{minipage}{0.48\linewidth}
\centering
\includegraphics[trim=0mm 0mm 0mm 0mm, clip, width=80mm]{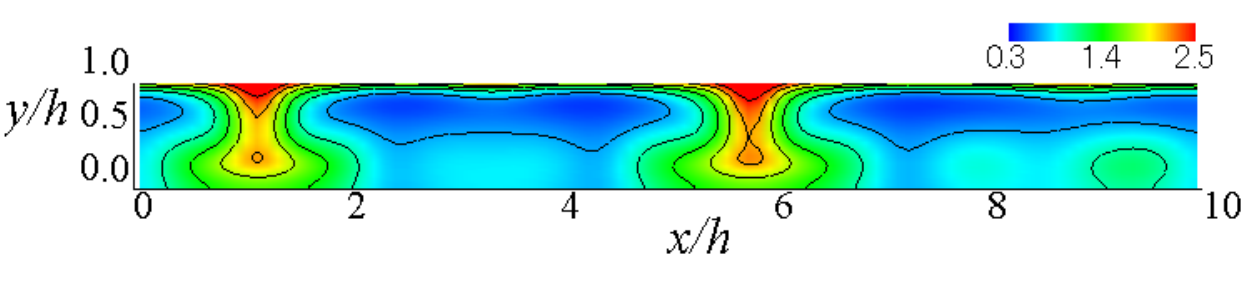} \\
\end{minipage}
\\ (c) Baceria \\
\begin{minipage}{0.48\linewidth}
\centering
\includegraphics[trim=0mm 0mm 0mm 0mm, clip, width=80mm]{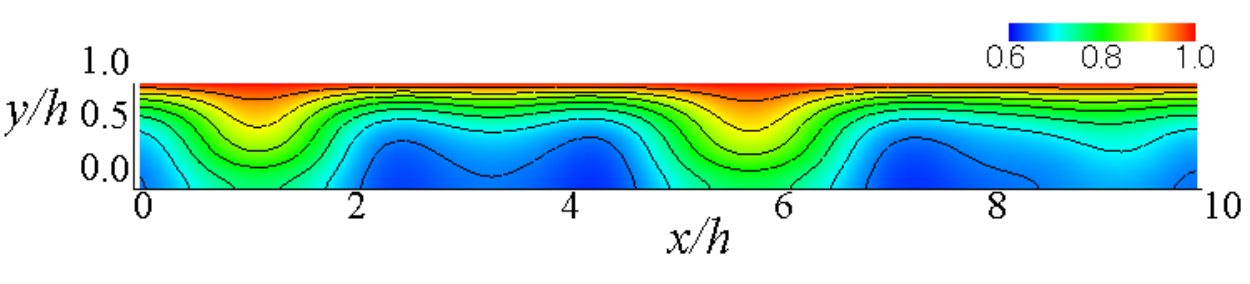} \\
\end{minipage}
\begin{minipage}{0.48\linewidth}
\centering
\includegraphics[trim=0mm 0mm 0mm 0mm, clip, width=80mm]{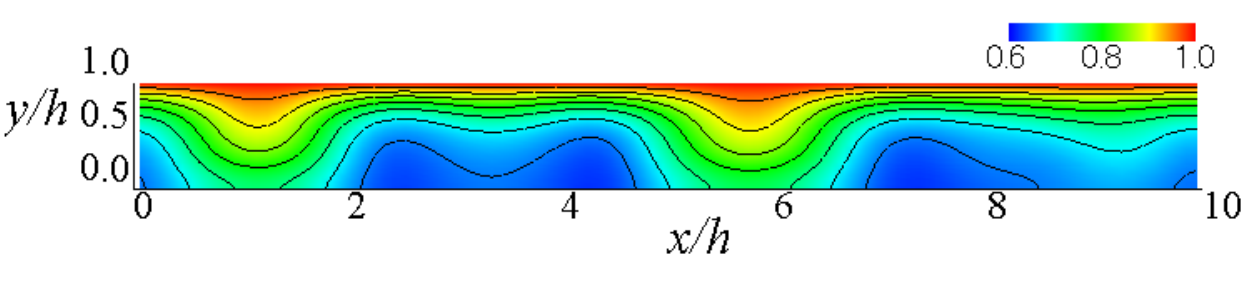} \\
\end{minipage}
\\ (d) Oxygen \\
\caption{Velocity vectors, and contours of bacterial concentration, oxygen and temperature
at $z/h = 2.75$ for $\Gamma = 1000$, $Ra = 500$, $\Theta_\mathrm{A} = 0.9$, and $F = 2000$ at $T = 2\pi/4$ (left) and $T = 6\pi/4$ (right): 
Contour intervals are 0.275 from 0.3 to 2.5 for bacteria, 0.04 from 0.6 to 1.0 for oxygen, and 0.19 from 0 to 1.9 for temperature.}
\label{time_vari_r1000rt500a90f2000}
\end{figure}

We consider the impact of temperature fluctuations at the lower wall on unsteady thermo-bioconvection. 
The calculation conditions are $\Gamma = 1000$, $Ra = 500$, 
$\Theta_\mathrm{A} = 0.9$, $F = 20$, and $2000$. 
First, the time evolution of the velocity, temperature, 
and concentration fields for $F = 20$ and $2000$ are plotted 
in Figs. \ref{time_vari_r1000rt500a90f20} and \ref{time_vari_r1000rt500a90f2000}, respectively. 
The left and right sides of the figure show the results at $T = 2 \pi/4$ and $T = 6 \pi/4$, respectively. 
At the low frequency of $F = 20$, 
the thermo-bioconvection responds well to the temperature fluctuation. 
At the time $T = 2 \pi/4$ when the temperature becomes maximum, 
the highest temperature gradient occurs, and the buoyancy increases. 
Hence, it is observed from the velocity field that intensive interference 
between thermal convection and bioconvection strengthens the convection. 
Additionally, the convective transport of substances strengthened 
so that the areas of high concentrations of bacteria and oxygen extend further 
from near the water surface toward the bottom. 
The regions of high bacterial and oxygen concentrations extend the most 
from near the water surface toward the bottom at $T = 3 \pi/4$, 
and a phase difference occurs between the time when convection is most promoted 
and the time when the regions of high concentrations extend the most. 
This is owing to a delay in the transport of bacteria and oxygen. 
In the temperature field, there is almost no heat transport by convection, 
and heat conduction is dominant. 
This is due to the high Lewis number $Le$. 
At the time $T = 6 \pi/4$ when the temperature reaches its minimum, 
as the temperature gradient is at its smallest and buoyancy force decreases, 
the interaction between thermal convection and bioconvection decays, 
and convection is the weakest. 
Compared to the results at $T = 2 \pi/4$, 
the regions of high bacterial and oxygen concentrations do not extend as far toward the bottom. 
The concentrations of bacteria and oxygen are higher near the bottom 
compared to the results for $T = 2 \pi/4$. 
This is due to a phase difference between the times when convection is most promoted 
and when the regions of high concentrations extend the most. 
Thus, at the low frequency of $F = 20$, 
the thermo-bioconvection responds well to the temperature fluctuation, 
leading to significant temporal changes in the velocity and concentration fields. 
At the high frequency of $F = 2000$, 
as the thermo-bioconvection does not respond well to the temperature fluctuation, 
the temperature changes at the wall do not diffuse sufficiently to the water surface; 
thus, it can be seen from the temperature field 
that the effect of temperature fluctuations on the suspension near the water surface is small. 
Additionally, the time variation in the convection velocity becomes low, 
and we can observe that the concentration distribution of bacteria and oxygen hardly changes with time.

\begin{figure}[!t]
\centering
\begin{minipage}{53mm}
\centering
\includegraphics[trim=0mm 10mm 0mm 0mm, clip, width=53mm]{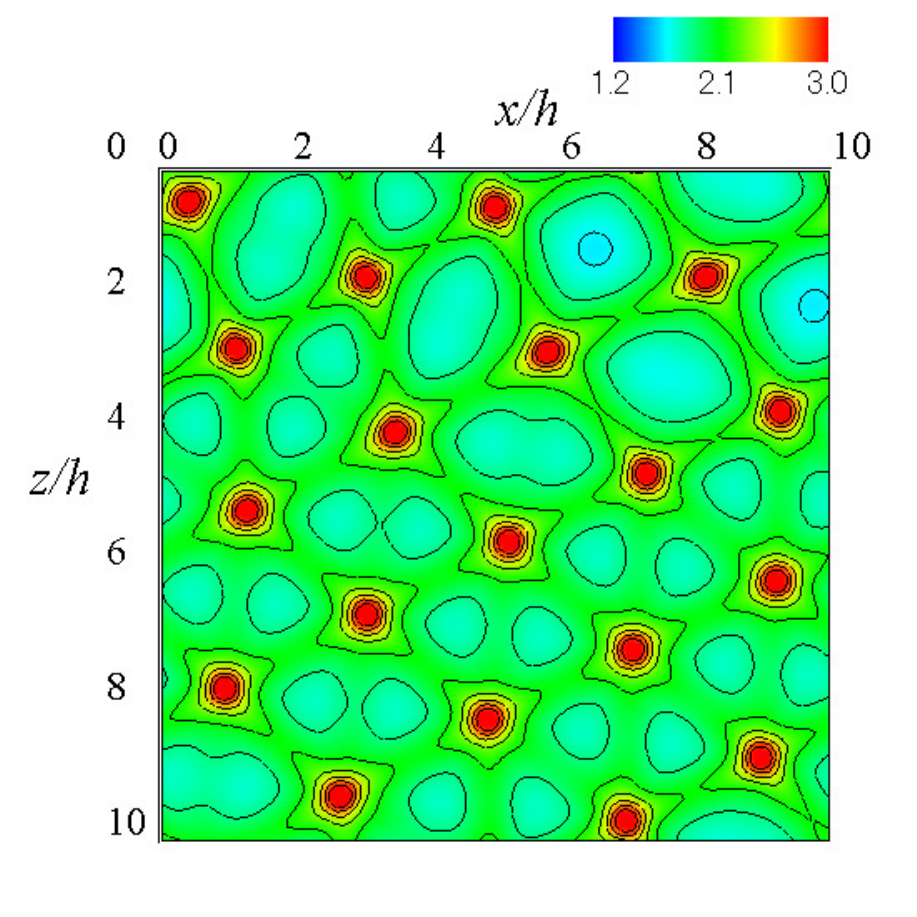} \\
(a) $F = 0$ \\
\end{minipage}
\begin{minipage}{53mm}
\centering
\includegraphics[trim=0mm 10mm 0mm 0mm, clip, width=53mm]{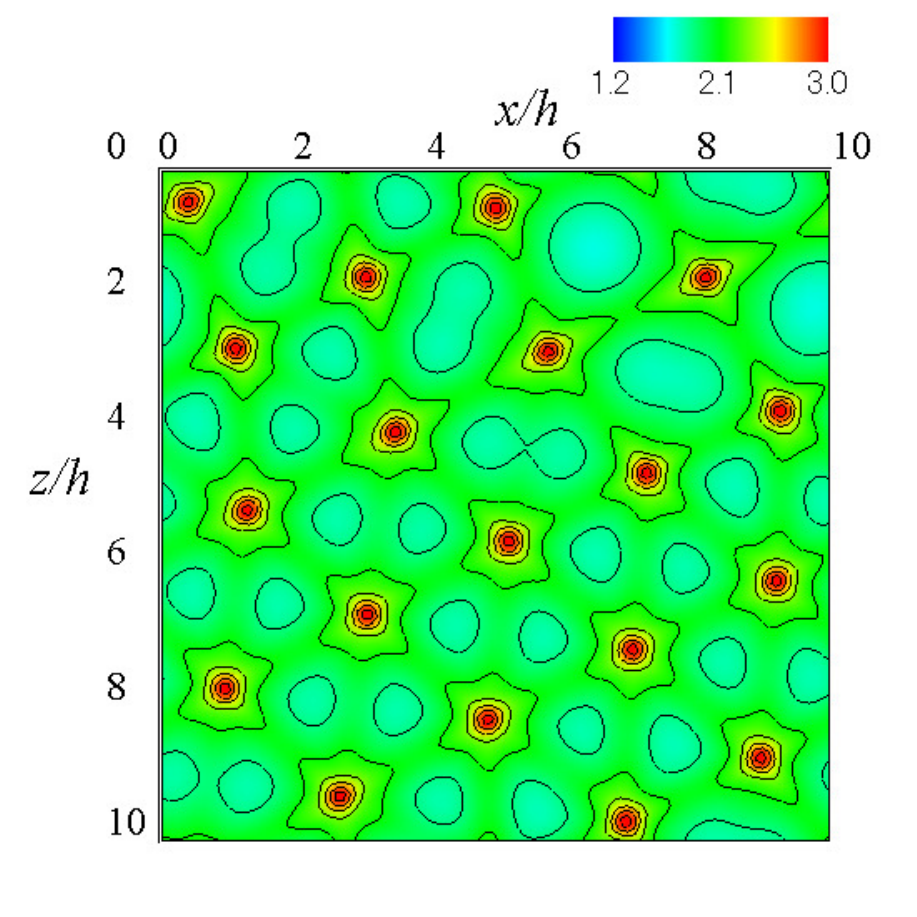} \\
(b) $F = 20$, $T = 2\pi/4$ \\
\end{minipage}
\begin{minipage}{53mm}
\centering
\includegraphics[trim=0mm 10mm 0mm 0mm, clip, width=53mm]{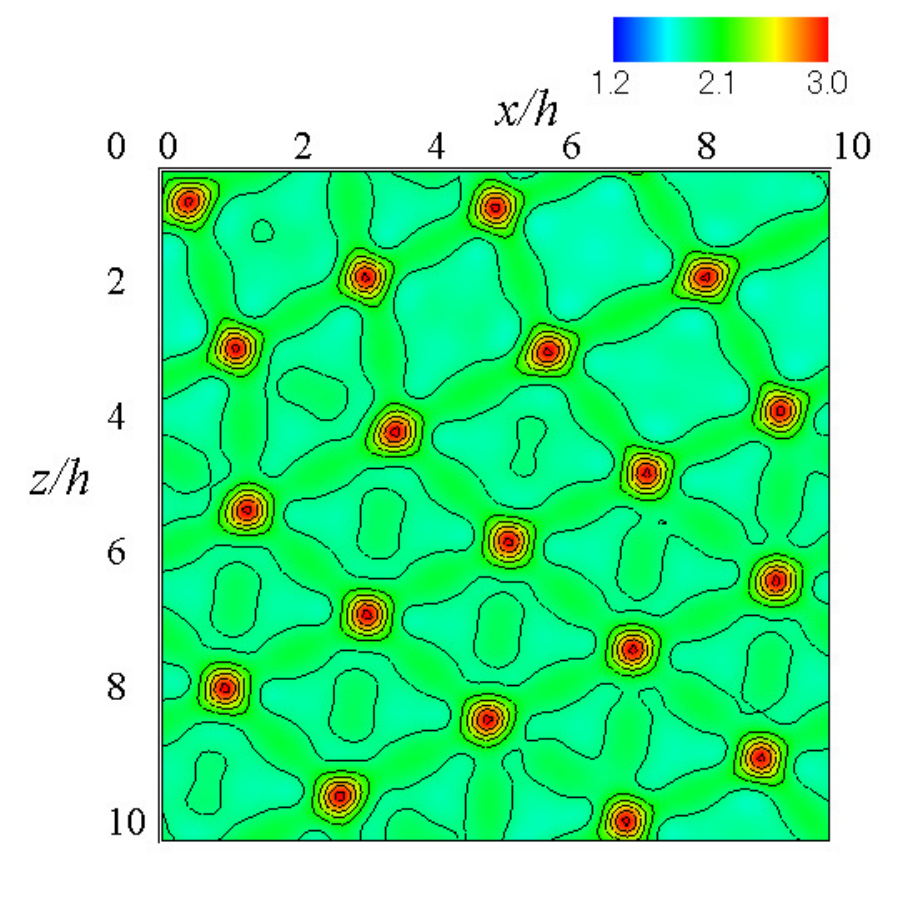} \\
(c) $F = 20$, $T = 6\pi/4$ \\
\end{minipage}

\vspace*{0.5\baselineskip}
\begin{minipage}{53mm}
\centering
\includegraphics[trim=0mm 10mm 0mm 0mm, clip, width=53mm]{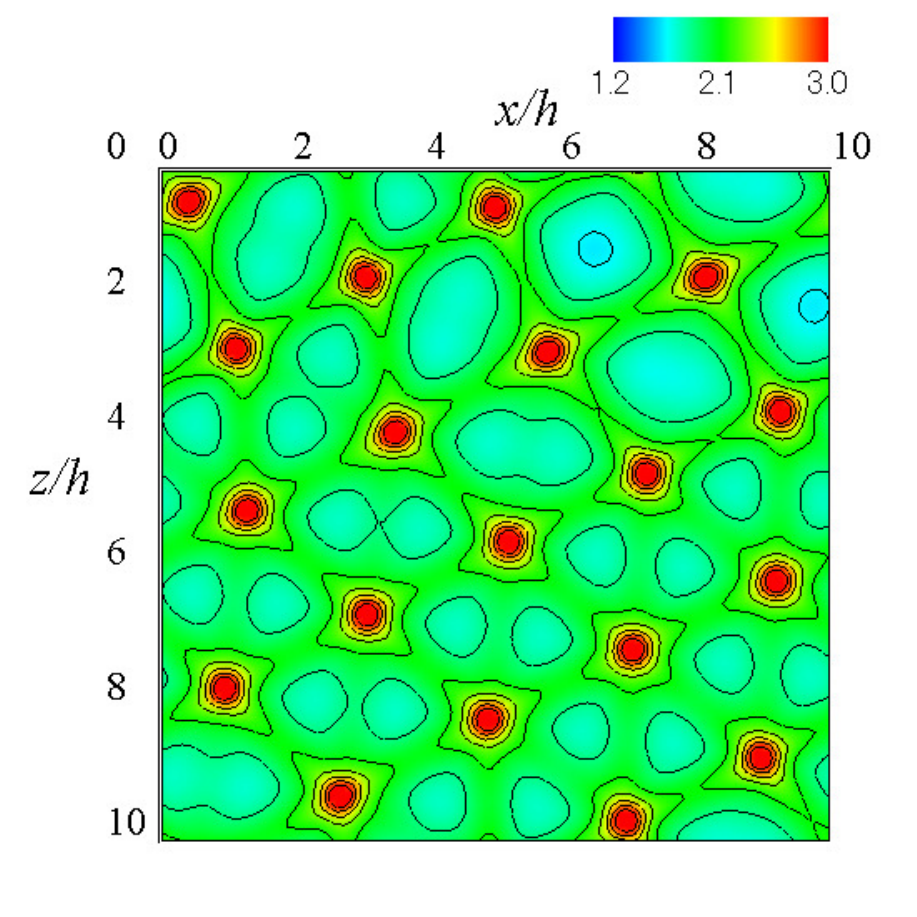} \\
(d) $F = 2000$, $T = 2\pi/4$ \\
\end{minipage}
\begin{minipage}{53mm}
\centering
\includegraphics[trim=0mm 10mm 0mm 0mm, clip, width=55mm]{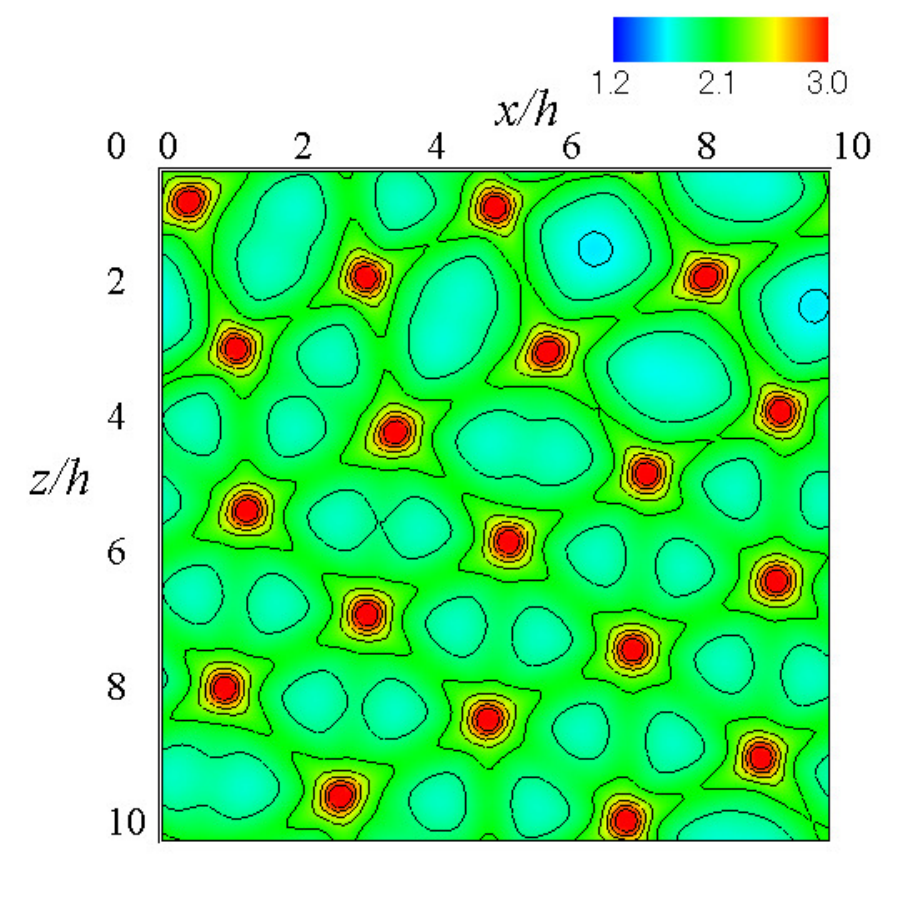} \\
(e) $F = 2000$, $T = 6\pi/4$ \\
\end{minipage}
\caption{Bacterial concentration contours in $x-z$ plane at $y/h=0.99$ for 
$\Gamma = 1000$, $Ra = 500$, and $\Theta_\mathrm{A} = 0.9$: 
Contour interval is 0.18 from 1.2 to 3.0.}
\label{pattern}
\end{figure}

Subsequently, we investigate the effect of temperature fluctuations on the thermo-bioconvection pattern. 
Figure \ref{pattern} plots the bacterial concentration distribution near the water surface 
for $\Gamma = 1000$, $Ra = 500$, and $\Theta_\mathrm{A} = 0.9$. 
Figure (a) shows the result for steady heating, 
while Figs. (b), (c), (d), and (e) show the results for unsteady heating. 
Here, we present the results for the times $T = 2 \pi/4$ and $6 \pi/4$ 
when the temperature reaches its maximum and minimum, respectively. 
Plumes appear in areas of high concentration. 
For the steady heating of $F = 0$, $19$ plumes occur randomly. 
For the unsteady heating of $F = 20$ and $2000$, $19$ plumes arise randomly regardless of time, 
and the positions of the plumes match the results for $F = 0$. 
From the above, it was found that for unsteady heating, 
the number and positions of plumes do not change with time and match the steady results. 
We also confirmed that the same tendency was observed for other frequencies.

\begin{figure}[!t]
\centering
\begin{minipage}{0.48\linewidth}
\centering
\includegraphics[trim=0mm 0mm 0mm 0mm, clip, width=75mm]{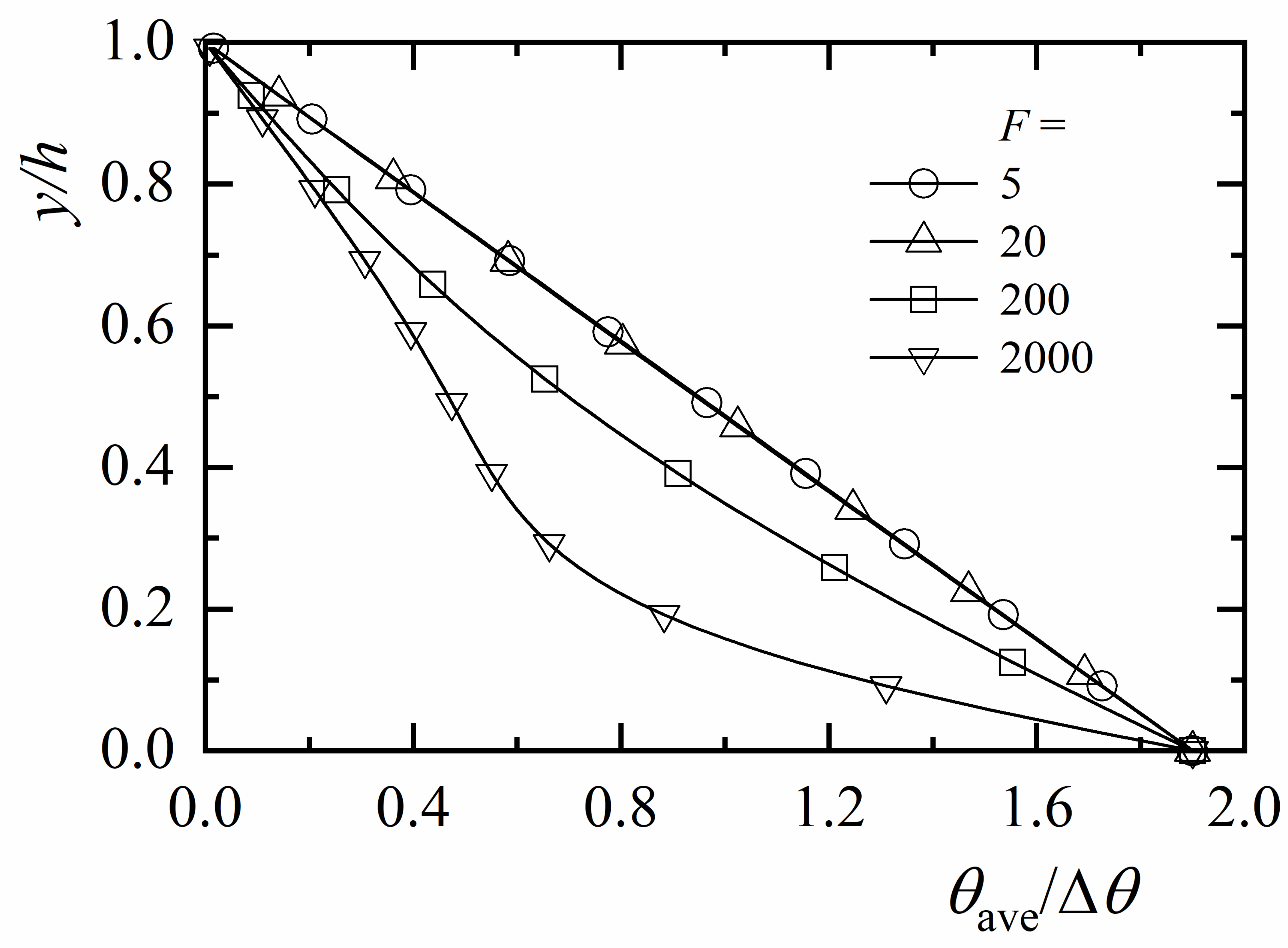}\\
\vspace*{-0.2\baselineskip}
(a) $T = 2\pi/4$ \\
\end{minipage}
\begin{minipage}{0.48\linewidth}
\centering
\includegraphics[trim=0mm 0mm 0mm 0mm, clip, width=75mm]{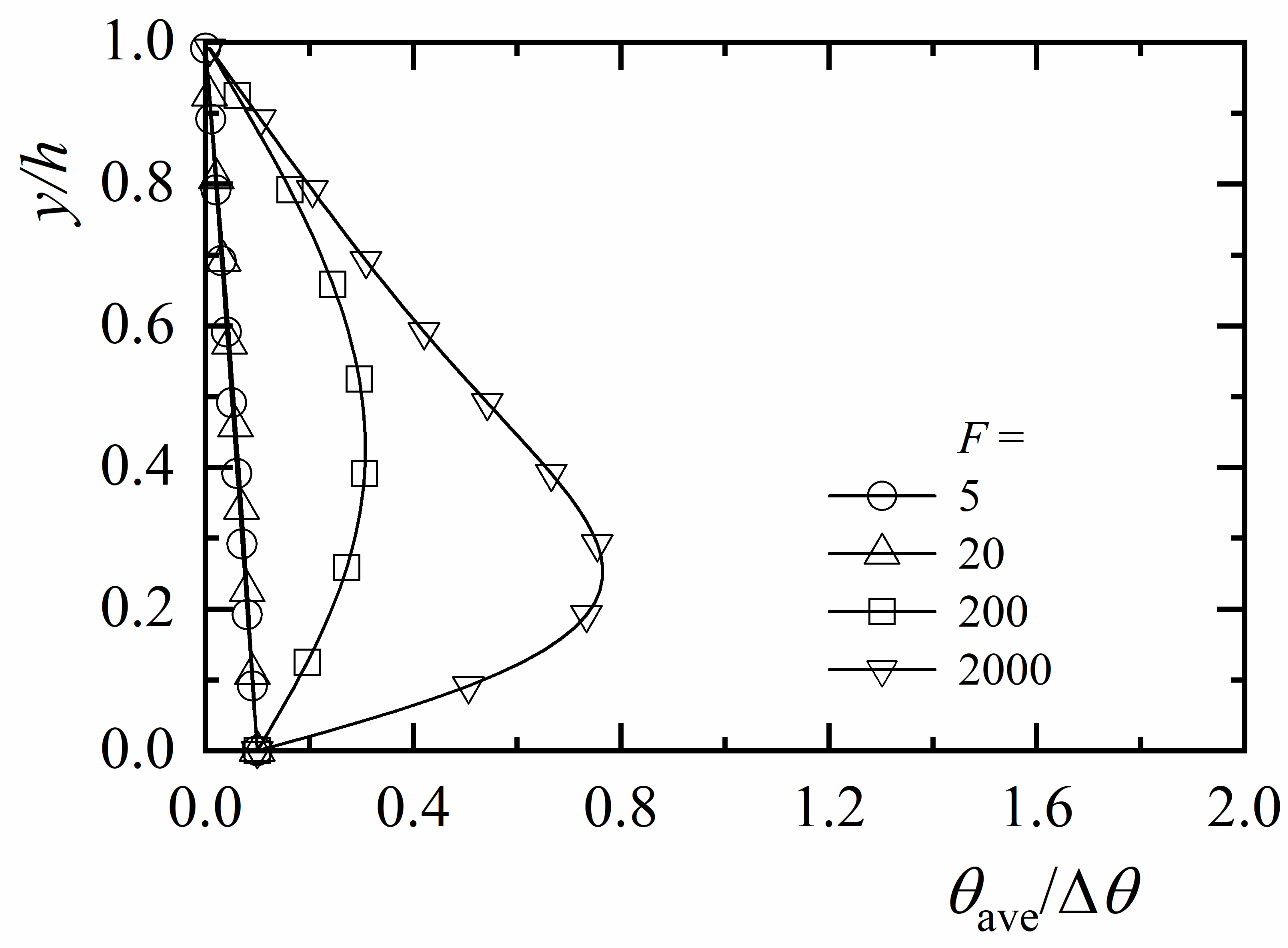}\\
\vspace*{-0.2\baselineskip}
(b) $T = 6\pi/4$ \\
\end{minipage}
\caption{Vertical distributions of temperature averaged in $x$--$z$ plane.}
\label{time_vari_dist_t_f5_2000}
\centering
\end{figure}

\subsubsection{The response of thermo-bioconvection to temperature fluctuations}

We clarify the variation in the following ability of thermo-bioconvection 
with differences in frequency $F$. 
The calculation conditions are set to be $\Gamma = 1000$, $Ra = 500$, 
$\Theta_\mathrm{A} = 0.9$, and $F = 5$, $20$, $200$, and $2000$. 
Figure \ref{time_vari_dist_t_f5_2000} shows the $y$-direction distribution 
of area-averaged temperature $\theta_{\mathrm{ave}}$ on the $x$--$z$ plane 
at $T$ = 2$\pi$/4 and $T = 6 \pi/4$. 
For $F = 5$ and $20$, one period of temperature fluctuation is $1152.0$ s and $288.0$ s, respectively, 
which is sufficiently longer than the thermal diffusion time scale of $\tau_{\theta} = 40.3$ s. 
Therefore, as the temperature change at the wall surface can diffuse to the upper surface during one period, 
the difference in temperature distribution near the water surface 
at $T = 2 \pi/4$ and $T = 6 \pi/4$ becomes large; 
hence, it can be seen that the temperature field in the entire region changes 
following the temperature fluctuation at the wall surface. 
When the frequency increases to $F = 200$ and $2000$, 
the period of each temperature fluctuation becomes shorter than the thermal diffusion time scale $\tau_{\theta}$. 
Consequently, there is not enough time for the temperature change 
at the wall surface to diffuse toward the upper surface, 
and the response of thermo-bioconvection to the temperature fluctuation worsens. 
The temperature over the entire area, excluding the water and bottom surfaces, 
decreases at $T = 2 \pi/4$ and increases at $T = 6 \pi/4$ 
compared to the results for $F = 5$ and $20$. 
This tendency is most prominent at the high frequency of $F = 2000$. 
In conclusion, it was found that the response of thermo-bioconvection to temperature fluctuations 
decreases as the frequency increases.

\begin{figure}[!t]
\centering
\begin{minipage}{0.48\linewidth}
\centering
\includegraphics[trim=0mm 0mm 0mm 0mm, clip, width=75mm]{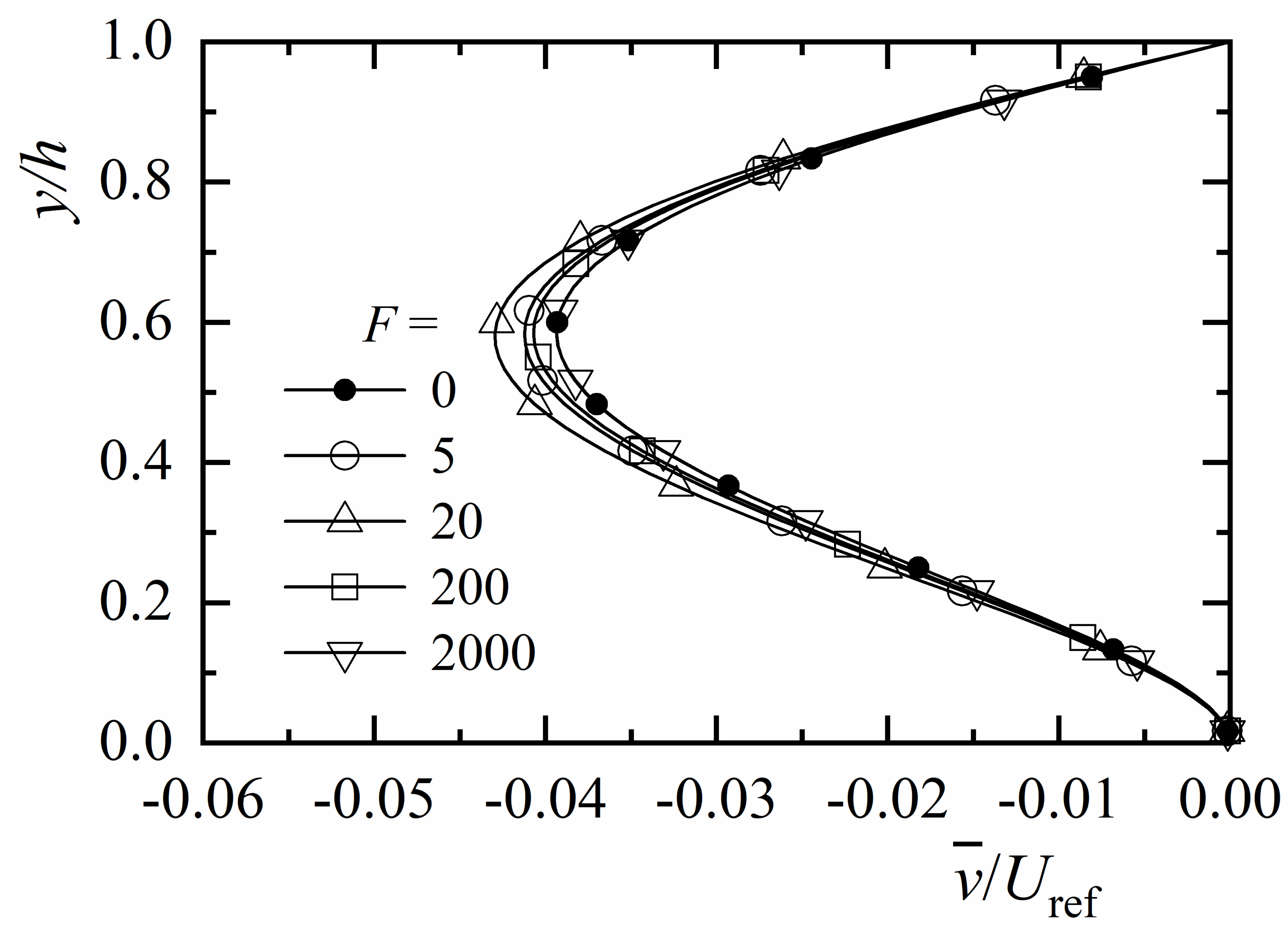} \\
\vspace*{-0.2\baselineskip}
(a) Center of plume \\
\end{minipage}
\begin{minipage}{0.48\linewidth}
\centering
\includegraphics[trim=0mm 0mm 0mm 0mm, clip, width=75mm]{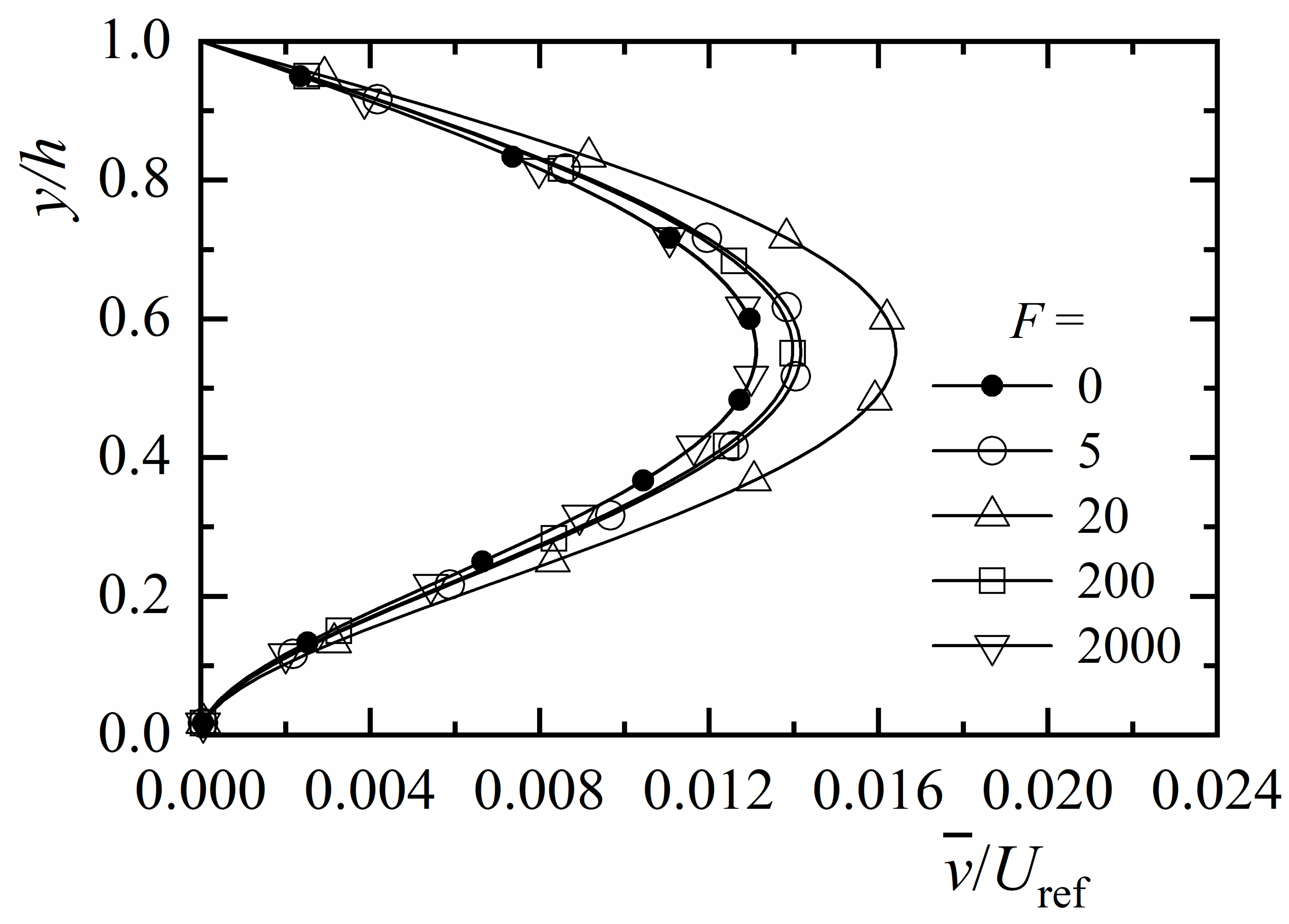}\\
\vspace*{-0.2\baselineskip}
(b) Between plumes \\
\end{minipage}
\caption{Time--averaged vertical velocity distributions in $y$-direction at the center of plume and between plumes 
for $\Gamma = 1000$, $Ra = 500$, and $\Theta_\mathrm{A} = 0.9$.}
\label{time_ave_v_dist}
\end{figure}

\begin{figure}[!t]
\centering
\begin{minipage}{0.48\linewidth}
\centering
\includegraphics[trim=0mm 0mm 0mm 0mm, clip, width=75mm]{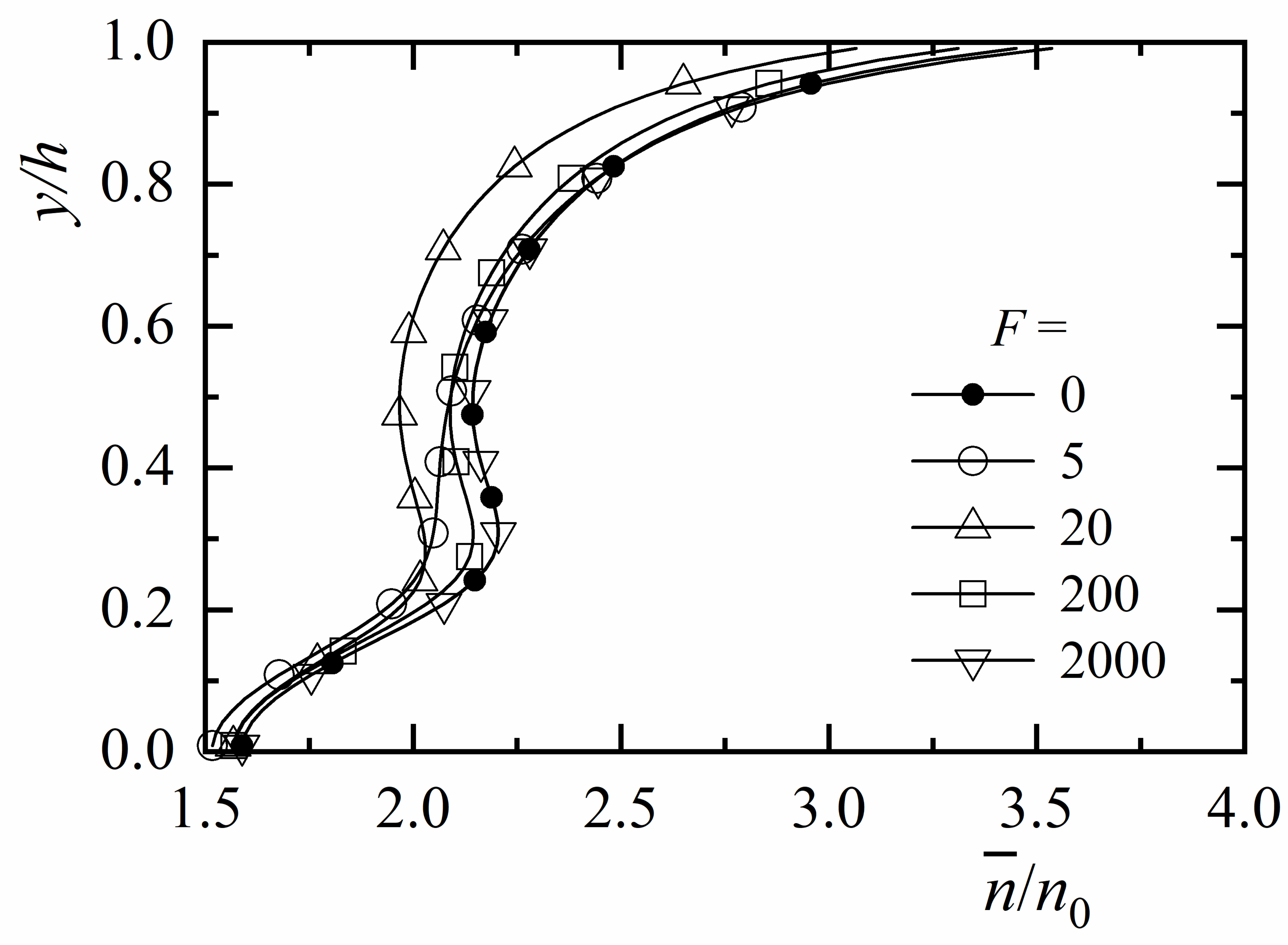} \\
\vspace*{-0.2\baselineskip}
(a) Center of plume \\
\end{minipage}
\begin{minipage}{.48\linewidth}
\centering
\includegraphics[trim=0mm 0mm 0mm 0mm, clip, width=75mm]{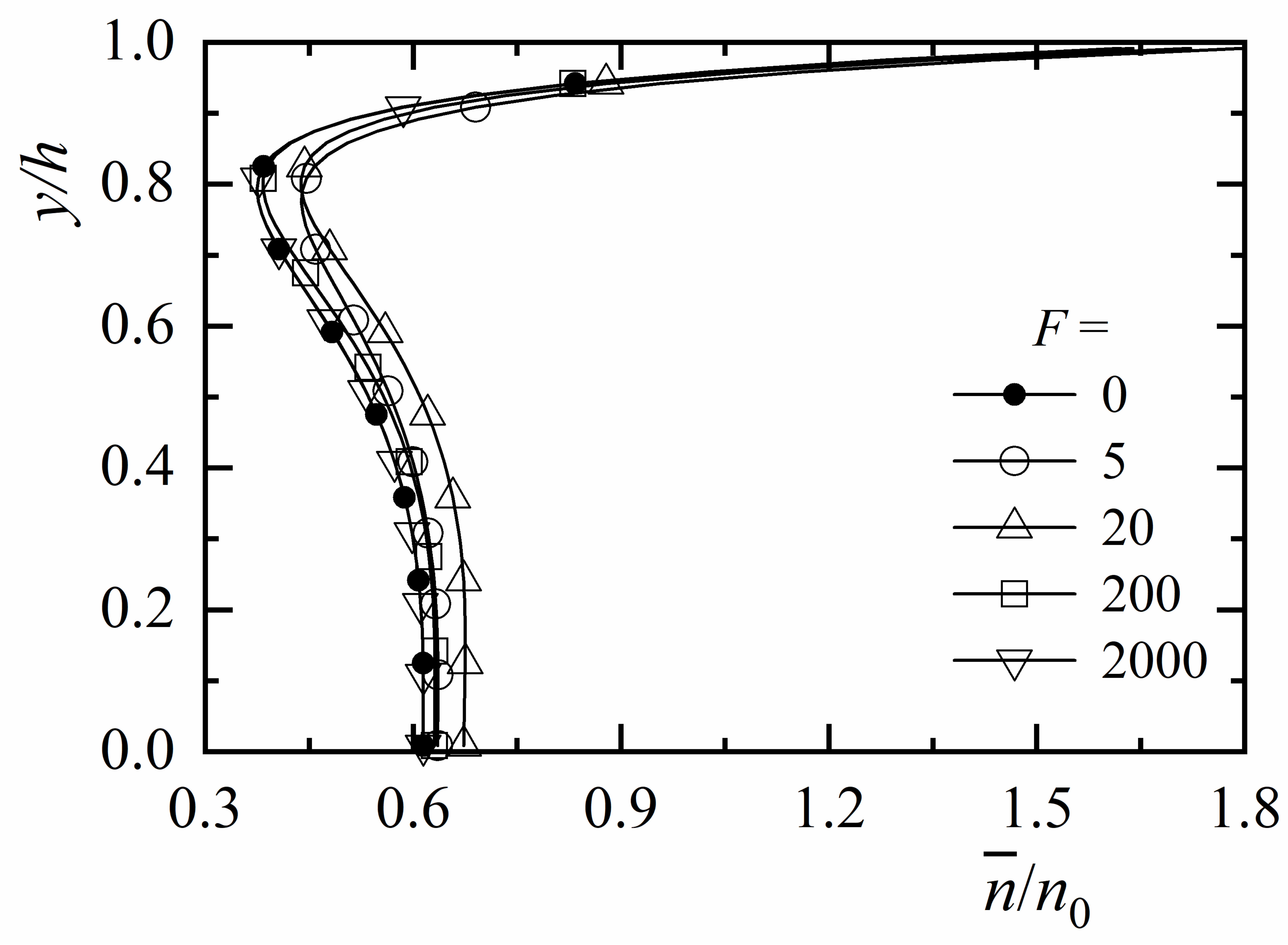}\\
\vspace*{-0.2\baselineskip}
(b) Between plumes \\
\end{minipage}
\caption{Time--averaged bacterial concentration distributions in $y$-direction at the center of plume and between plumes 
for $\Gamma = 1000$, $Ra = 500$, and $\Theta_\mathrm{A} = 0.9$.}
\label{time_ave_n_dist}
\centering
\end{figure}

\begin{figure}[!t]
\centering
\begin{minipage}{0.48\linewidth}
\centering
\includegraphics[trim=0mm 0mm 0mm 0mm, clip, width=75mm]{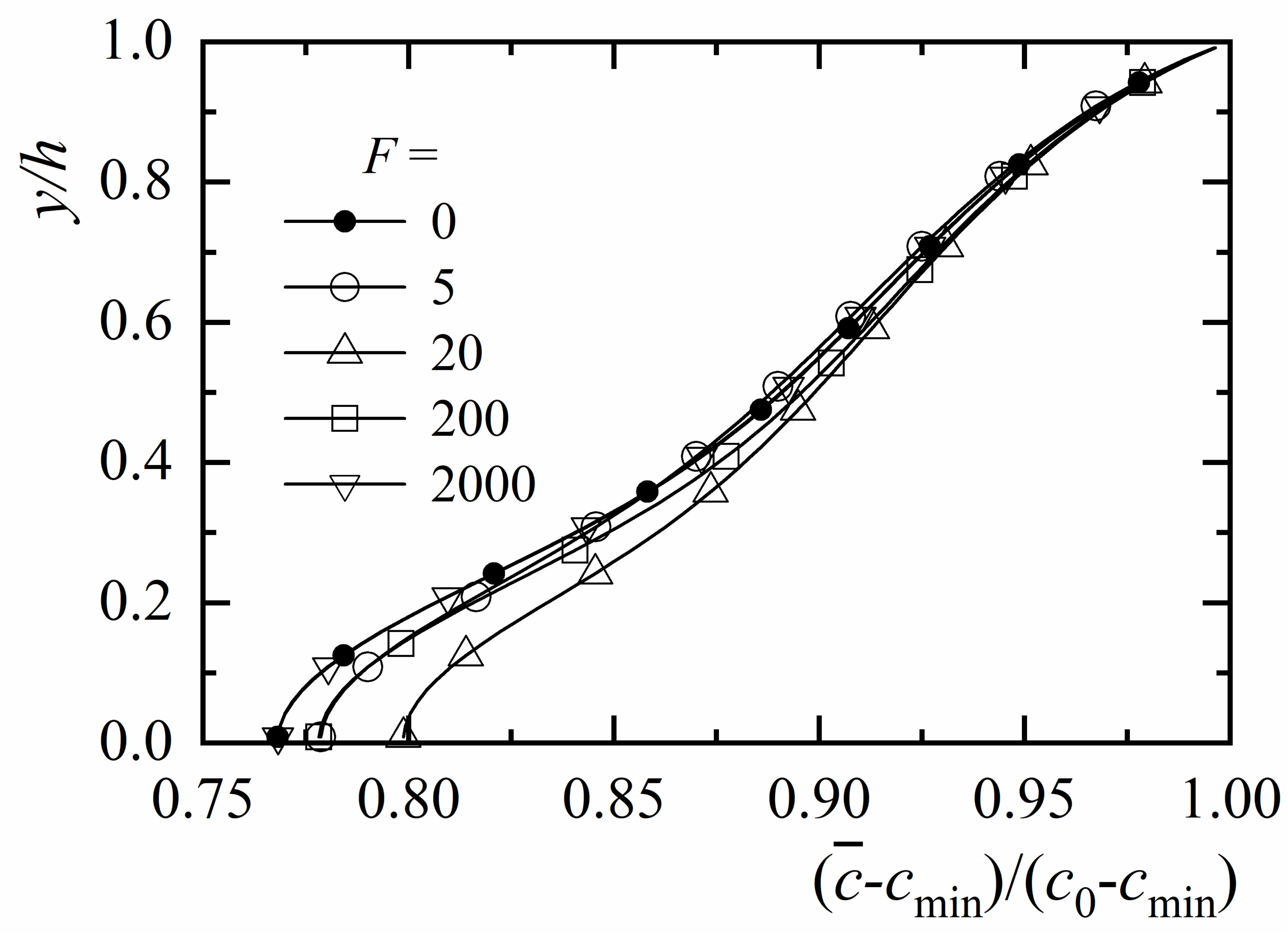} \\
\vspace*{-0.2\baselineskip}
(a) Center of plume \\
\end{minipage}
\begin{minipage}{0.48\linewidth}
\centering
\includegraphics[trim=0mm 0mm 0mm 0mm, clip, width=75mm]{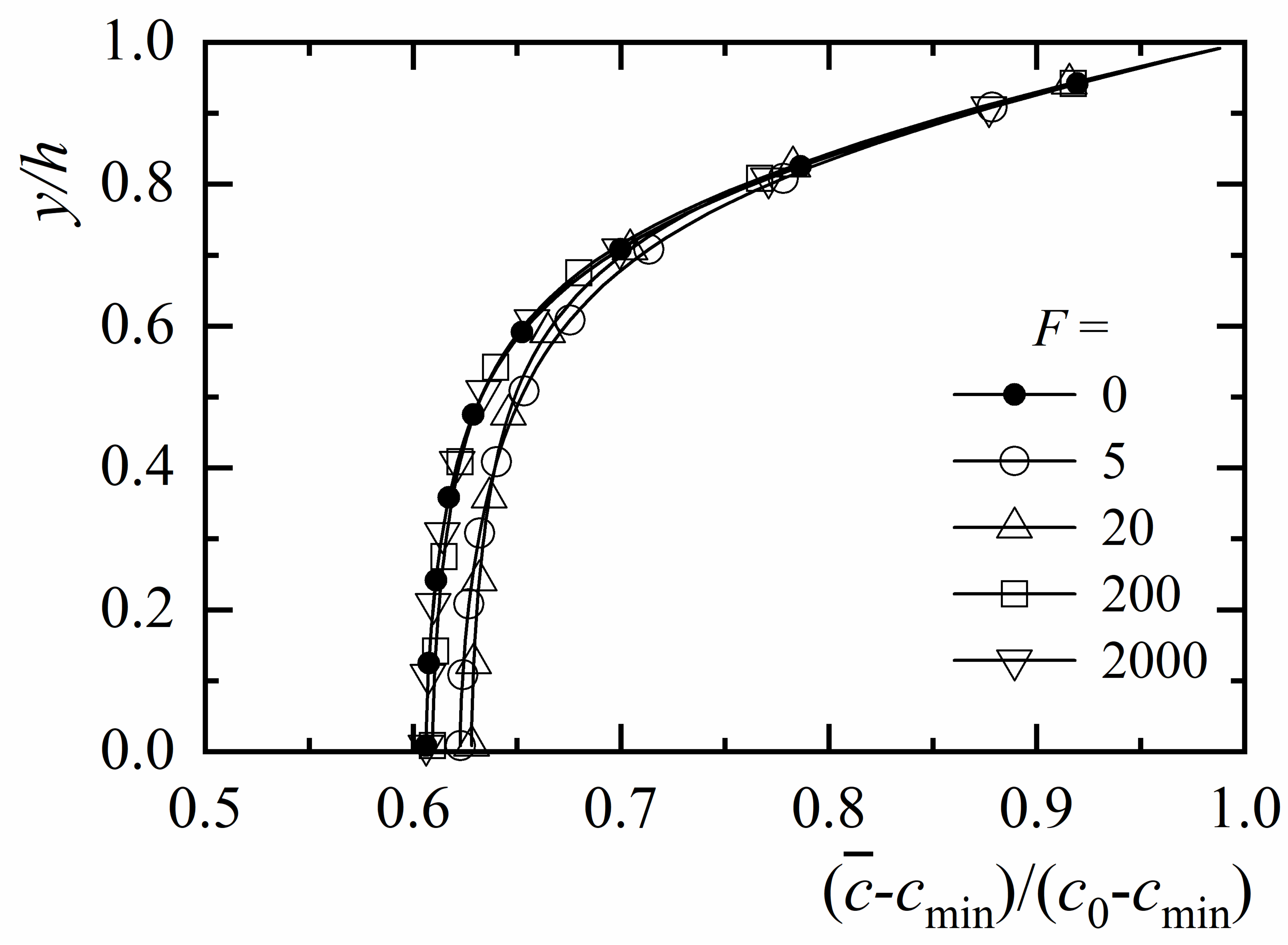}\\
\vspace*{-0.2\baselineskip}
(b) Between plumes \\
\end{minipage}
\caption{Time--averaged oxygen concentration distributions in $y$-direction at the center of plume and between plumes 
for $\Gamma = 1000$, $Ra = 500$, and $\Theta_\mathrm{A} = 0.9$.}
\label{time_ave_c_dist}
\centering
\end{figure}

\begin{figure}[!t]
\centering
\begin{minipage}{0.48\linewidth}
\centering
\includegraphics[trim=0mm 0mm 0mm 0mm, clip, width=75mm]{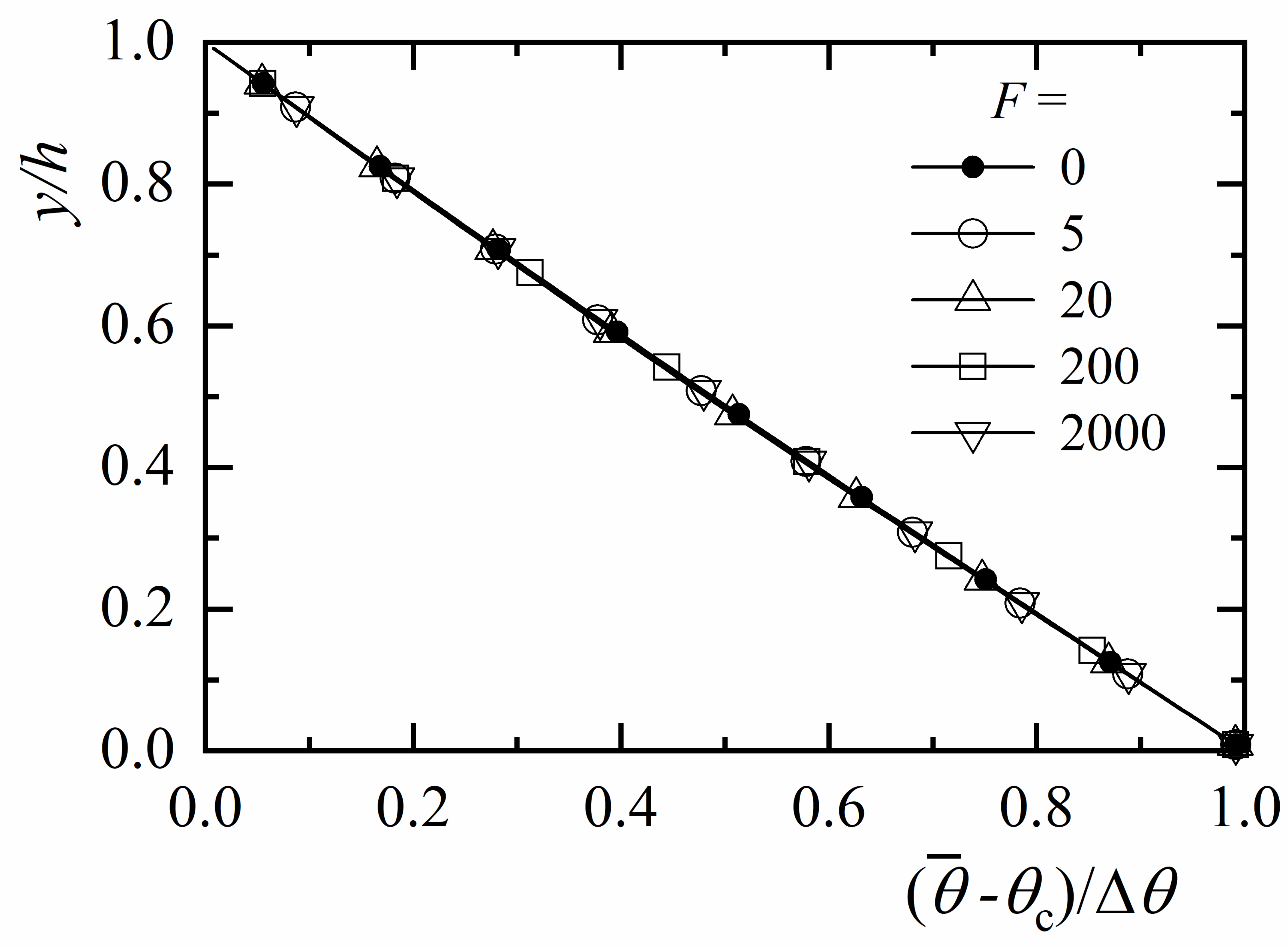} \\
\vspace*{-0.2\baselineskip}
(a) Center of plume \\
\end{minipage}
\begin{minipage}{0.48\linewidth}
\centering
\includegraphics[trim=0mm 0mm 0mm 0mm, clip, width=75mm]{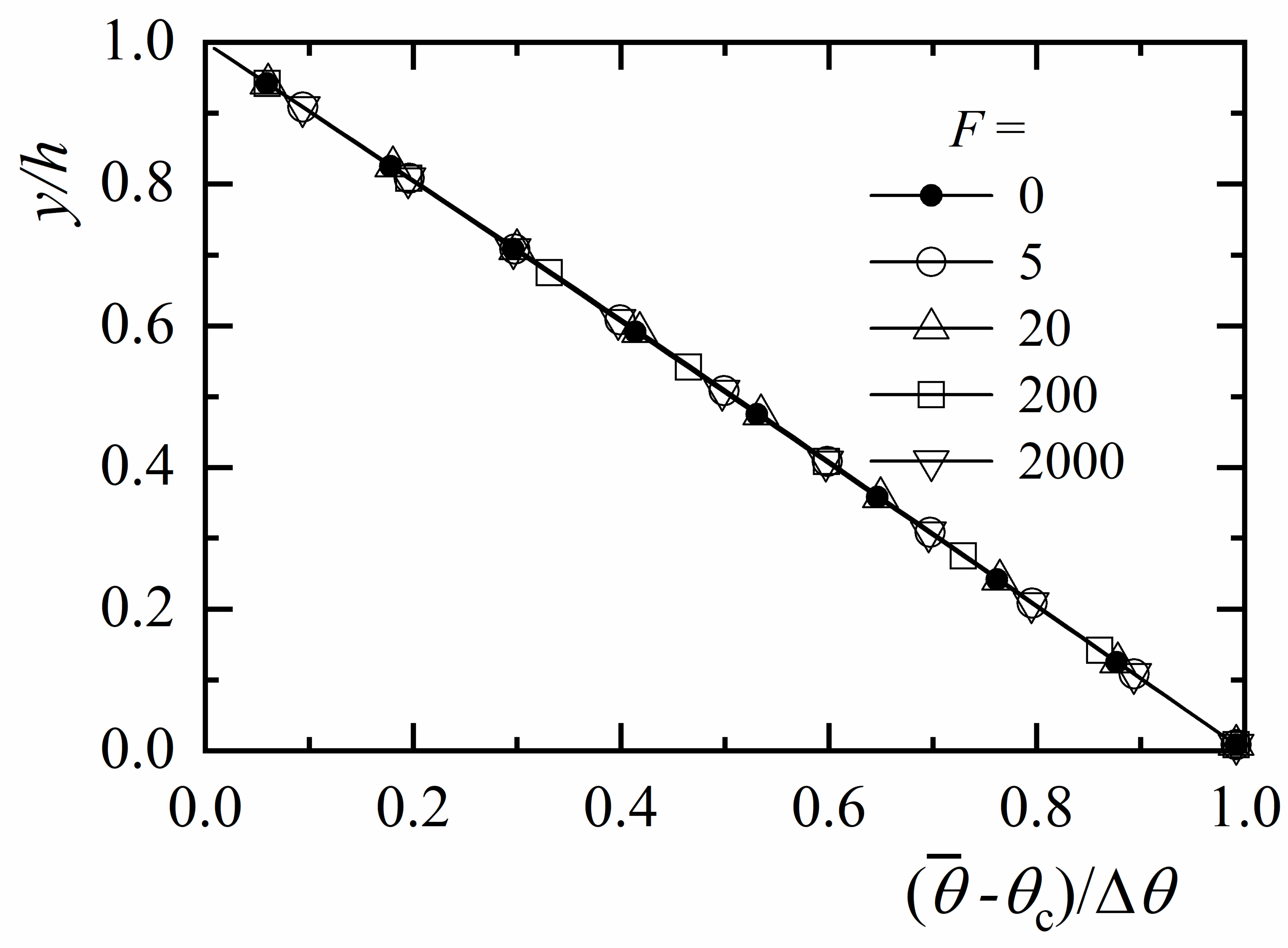} \\
\vspace*{-0.2\baselineskip}
(b) Between plumes \\
\end{minipage}
\caption{Time-averaged temperature distributions in $y$-direction at the center of plume and between plumes 
for $\Gamma = 1000$, $Ra = 500$, and $\Theta_\mathrm{A} = 0.9$.}
\label{time_ave_t_dist}
\end{figure}

\subsubsection{Time-averaged characteristics}

We investigate the impact of frequency $F$ on the time-averaged characteristics 
of thermo-bioconvection. 
For $\Gamma = 1000$, $Ra = 500$, and $\Theta_\mathrm{A} = 0.9$, 
the $y$-direction distributions of time-averaged vertical velocity $\overline{v}$, 
bacterial concentration $\overline{n}$, oxygen concentration $\overline{c}$, 
and temperature $\overline{\theta}$ at the plume center and between the plumes are shown 
in Figs. \ref{time_ave_v_dist}, \ref{time_ave_n_dist}, \ref{time_ave_c_dist}, 
and \ref{time_ave_t_dist}, respectively. 
The distributions of $\overline{v}$ at $F = 5$, $20$, and $200$ are faster than the result at $F = 0$ 
regardless of the location. 
Espetially, $\overline{v}$ is the fastest at the resonance frequency $F = 20$ 
at which $|v_\mathrm{down}^\mathrm{int}|$ is the fastest instantaneously. 
The magnitude of $\overline{v}$ at $F = 20$ increases up to a value of $9.12\%$ near $y/h = 0.58$ at the plume center 
and up to a value of $25.08\%$ at $y/h = 0.55$ between the plumes, 
compared to the result at $F = 0$. 
This indicates that temperature fluctuations at the wall significantly affect 
convective velocity between the plumes where suspension flows from multiple plumes. 

The distributions $\overline{n}$ at $F = 5$, $20$, and $200$ decrease at the plume center 
and increase between the plumes, compared to the result for $F = 0$. 
This suggests that the high concentration of bacteria in the plume center is transported 
to the area between the plumes by temperature fluctuations, 
enhancing the convective transport. 
The distributions $\overline{c}$ at $F = 5$, $20$, and $200$ increase near the bottom wall 
compared to the result for $F = 0$, regardless of the location. 
This is because convection enhancement caused by temperature fluctuations transports 
a large amount of oxygen near the water surface downward. 
As for the transport of bacteria and oxygen, 
the difference from the result for $F = 0$ is most noticeable 
at the resonant frequency $F = 20$. 

At $F = 2000$, the response of thermo-bioconvection to temperature fluctuations worsens, 
and the time change in $|v_\mathrm{down}^\mathrm{int}|$ is the lowest; 
hence, $\overline{v}$, $\overline{n}$, and $\overline{c}$ are consistent with the results for $F = 0$. 
The distribution of $\overline{\theta}$ is consistent at all frequencies, regardless of the location. 
This is because $Le$ is high and thermal diffusion is large relative to convection; 
hence, the influence of convection enhancement hardly appears.

\begin{figure}[!t]
\vspace*{-0.2\baselineskip}
\centering
\begin{minipage}{0.48\linewidth}
\centering
\includegraphics[trim=0mm 0mm 0mm 0mm, clip, width=65mm]{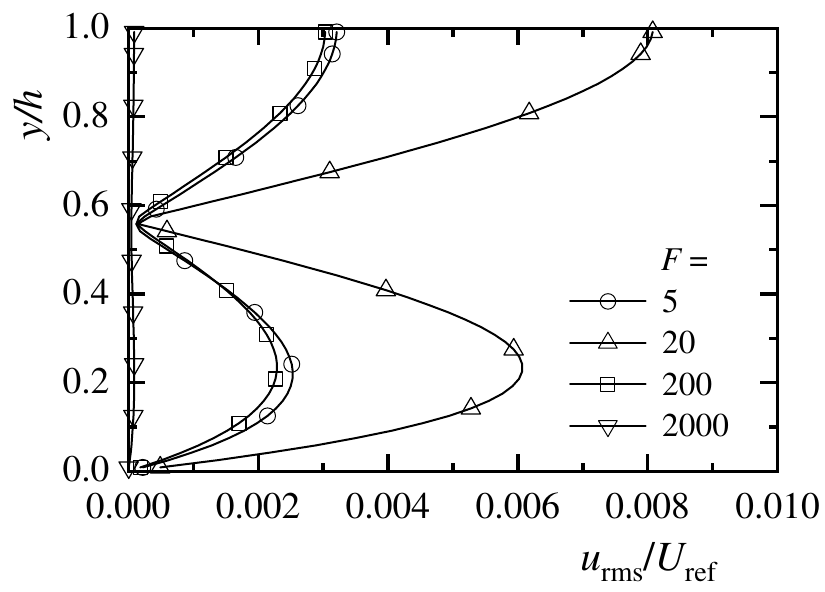} \\
\vspace*{-3mm}
(a) $u_\mathrm{rms}/U_\mathrm{ref}$ \\
\end{minipage}
\begin{minipage}{0.48\linewidth}
\centering
\includegraphics[trim=0mm 0mm 0mm 0mm, clip, width=65mm]{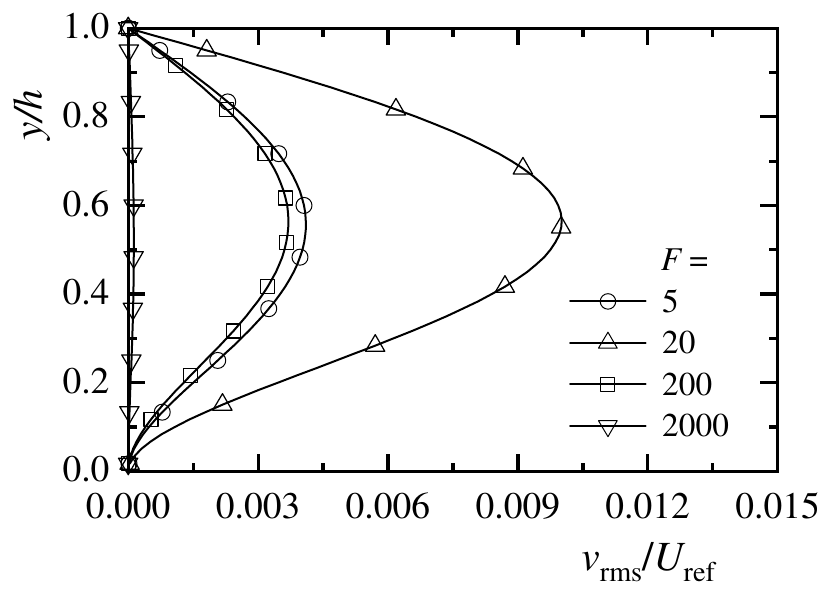} \\
\vspace*{-3mm}
(b) $v_\mathrm{rms}/U_\mathrm{ref}$ \\
\end{minipage}
\begin{minipage}{0.48\linewidth}
\centering
\includegraphics[trim=0mm 0mm 0mm 0mm, clip, width=65mm]{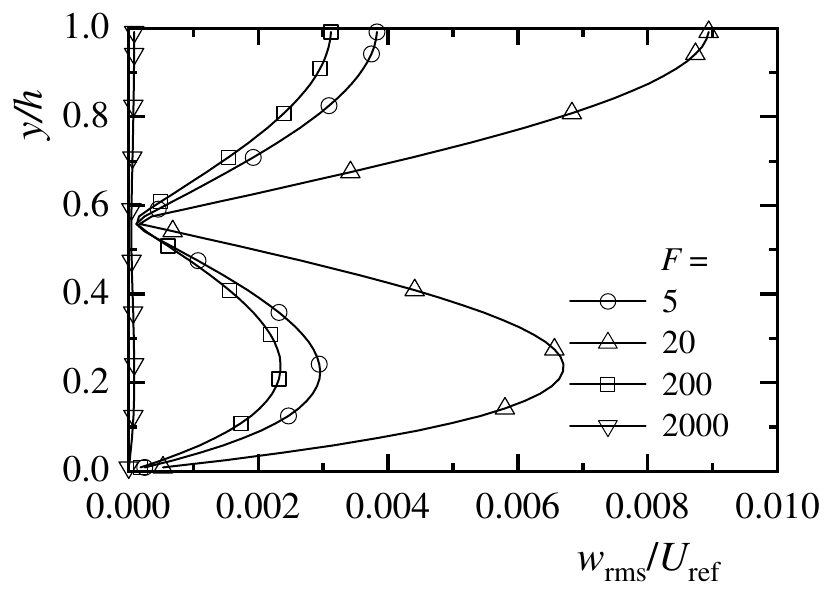} \\
\vspace*{-3mm}
(c) $w_\mathrm{rms}/U_\mathrm{ref}$ \\
\end{minipage}
\begin{minipage}{0.48\linewidth}
\centering
\includegraphics[trim=0mm 0mm 0mm 0mm, clip, width=65mm]{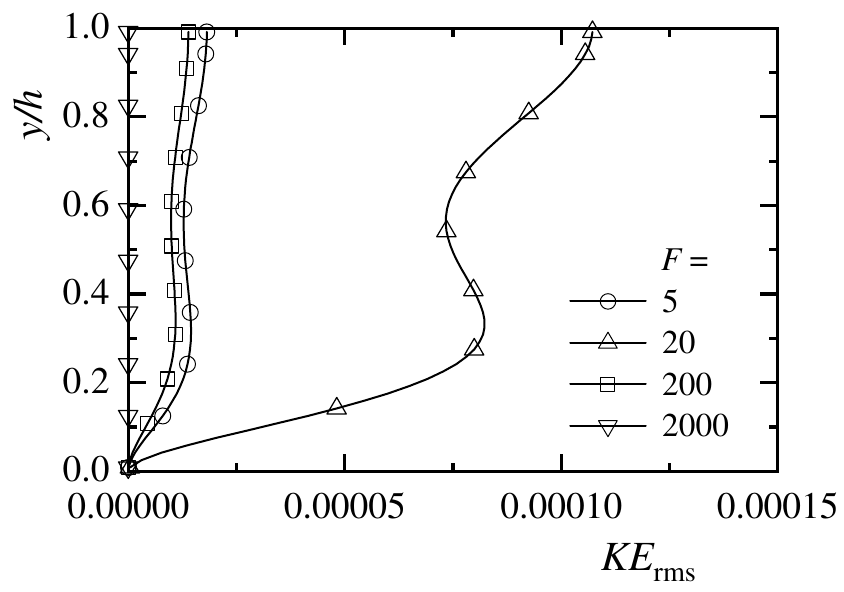} \\
\vspace*{-3mm}
(d) $KE_\mathrm{rms}$ \\
\end{minipage}
\begin{minipage}{0.48\linewidth}
\centering
\includegraphics[trim=0mm 0mm 0mm 0mm, clip, width=65mm]{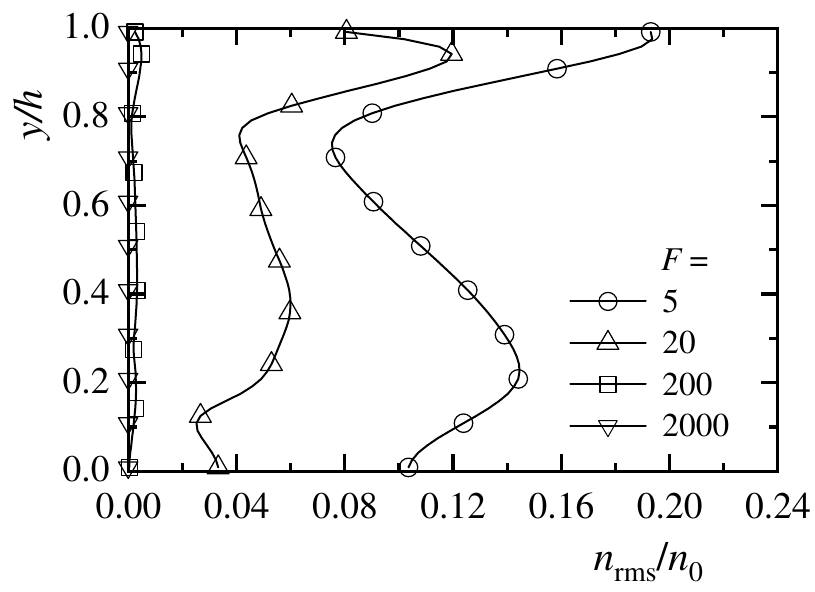} \\
\vspace*{-3mm}
(e) $n_\mathrm{rms}/n_0$ \\
\end{minipage}
\begin{minipage}{0.48\linewidth}
\centering
\includegraphics[trim=0mm 0mm 0mm 0mm, clip, width=65mm]{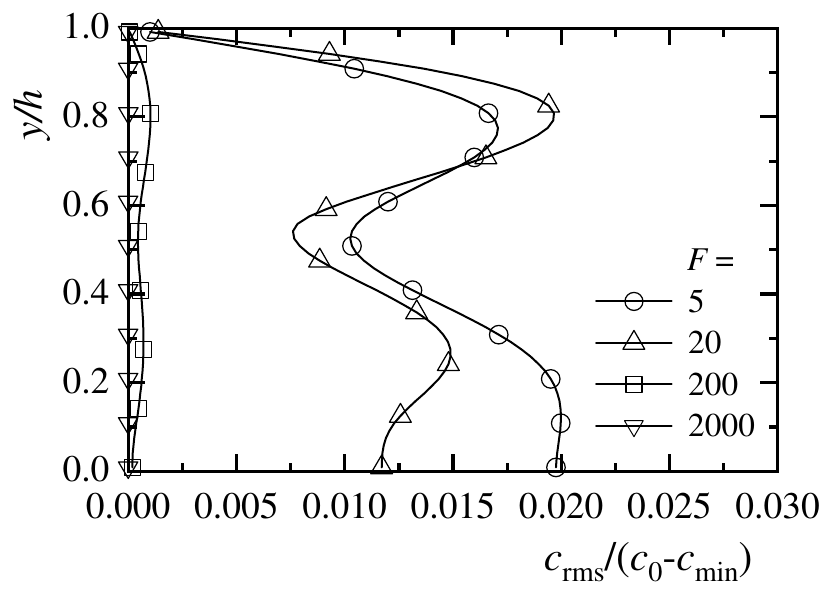} \\
\vspace*{-3mm}
(f) $c_\mathrm{rms}/(c_0-c_\mathrm{min})$ \\
\end{minipage}
\begin{minipage}{0.48\linewidth}
\centering
\includegraphics[trim=0mm 0mm 0mm 0mm, clip, width=65mm]{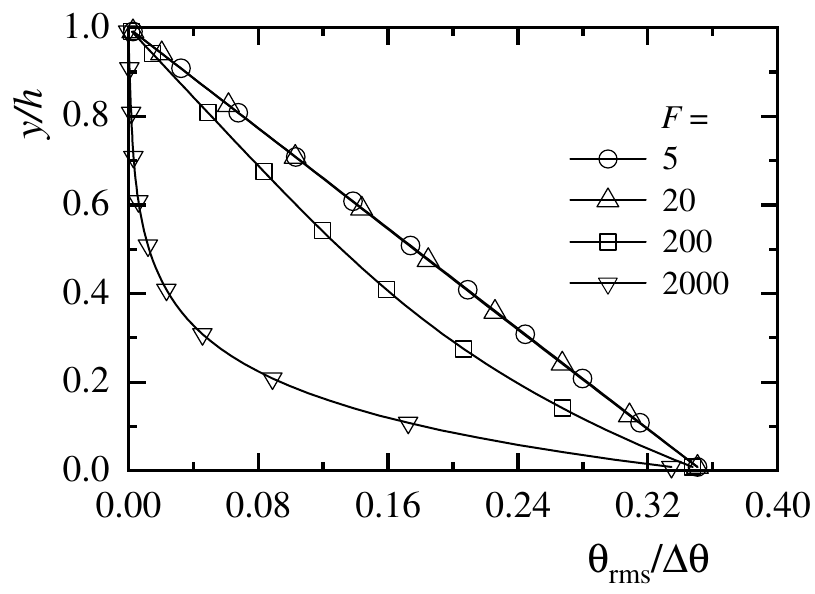} \\
\vspace*{-3mm}
(g) $\theta_\mathrm{rms}/\Delta\theta$ \\
\end{minipage}
\caption{Turbulence intensity distributions of velocities, 
concentrations and temperature averaged in the $x$--$z$ plane.}
\label{rms_xz}
\end{figure}

\subsubsection{Turbulence characteristics}

To investigate the effect of frequency $F$ on the turbulence characteristics of thermo-bioconvection, 
Fig. \ref{rms_xz} shows the area-averaged distribution at the $y$--$z$ cross-section 
for the turbulence intensities $u_\mathrm{rms}$, $v_\mathrm{rms}$, 
$w_\mathrm{rms}$ of velocity fluctuations in the $x$, $y$, and $z$-directions, 
turbulence kinetic energy $KE_\mathrm{rms}$, 
and turbulence intensities $n_\mathrm{rms}$, $c_\mathrm{rms}$, 
$\theta_\mathrm{rms}$ of bacterial concentration, oxygen concentration, 
and temperature fluctuations. 
The distributions $u_\mathrm{rms}$ and $w_\mathrm{rms}$ are high near the water surface regardless of $F$ 
and reach maximums near $y/h = 0.24$. 
This indicates plumes change over time owing to temperature fluctuations at the lower wall, 
causing fluctuations in the horizontal flow near the water surface and lower wall. 
The distributions $v_\mathrm{rms}$ for $F = 5$, $20$, and $200$ reach maximums near $y/h = 0.54$ 
because fluctuations due to periodically changing upward and downward flows within the suspension appear. At $F = 5$, $20$, and $200$, $KE_\mathrm{rms}$ is high in the areas where $u_\mathrm{rms}$, $v_\mathrm{rms}$, and $w_\mathrm{rms}$ are high. Furthermore, $u_\mathrm{rms}$, $v_\mathrm{rms}$, $w_\mathrm{rms}$, and $KE_\mathrm{rms}$ are the highest at the resonant frequency $F = 20$ where the time changes of $|v_\mathrm{down}^\mathrm{int}|$ and $|v_\mathrm{up}^\mathrm{int}|$ are maximum and are low at frequencies $F = 5$ and $200$ where the time changes of $|v_\mathrm{down}^\mathrm{int}|$ and $|v_\mathrm{up}^\mathrm{int}|$ are low. At $F = 2000$, the time changes of $|v_\mathrm{down}^\mathrm{int}|$ and $|v_\mathrm{up}^\mathrm{int}|$ are the lowest; hence, $u_\mathrm{rms}$, $v_\mathrm{rms}$, and $w_\mathrm{rms}$ are almost zero.

Subsequently, we consider the distributions of $n_\mathrm{rms}$, $c_\mathrm{rms}$, 
and $\theta_\mathrm{rms}$ shown in Figs. \ref{rms_xz}(e), (f), and (g). 
Both $n_\mathrm{rms}$ and $c_\mathrm{rms}$ at $F = 5$ and $20$ exhibit high values 
near the water surface and around $y/h = 0.24$. 
The coordinates, where turbulence intensities become high, approximately coincide with 
the positions where $u_\mathrm{rms}$, $w_\mathrm{rms}$, and $KE_\mathrm{rms}$ are high, 
which suggests that the concentrations of bacteria and oxygen are fluctuating owing to convection. 
Additionally, the turbulence intensity for $F = 20$ is lower than that for $F = 5$, 
and the effect of the resonance phenomenon does not appear 
even though $F = 20$ generates high turbulence intensities of velocities. 
This is because at the low frequency of $F = 5$, 
bacteria and oxygen have enough time to be transported, 
causing large fluctuations in the concentration distribution. 
For $\theta_\mathrm{rms}$, as $Le$ is high, 
the impact of convection enhancement caused by temperature fluctuations does not appear. 
Hence, $\theta_\mathrm{rms}$ is high at the bottom where the temperature fluctuates 
and low near the water surface. 
For $F = 200$ and $2000$, 
the following ability of thermo-bioconvection decreases, 
and the suspension approaches a steady field; 
hence, $n_\mathrm{rms}$, $c_\mathrm{rms}$, and $\theta_\mathrm{rms}$ decrease.

\subsubsection{Grid dependency of calculation results}

Finally, we confirmed the grid dependency on the calculation results for the unsteady field. 
The calculation conditions were set to be $\Gamma$ = 1000, $Ra = 700$, 
$\Theta_\mathrm{A} = 0.5$, and $F = 5$, $2000$. 
The Rayleigh number is the highest combination in the calculation of the unsteady field. 
Again, we compared the results in which the same thermo-bioconvection pattern occurred.

\begin{figure}[!t]
\centering
\begin{minipage}{0.48\linewidth}
\centering
\includegraphics[trim=0mm 0mm 0mm 0mm, clip, width=80mm]{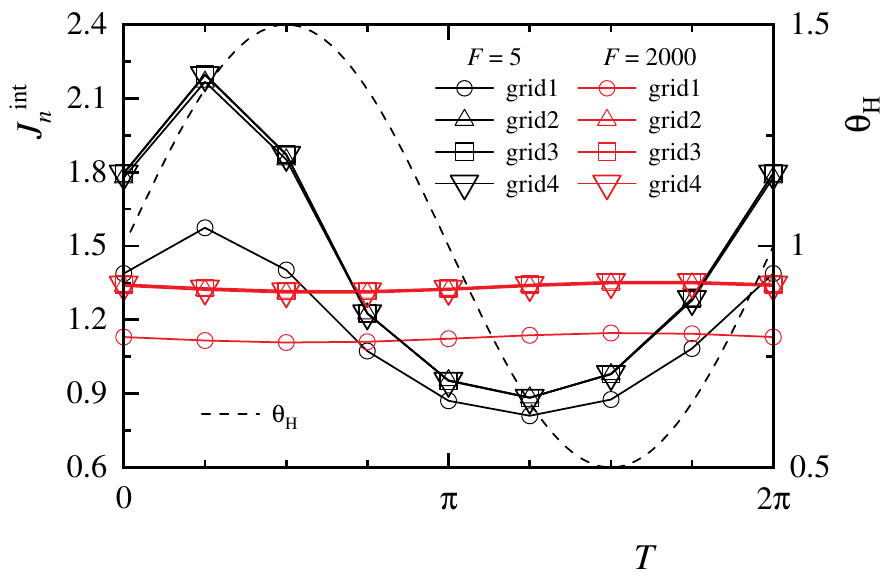}\\
\vspace*{-0.2\baselineskip}
(a) Bacteria \\
\end{minipage}
\begin{minipage}{0.48\linewidth}
\centering
\includegraphics[trim=0mm 0mm 0mm 0mm, clip, width=80mm]{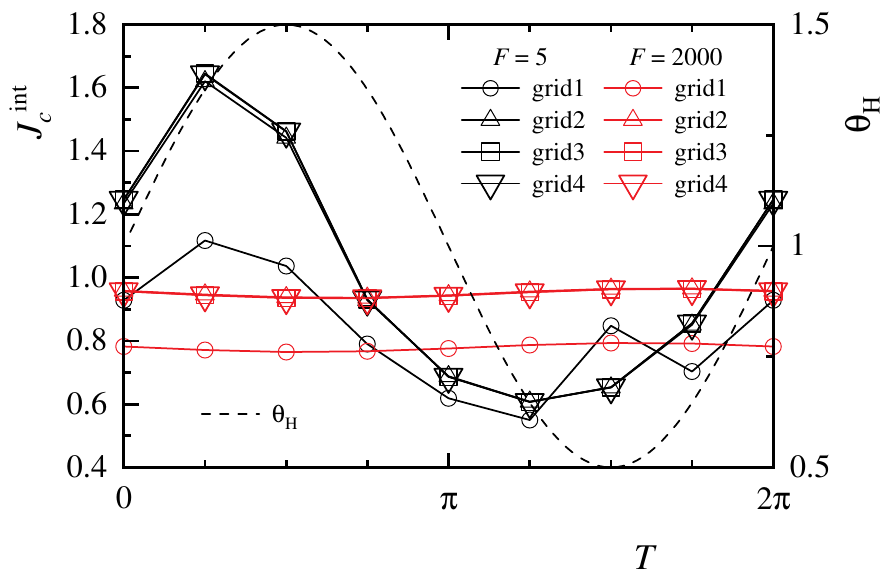} \\
\vspace*{-0.2\baselineskip}
(b) Oxygen \\
\end{minipage}
\caption{Time variations of integral values of total flux of bacteria and oxygen 
for $\Gamma = 1000$, $Ra = 700$, and $\Theta_\mathrm{A} = 0.5$ 
using four grids.}
\label{flux_time_ver_grid}
\end{figure}

Figure \ref{flux_time_ver_grid} shows the time evolution of the integrals 
$J_{n}^\mathrm{int}$ and $J_{c}^\mathrm{int}$ of the magnitudes of the total bacterial and oxygen flux vectors 
obtained using each grid. 
At low and high frequencies, 
the results of grid1 and grid2 are different, 
while the results of grid2, grid3, and grid4 agree well.

Table \ref{unstfluxav_tbio} shows the integrals 
$\overline{J}_{n}^\mathrm{int}$ and $\overline{J}_{c}^\mathrm{int}$ 
of the magnitudes of the time-averaged total bacterial and oxygen flux vectors. 
The relative differences in $\overline{J}_{n}^\mathrm{int}$ 
for grid1, grid2, and grid3 compared to grid4 are approximately 
$-17.194\%$, $2.713\%$, and $0.066\%$ for $F = 5$, 
and approximately $-15.582\%$, $0.285\%$, and $0.105\%$ for $F = 2000$, respectively. 
The relative differences in $\overline{J}_{c}^\mathrm{int}$ are approximately 
$-21.950\%$, $0.168\%$, and $0.059\%$ for $F = 5$, 
and approximately $-18.102\%$, $0.210\%$, and $0.073\%$ for $F = 2000$, respectively. 
Based on these results, 
it can be concluded that the calculation results obtained using grid2 are valid 
even in unsteady fields.

\begin{table}[!t]
\begin{center}
\caption{Time--averaged integral values of total flux of bacteria and oxygen using different computational grids: 
$\Gamma = 1000$, $Ra = 1200$, $\Theta_\mathrm{A} = 0.5$, and $F = 5, 2000$.}
\label{unstfluxav_tbio}
\centering
\vspace{1.5mm}
\begin{tabular}{|c|c|c|c|c|}
\hline
 & \multicolumn{2}{|c|}{$F = 5$} & \multicolumn{2}{|c|}{$F = 2000$} \\
\cline{2-5}
       & Bacteria: $\overline{J}_{n}^\mathrm{int}$ & Oxygen: $\overline{J}_{c}^\mathrm{int}$ & Bacteria: $\overline{J}_{n}^{\mathrm{int}}$ & Oxygen: $\overline{J}_{c}^\mathrm{int}$ \\ \hline
 grid1 & 1.1322       & 0.7908    & 1.1269       & 0.7809     \\ \hline
 grid2 & 1.4044       & 1.0149    & 1.3387       & 0.9555     \\ \hline
 grid3 & 1.3682       & 1.0138    & 1.3363       & 0.9542     \\ \hline
 grid4 & 1.3673       & 1.0132    & 1.3349       & 0.9535     \\ \hline
\end{tabular}
\end{center}
\end{table}

\section{CONCLUSION}
\label{summary}

In this study, we performed a three-dimensional numerical analysis of the thermo-bioconvection 
generated by a suspension of chemotactic bacteria 
under unsteady bottom heating conditions 
to investigate the effects of the amplitude and frequency of temperature fluctuations 
on the thermo-bioconvection pattern, 
as well as the transport characteristics of bacteria and oxygen.

Under steady heating conditions, 
bioconvection and thermal convection coexist, 
and thermo-bioconvection occurs around plumes. 
Increasing the thermal Rayleigh number increases the effect of interference 
between thermal convection and bioconvection, 
and the convective velocity increases. 
Then, the transport of substances is promoted by convection, 
and the transport characteristics of bacteria and oxygen are improved 
throughout the entire region.

Temperature fluctuations at the wall surface cause unsteady thermo-bioconvection. 
At low frequencies, the thermo-bioconvection follows temperature fluctuations, 
and the interaction between thermal convection and bioconvection strengthens. 
Thus, the transport properties of bacteria and oxygen throughout the domain are improved. 
As the frequency increases, 
the ability of thermo-bioconvection to follow temperature fluctuations deteriorates, 
and the interference between bioconvection and thermal convection weakens, 
slowing the convective velocity and reducing the transport characteristics of bacteria and oxygen. 
A resonance phenomenon occurs at the frequency 
where the instability of the suspension owing to the density difference between the bacteria and water is maintained 
and where thermo-bioconvection can follow temperature fluctuations. 
Furthermore, regardless of frequency, 
the number and position of plumes do not change with time 
and are not affected by temperature fluctuations.

Focusing on the time-averaged characteristics of unsteady thermo-bioconvection, 
as a resonance phenomenon occurs, at the resonance frequency, 
the transport characteristics of bacteria and oxygen throughout the entire region 
within the suspension improve significantly. 
Furthermore, as the amplitude of temperature fluctuations and thermal Rayleigh number increase, 
the interference between thermal convection and bioconvection intensifies, 
leading to a noticeable improvement in transport characteristics 
owing to the resonance phenomenon. 
At this time, the amplitude of temperature fluctuations and thermal Rayleigh number 
do not almost affect the resonance frequency.


\vspace*{1.0\baselineskip}
\noindent
{\bf Acknowledgements.}
The numerical results in this research were obtained 
using supercomputing resources at Cyberscience Center, Tohoku University. 
We would like to express our gratitude to Associate Professor Yosuke Suenaga 
of Iwate University for his support of our laboratory. 
The authors wish to acknowledge the time and effort of everyone involved in this study.

\vspace*{1.0\baselineskip}
\noindent
{\bf Declaration of interests.}
The authors have no conflicts to disclose. \\

\vspace*{1.0\baselineskip}
\noindent
{\bf Author ORCID.} \\
H. Yanaoka \url{https://orcid.org/0000-0002-4875-8174}.

\vspace*{1.0\baselineskip}
\noindent
{\bf Author contributions.} \\
{\bf Hideki Yanaoka}: 
Conceptualization (lead); Investigation (equal); 
Methodology (lead); Visualization (equal); Writing - original draft (equal); 
Writing - review \& editing (lead). 
{\bf Daichi Koizumi}: 
Formal analysis (lead); Investigation (equal); 
Validation (lead); Visualization (equal); Writing - original draft (equal). \\

\noindent
{\bf Funding}: 
This work was supported by JSPS KAKENHI Grant Number JP22K03918.


\bibliographystyle{arXiv_elsarticle-harv}
\bibliography{arXiv2025_koizumi_bibfile}

\begin{thebibliography}{53}
\expandafter\ifx\csname natexlab\endcsname\relax\def\natexlab#1{#1}\fi
\providecommand{\url}[1]{\texttt{#1}}
\providecommand{\href}[2]{#2}
\providecommand{\path}[1]{#1}
\providecommand{\DOIprefix}{doi:}
\providecommand{\ArXivprefix}{arXiv:}
\providecommand{\URLprefix}{URL: }
\providecommand{\Pubmedprefix}{pmid:}
\providecommand{\doi}[1]{\href{http://dx.doi.org/#1}{\path{#1}}}
\providecommand{\Pubmed}[1]{\href{pmid:#1}{\path{#1}}}
\providecommand{\bibinfo}[2]{#2}
\ifx\xfnm\relax \def\xfnm[#1]{\unskip,\space#1}\fi
\bibitem[{Alloui et~al.(2006)Alloui, Nguyen and Bilgen}]{Alloui_et_al_2006}
\bibinfo{author}{Alloui, Z.}, \bibinfo{author}{Nguyen, T.H.},
  \bibinfo{author}{Bilgen, E.}, \bibinfo{year}{2006}.
\newblock \bibinfo{title}{Stability analysis of thermo-bioconvection in
  suspensions of gravitactic microorganisms in a fluid layer}.
\newblock \bibinfo{journal}{Int. Commun. Heat Mass Transf.}
  \bibinfo{volume}{33}, \bibinfo{pages}{1198--1206}.
\newblock
  \DOIprefix\doi{https://doi.org/10.1016/j.icheatmasstransfer.2006.08.012}.
\bibitem[{Amsden and Harlow(1970)}]{Amsden&Harlow_1970}
\bibinfo{author}{Amsden, A.A.}, \bibinfo{author}{Harlow, F.H.},
  \bibinfo{year}{1970}.
\newblock \bibinfo{title}{A simplified {MAC} technique for incompressible fluid
  flow calculations}.
\newblock \bibinfo{journal}{J. Comput. Phys.} \bibinfo{volume}{6},
  \bibinfo{pages}{322--325}.
\newblock \DOIprefix\doi{https://doi.org/10.1016/0021-9991(70)90029-X}.
\bibitem[{Bees and Croze(2014)}]{Bees&Croze_2014}
\bibinfo{author}{Bees, M.A.}, \bibinfo{author}{Croze, O.A.},
  \bibinfo{year}{2014}.
\newblock \bibinfo{title}{Mathematics for streamlined biofuel production from
  unicellular algae}.
\newblock \bibinfo{journal}{Biofuels} \bibinfo{volume}{5},
  \bibinfo{pages}{53--65}.
\newblock \DOIprefix\doi{https://doi.org/10.4155/bfs.13.66}.
\bibitem[{Bees and Hill(1997)}]{Bees&Hill_1997}
\bibinfo{author}{Bees, M.A.}, \bibinfo{author}{Hill, N.A.},
  \bibinfo{year}{1997}.
\newblock \bibinfo{title}{Wavelengths of bioconvection patterns}.
\newblock \bibinfo{journal}{J. Exp. Biol.} \bibinfo{volume}{200},
  \bibinfo{pages}{1515--1526}.
\newblock \DOIprefix\doi{https://doi.org/10.1242/jeb.200.10.1515}.
\bibitem[{Berg and Brown(1972)}]{Berg&Brown_1972}
\bibinfo{author}{Berg, H.C.}, \bibinfo{author}{Brown, D.A.},
  \bibinfo{year}{1972}.
\newblock \bibinfo{title}{Chemotaxis in \textit{{E}scherichia coli} analysed by
  three-dimensional tracking}.
\newblock \bibinfo{journal}{Nature} \bibinfo{volume}{239},
  \bibinfo{pages}{500--504}.
\newblock \DOIprefix\doi{https://doi.org/10.1038/239500a0}.
\bibitem[{Biswas et~al.(2022)Biswas, Mandal, Manna and
  Benim}]{Biswas_et_al_2022}
\bibinfo{author}{Biswas, N.}, \bibinfo{author}{Mandal, D.K.},
  \bibinfo{author}{Manna, N.K.}, \bibinfo{author}{Benim, A.C.},
  \bibinfo{year}{2022}.
\newblock \bibinfo{title}{Magneto-hydrothermal triple-convection in a
  {W}-shaped porous cavity containing oxytactic bacteria}.
\newblock \bibinfo{journal}{Sci. Rep.} \bibinfo{volume}{12}, \bibinfo{pages}{30
  pages}.
\newblock \DOIprefix\doi{https://doi.org/10.1038/s41598-022-18401-7}.
\bibitem[{Chertock et~al.(2012)Chertock, Fellner, Kurganov, Lorz and
  Markowich}]{Chertock_et_al_2012}
\bibinfo{author}{Chertock, A.}, \bibinfo{author}{Fellner, K.},
  \bibinfo{author}{Kurganov, A.}, \bibinfo{author}{Lorz, A.},
  \bibinfo{author}{Markowich, P.A.}, \bibinfo{year}{2012}.
\newblock \bibinfo{title}{Sinking, merging and stationary plumes in a coupled
  chemotaxis-fluid model: {A} high-resolution numerical approach}.
\newblock \bibinfo{journal}{J. Fluid Mech.} \bibinfo{volume}{694},
  \bibinfo{pages}{155--190}.
\newblock \DOIprefix\doi{https://doi.org/10.1017/jfm.2011.534}.
\bibitem[{Czir$\acute{\mbox{o}}$k et~al.(2000)Czir$\acute{\mbox{o}}$k,
  J$\acute{\mbox{a}}$nosi and Kessler}]{Czirok_et_al_2000}
\bibinfo{author}{Czir$\acute{\mbox{o}}$k, A.},
  \bibinfo{author}{J$\acute{\mbox{a}}$nosi, I.M.}, \bibinfo{author}{Kessler,
  J.O.}, \bibinfo{year}{2000}.
\newblock \bibinfo{title}{Bioconvective dynamics : dependence on organism
  behaviour}.
\newblock \bibinfo{journal}{J. Exp. Biol.} \bibinfo{volume}{203},
  \bibinfo{pages}{3345--3354}.
\newblock \DOIprefix\doi{https://doi.org/10.1242/jeb.203.21.3345}.
\bibitem[{Geng and Kuznetsov(2005)}]{Geng&Kuznetsov_2005}
\bibinfo{author}{Geng, P.}, \bibinfo{author}{Kuznetsov, A.V.},
  \bibinfo{year}{2005}.
\newblock \bibinfo{title}{Settling of bidispersed small solid particles in a
  dilute suspension containing gyrotactic micro-organisms}.
\newblock \bibinfo{journal}{Int. J. Eng. Sci.} \bibinfo{volume}{43},
  \bibinfo{pages}{992--1010}.
\newblock \DOIprefix\doi{https://doi.org/10.1016/j.ijengsci.2005.03.002}.
\bibitem[{Ghorai and Hill(2000)}]{Ghorai&Hill_2000}
\bibinfo{author}{Ghorai, S.}, \bibinfo{author}{Hill, N.A.},
  \bibinfo{year}{2000}.
\newblock \bibinfo{title}{Wavelengths of gyrotactic plumes in bioconvection}.
\newblock \bibinfo{journal}{Bull. Math. Biol.} \bibinfo{volume}{62},
  \bibinfo{pages}{429--450}.
\newblock \DOIprefix\doi{https://doi.org/10.1006/bulm.1999.0160}.
\bibitem[{Ghorai and Hill(2002)}]{Ghorai&Hill_2002}
\bibinfo{author}{Ghorai, S.}, \bibinfo{author}{Hill, N.A.},
  \bibinfo{year}{2002}.
\newblock \bibinfo{title}{Axisymmetric bioconvection in a cylinder}.
\newblock \bibinfo{journal}{J. Theor. Biol.} \bibinfo{volume}{219},
  \bibinfo{pages}{137--152}.
\newblock \DOIprefix\doi{https://doi.org/10.1006/jtbi.2002.3077}.
\bibitem[{Hart and Edwards(1987)}]{Hart&Edwards_1987}
\bibinfo{author}{Hart, A.}, \bibinfo{author}{Edwards, C.},
  \bibinfo{year}{1987}.
\newblock \bibinfo{title}{Buoyant density fluctuations during the cell cycle of
  \textit{{B}acillus subtilis}}.
\newblock \bibinfo{journal}{Arch. Microbiol.} \bibinfo{volume}{147},
  \bibinfo{pages}{68--72}.
\newblock \DOIprefix\doi{https://doi.org/10.1007/BF00492907}.
\bibitem[{Hillesdon and Pedley(1996)}]{Hillesdon&Pedley_1996}
\bibinfo{author}{Hillesdon, A.J.}, \bibinfo{author}{Pedley, T.J.},
  \bibinfo{year}{1996}.
\newblock \bibinfo{title}{Bioconvection in suspensions of oxytactic bacteria:
  {L}inear theory}.
\newblock \bibinfo{journal}{J. Fluid Mech.} \bibinfo{volume}{324},
  \bibinfo{pages}{223--259}.
\newblock \DOIprefix\doi{https://doi.org/10.1017/S0022112096007902}.
\bibitem[{Hillesdon et~al.(1995)Hillesdon, Pedley and
  Kessler}]{Hillesdon_et_al_1995}
\bibinfo{author}{Hillesdon, A.J.}, \bibinfo{author}{Pedley, T.J.},
  \bibinfo{author}{Kessler, J.O.}, \bibinfo{year}{1995}.
\newblock \bibinfo{title}{The development of concentration gradients in a
  suspension of chemotactic bacteria}.
\newblock \bibinfo{journal}{Bull. Math. Biol.} \bibinfo{volume}{57},
  \bibinfo{pages}{299--344}.
\newblock \DOIprefix\doi{https://doi.org/10.1007/BF02460620}.
\bibitem[{Hirooka and Nagase(2003)}]{Hirooka&Nagase_2003}
\bibinfo{author}{Hirooka, T.}, \bibinfo{author}{Nagase, H.},
  \bibinfo{year}{2003}.
\newblock \bibinfo{title}{Biodegradation of endocrine disrupting chemicals and
  its application for bioremediation --utilization of photoautotrophic
  microorganisms--}.
\newblock \bibinfo{journal}{J. Environ. Biotechnol.} \bibinfo{volume}{3},
  \bibinfo{pages}{23--32}.
\bibitem[{Itoh et~al.(2001)Itoh, Toida and Saotome}]{Itoh_et_al_2001}
\bibinfo{author}{Itoh, A.}, \bibinfo{author}{Toida, H.},
  \bibinfo{author}{Saotome, Y.}, \bibinfo{year}{2001}.
\newblock \bibinfo{title}{Control of bioconvection and it's application for
  mechanical system (1st report, basic effects of electrical field on
  bioconvection)}.
\newblock \bibinfo{journal}{JSME, Ser. B} \bibinfo{volume}{67},
  \bibinfo{pages}{2449--2454}.
\newblock \DOIprefix\doi{https://doi.org/10.1299/kikaib.67.2449}.
\bibitem[{Itoh et~al.(2006)Itoh, Toida and Saotome}]{Itoh_et_al_2006}
\bibinfo{author}{Itoh, A.}, \bibinfo{author}{Toida, H.},
  \bibinfo{author}{Saotome, Y.}, \bibinfo{year}{2006}.
\newblock \bibinfo{title}{Control of bioconvection and it's application for
  mechanical system (2nd report, pitch optimization of electrodes array and
  driving of mechanical system by controlled bioconvection)}.
\newblock \bibinfo{journal}{JSME, Ser. B} \bibinfo{volume}{72},
  \bibinfo{pages}{972--978}.
\newblock \DOIprefix\doi{https://doi.org/10.1299/kikaib.72.972}.
\bibitem[{J$\acute{\mbox{a}}$nosi et~al.(1998)J$\acute{\mbox{a}}$nosi, Kessler
  and Horv$\acute{\mbox{a}}$th}]{Janosi_et_al_1998}
\bibinfo{author}{J$\acute{\mbox{a}}$nosi, I.M.}, \bibinfo{author}{Kessler,
  J.O.}, \bibinfo{author}{Horv$\acute{\mbox{a}}$th, V.K.},
  \bibinfo{year}{1998}.
\newblock \bibinfo{title}{Onset of bioconvection in suspensions of
  \textit{{B}acillus subtilis}}.
\newblock \bibinfo{journal}{Phys. Rev. E} \bibinfo{volume}{58},
  \bibinfo{pages}{4793--4800}.
\newblock \DOIprefix\doi{https://link.aps.org/doi/10.1103/PhysRevE.58.4793}.
\bibitem[{Kage et~al.(2013)Kage, Hosoya, Baba and Mogami}]{Kage_et_al_2013}
\bibinfo{author}{Kage, A.}, \bibinfo{author}{Hosoya, C.},
  \bibinfo{author}{Baba, S.A.}, \bibinfo{author}{Mogami, Y.},
  \bibinfo{year}{2013}.
\newblock \bibinfo{title}{Drastic reorganization of bioconvection pattern of
  \textit{{C}hlamydomonas}: {Q}uantitative analysis of the pattern transition
  response}.
\newblock \bibinfo{journal}{J. Exp. Biol.} \bibinfo{volume}{216},
  \bibinfo{pages}{4557--4566}.
\newblock \DOIprefix\doi{https://doi.org/10.1242/jeb.092791}.
\bibitem[{Karimi and Paul(2013)}]{Karimi&Paul_2013}
\bibinfo{author}{Karimi, A.}, \bibinfo{author}{Paul, M.R.},
  \bibinfo{year}{2013}.
\newblock \bibinfo{title}{Bioconvection in spatially extended domains}.
\newblock \bibinfo{journal}{Phys. Rev. E} \bibinfo{volume}{87},
  \bibinfo{pages}{053016}.
\newblock \DOIprefix\doi{https://link.aps.org/doi/10.1103/PhysRevE.87.053016}.
\bibitem[{Kazmierczak and Chinoda(1992)}]{Kazmierczak&Chinoda_1992}
\bibinfo{author}{Kazmierczak, M.}, \bibinfo{author}{Chinoda, Z.},
  \bibinfo{year}{1992}.
\newblock \bibinfo{title}{Buoyancy-driven flow in an enclosure with time
  periodic boundary conditions}.
\newblock \bibinfo{journal}{Int. J. Heat Mass Transf.} \bibinfo{volume}{35},
  \bibinfo{pages}{1507--1518}.
\newblock \DOIprefix\doi{https://doi.org/10.1016/0017-9310(92)90040-Y}.
\bibitem[{Khan et~al.(2020)Khan, Salahuddin, Malik, Alqarni and
  Alqahtani}]{Khan_et_al_2020}
\bibinfo{author}{Khan, M.}, \bibinfo{author}{Salahuddin, T.},
  \bibinfo{author}{Malik, M.Y.}, \bibinfo{author}{Alqarni, M.S.},
  \bibinfo{author}{Alqahtani, A.M.}, \bibinfo{year}{2020}.
\newblock \bibinfo{title}{Numerical modeling and analysis of bioconvection on
  {MHD} flow due to an upper paraboloid surface of revolution}.
\newblock \bibinfo{journal}{Physica A} \bibinfo{volume}{553},
  \bibinfo{pages}{124231}.
\newblock \DOIprefix\doi{https://doi.org/10.1016/j.physa.2020.124231}.
\bibitem[{Kuznetsov(2005a)}]{Kuznetsov_2005-3}
\bibinfo{author}{Kuznetsov, A.V.}, \bibinfo{year}{2005}a.
\newblock \bibinfo{title}{Investigation of the onset of thermo-bioconvection in
  a suspension of oxytactic microorganisms in a shallow fluid layer heated from
  below}.
\newblock \bibinfo{journal}{Theor. Comput. Fluid Dyn.} \bibinfo{volume}{19},
  \bibinfo{pages}{287--299}.
\newblock \DOIprefix\doi{https://doi.org/10.1007/s00162-005-0167-3}.
\bibitem[{Kuznetsov(2005b)}]{Kuznetsov_2005-10}
\bibinfo{author}{Kuznetsov, A.V.}, \bibinfo{year}{2005}b.
\newblock \bibinfo{title}{The onset of bioconvection in a suspension of
  negatively geotactic microorganisms with highfrequency vertical vibration}.
\newblock \bibinfo{journal}{Int. Commun. Heat Mass Transf.}
  \bibinfo{volume}{32}, \bibinfo{pages}{1119--1127}.
\newblock
  \DOIprefix\doi{https://doi.org/10.1016/j.icheatmasstransfer.2005.05.004}.
\bibitem[{Kuznetsov(2005c)}]{Kuznetsov_2005-8}
\bibinfo{author}{Kuznetsov, A.V.}, \bibinfo{year}{2005}c.
\newblock \bibinfo{title}{Thermo-bioconvection in a suspension of oxytactic
  bacteria}.
\newblock \bibinfo{journal}{Int. Commun. Heat Mass Transf.}
  \bibinfo{volume}{32}, \bibinfo{pages}{991--999}.
\newblock
  \DOIprefix\doi{https://doi.org/10.1016/j.icheatmasstransfer.2004.11.005}.
\bibitem[{Kuznetsov(2011)}]{Kuznetsov_2011}
\bibinfo{author}{Kuznetsov, A.V.}, \bibinfo{year}{2011}.
\newblock \bibinfo{title}{Nanofluid bioconvection: interaction of
  microorganisms oxytactic upswimming, nanoparticle distribution, and
  heating/cooling from below}.
\newblock \bibinfo{journal}{Theor. Comput. Fluid Dyn.} \bibinfo{volume}{26},
  \bibinfo{pages}{291--310}.
\newblock \DOIprefix\doi{https://doi.org/10.1007/s00162-011-0230-1}.
\bibitem[{Kwak and Hyun(1996)}]{Kwak&Hyun_1996}
\bibinfo{author}{Kwak, H.S.}, \bibinfo{author}{Hyun, J.M.},
  \bibinfo{year}{1996}.
\newblock \bibinfo{title}{Natural convection in an enclosure having a vertical
  sidewall with time-varying temperature}.
\newblock \bibinfo{journal}{J. Fluid Mech.} \bibinfo{volume}{329},
  \bibinfo{pages}{65--88}.
\newblock \DOIprefix\doi{https://doi.org/10.1017/S0022112096008828}.
\bibitem[{Kwak et~al.(1998)Kwak, Kuwahara and Hyun}]{Kwak_et_al_1998}
\bibinfo{author}{Kwak, H.S.}, \bibinfo{author}{Kuwahara, K.},
  \bibinfo{author}{Hyun, J.M.}, \bibinfo{year}{1998}.
\newblock \bibinfo{title}{Resonant enhancement of natural convection heat
  transfer in a square enclosure}.
\newblock \bibinfo{journal}{Int. J. Heat Mass Transf.} \bibinfo{volume}{41},
  \bibinfo{pages}{2837--2846}.
\newblock \DOIprefix\doi{https://doi.org/10.1016/S0017-9310(98)00018-0}.
\bibitem[{Mantle et~al.(1994)Mantle, Kazmierczak and Hiawy}]{Mantle_et_al_1994}
\bibinfo{author}{Mantle, J.}, \bibinfo{author}{Kazmierczak, M.},
  \bibinfo{author}{Hiawy, B.}, \bibinfo{year}{1994}.
\newblock \bibinfo{title}{The effect of temperature modulation on natural
  convection in a horizontal layer heated from below: {H}igh-{R}ayleigh-number
  experiments}.
\newblock \bibinfo{journal}{J. Heat Transfer} \bibinfo{volume}{116},
  \bibinfo{pages}{614--620}.
\newblock \DOIprefix\doi{https://doi.org/10.1115/1.2910913}.
\bibitem[{Matsumoto and Nishimura(1998)}]{Matsumoto&Nishimura_1998}
\bibinfo{author}{Matsumoto, M.}, \bibinfo{author}{Nishimura, T.},
  \bibinfo{year}{1998}.
\newblock \bibinfo{title}{Mersenne twister: {A} 623-dimensionally
  equidistributed uniform pseudo-random number generator}.
\newblock \bibinfo{journal}{ACM Trans. Model. Comput. Simul.}
  \bibinfo{volume}{8}, \bibinfo{pages}{3--30}.
\newblock \DOIprefix\doi{https://doi.org/10.1145/272991.272995}.
\bibitem[{Metcalfe and Pedley(1998)}]{Metcalfe&Pedley_1998}
\bibinfo{author}{Metcalfe, A.M.}, \bibinfo{author}{Pedley, T.J.},
  \bibinfo{year}{1998}.
\newblock \bibinfo{title}{Bacterial bioconvection: {W}eakly nonlinear theory
  for pattern selection}.
\newblock \bibinfo{journal}{J. Fluid Mech.} \bibinfo{volume}{370},
  \bibinfo{pages}{249--270}.
\newblock \DOIprefix\doi{https://doi.org/10.1017/s0022112098001979}.
\bibitem[{Metcalfe and Pedley(2001)}]{Metcalfe&Pedley_2001}
\bibinfo{author}{Metcalfe, A.M.}, \bibinfo{author}{Pedley, T.J.},
  \bibinfo{year}{2001}.
\newblock \bibinfo{title}{Falling plumes in bacterial bioconvection}.
\newblock \bibinfo{journal}{J. Fluid Mech.} \bibinfo{volume}{445},
  \bibinfo{pages}{121--149}.
\newblock \DOIprefix\doi{https://doi.org/10.1017/s0022112001005547}.
\bibitem[{Naseem et~al.(2017)Naseem, Shafiq, Zhao and
  Naseem}]{Naseem_et_al_2017}
\bibinfo{author}{Naseem, F.}, \bibinfo{author}{Shafiq, A.},
  \bibinfo{author}{Zhao, L.}, \bibinfo{author}{Naseem, A.},
  \bibinfo{year}{2017}.
\newblock \bibinfo{title}{{MHD} biconvective flow of powell eyring nanofluid
  over stretched surface}.
\newblock \bibinfo{journal}{AIP Adv.} \bibinfo{volume}{7},
  \bibinfo{pages}{065013}.
\newblock \DOIprefix\doi{https://doi.org/10.1063/1.4983014}.
\bibitem[{Noever and Matsos(1991a)}]{Noever&Matsos_1991a}
\bibinfo{author}{Noever, D.A.}, \bibinfo{author}{Matsos, H.C.},
  \bibinfo{year}{1991}a.
\newblock \bibinfo{title}{A bioassay for monitoring cadmium based on
  bioconvective patterns}.
\newblock \bibinfo{journal}{J. Environ. Sci. Health A} \bibinfo{volume}{26},
  \bibinfo{pages}{273--286}.
\newblock \DOIprefix\doi{https://doi.org/10.1080/10934529109375633}.
\bibitem[{Noever and Matsos(1991b)}]{Noever&Matsos_1991b}
\bibinfo{author}{Noever, D.A.}, \bibinfo{author}{Matsos, H.C.},
  \bibinfo{year}{1991}b.
\newblock \bibinfo{title}{Calcium protection from cadmium poisoning:
  {B}ioconvective indicators in \textit{{T}etrahymena}}.
\newblock \bibinfo{journal}{J. Environ. Sci. Health A} \bibinfo{volume}{26},
  \bibinfo{pages}{1105--1113}.
\newblock \DOIprefix\doi{https://doi.org/10.1080/10934529109375689}.
\bibitem[{Noever et~al.(1992)Noever, Matsos and Looger}]{Noever_et_al_1992}
\bibinfo{author}{Noever, D.A.}, \bibinfo{author}{Matsos, H.C.},
  \bibinfo{author}{Looger, L.L.}, \bibinfo{year}{1992}.
\newblock \bibinfo{title}{Bioconvective indicators in \textit{{T}etrahymena}:
  {N}ickel and copper protection from cadmium poisoning}.
\newblock \bibinfo{journal}{J. Environ. Sci. Health A} \bibinfo{volume}{27},
  \bibinfo{pages}{403--417}.
\newblock \DOIprefix\doi{https://doi.org/10.1080/10934529209375735}.
\bibitem[{Omori et~al.(2003)Omori, Habe, Yoshida, Horinouchi, Saiki and
  Nojiri}]{Omori_et_al_2003}
\bibinfo{author}{Omori, T.}, \bibinfo{author}{Habe, H.},
  \bibinfo{author}{Yoshida, T.}, \bibinfo{author}{Horinouchi, M.},
  \bibinfo{author}{Saiki, Y.}, \bibinfo{author}{Nojiri, H.},
  \bibinfo{year}{2003}.
\newblock \bibinfo{title}{Applied ecological microbiology}.
\newblock \bibinfo{publisher}{Shoko--do}.
\bibitem[{Omori et~al.(2002)Omori, Nojiri, Horinouchi and
  Kasuga}]{Omori_et_al_2000}
\bibinfo{author}{Omori, T.}, \bibinfo{author}{Nojiri, H.},
  \bibinfo{author}{Horinouchi, M.}, \bibinfo{author}{Kasuga, K.},
  \bibinfo{year}{2002}.
\newblock \bibinfo{title}{Environmental {B}iotechnology}.
\newblock \bibinfo{edition}{3} ed., \bibinfo{publisher}{Shoko--do}.
\bibitem[{Paolucci and Chenoweth(1989)}]{Paolucci&Chenoweth_1989}
\bibinfo{author}{Paolucci, S.}, \bibinfo{author}{Chenoweth, D.R.},
  \bibinfo{year}{1989}.
\newblock \bibinfo{title}{Transition to chaos in a differentially heated
  vertical cavity}.
\newblock \bibinfo{journal}{J. Fluid Mech.} \bibinfo{volume}{201},
  \bibinfo{pages}{379--410}.
\newblock \DOIprefix\doi{https://doi.org/10.1017/S0022112089000984}.
\bibitem[{Pedley et~al.(1988)Pedley, Hill and Kessler}]{Pedley_et_al_1988}
\bibinfo{author}{Pedley, T.J.}, \bibinfo{author}{Hill, N.A.},
  \bibinfo{author}{Kessler, J.O.}, \bibinfo{year}{1988}.
\newblock \bibinfo{title}{The growth of bioconvection patterns in a uniform
  suspension of gyrotactic micro-organisms}.
\newblock \bibinfo{journal}{J. Fluid Mech.} \bibinfo{volume}{195},
  \bibinfo{pages}{223--237}.
\newblock \DOIprefix\doi{https://doi.org/10.1017/s0022112088002393}.
\bibitem[{Pedley and Kessler(1992)}]{Pedley&Kessler_1992}
\bibinfo{author}{Pedley, T.J.}, \bibinfo{author}{Kessler, J.O.},
  \bibinfo{year}{1992}.
\newblock \bibinfo{title}{Hydrodynamic phenomena in suspensions of swimming
  microorganisms}.
\newblock \bibinfo{journal}{Annu. Rev. Fluid Mech.} \bibinfo{volume}{24},
  \bibinfo{pages}{313--358}.
\newblock
  \DOIprefix\doi{https://www.annualreviews.org/doi/abs/10.1146/annurev.fl.24.010192.001525}.
\bibitem[{Platt(1961)}]{Platt_1961}
\bibinfo{author}{Platt, J.R.}, \bibinfo{year}{1961}.
\newblock \bibinfo{title}{Bioconvection patterns in cultures of free-swimming
  organisms}.
\newblock \bibinfo{journal}{Science} \bibinfo{volume}{133},
  \bibinfo{pages}{1766--1767}.
\newblock \DOIprefix\doi{https://doi.org/10.1126/science.133.3466.1766}.
\bibitem[{Shi et~al.(2021)Shi, Hamid, Khan, Kumar, Gowda, Prasannakumara, Shah,
  Khan,  and Chung}]{Shi_et_al_2021}
\bibinfo{author}{Shi, Q.H.}, \bibinfo{author}{Hamid, A.},
  \bibinfo{author}{Khan, M.I.}, \bibinfo{author}{Kumar, R.N.},
  \bibinfo{author}{Gowda, R.J.P.}, \bibinfo{author}{Prasannakumara, B.C.},
  \bibinfo{author}{Shah, N.A.}, \bibinfo{author}{Khan, S.U.}, ,
  \bibinfo{author}{Chung, J.D.}, \bibinfo{year}{2021}.
\newblock \bibinfo{title}{Numerical study of bio-convection flow of
  magneto-cross nanofluid containing gyrotactic microorganisms with activation
  energy}.
\newblock \bibinfo{journal}{Sci. Rep.} \bibinfo{volume}{11},
  \bibinfo{pages}{16030}.
\newblock \DOIprefix\doi{https://doi.org/10.1038/s41598-021-95587-2}.
\bibitem[{Tanaka et~al.(2010)Tanaka, Nariai, Nakamura, Sato and
  Matsuzawa}]{Tanaka_et_al_2010}
\bibinfo{author}{Tanaka, H.}, \bibinfo{author}{Nariai, K.},
  \bibinfo{author}{Nakamura, S.}, \bibinfo{author}{Sato, K.},
  \bibinfo{author}{Matsuzawa, Y.}, \bibinfo{year}{2010}.
\newblock \bibinfo{title}{Development of the high-performance bioethanol
  fermenting reactor}.
\newblock \bibinfo{journal}{IHI Engineering Review} \bibinfo{volume}{43},
  \bibinfo{pages}{63--69}.
\newblock \DOIprefix\doi{https://doi.org/10.11501/11023114}.
\bibitem[{Tuval et~al.(2005)Tuval, Cisneros, Dombrowski, Wolgemuth, Kessler and
  Goldstein}]{Tuval_et_al_2005}
\bibinfo{author}{Tuval, I.}, \bibinfo{author}{Cisneros, L.},
  \bibinfo{author}{Dombrowski, C.}, \bibinfo{author}{Wolgemuth, C.W.},
  \bibinfo{author}{Kessler, J.O.}, \bibinfo{author}{Goldstein, R.E.},
  \bibinfo{year}{2005}.
\newblock \bibinfo{title}{Bacterial swimming and oxygen transport near contact
  lines}.
\newblock \bibinfo{journal}{Proc. Natl. Acad. Sci. U.S.A.}
  \bibinfo{volume}{102}, \bibinfo{pages}{2277--2282}.
\newblock \DOIprefix\doi{https://doi.org/10.1073/pnas.0406724102}.
\bibitem[{Uddin et~al.(2016)Uddin, Kabir and B\'{e}g}]{Uddin_et_al_2016}
\bibinfo{author}{Uddin, M.J.}, \bibinfo{author}{Kabir, M.N.},
  \bibinfo{author}{B\'{e}g, O.A.}, \bibinfo{year}{2016}.
\newblock \bibinfo{title}{Computational investigation of stefan blowing and
  multiple-slip effects on buoyancy-driven bioconvection nanofluid flow with
  microorganisms}.
\newblock \bibinfo{journal}{Int. J. Heat Mass Transf.} \bibinfo{volume}{95},
  \bibinfo{pages}{116--130}.
\newblock
  \DOIprefix\doi{http://dx.doi.org/10.1016/j.ijheatmasstransfer.2015.11.015}.
\bibitem[{Williams and Bees(2011)}]{Williams&Bees_2011}
\bibinfo{author}{Williams, C.R.}, \bibinfo{author}{Bees, M.A.},
  \bibinfo{year}{2011}.
\newblock \bibinfo{title}{A tale of three taxes: {P}hoto-gyro-gravitactic
  bioconvection}.
\newblock \bibinfo{journal}{J. Exp. Biol.} \bibinfo{volume}{214},
  \bibinfo{pages}{2398--2408}.
\newblock \DOIprefix\doi{https://doi.org/10.1242/jeb.051094}.
\bibitem[{Yanaoka(2023)}]{Yanaoka_2023}
\bibinfo{author}{Yanaoka, H.}, \bibinfo{year}{2023}.
\newblock \bibinfo{title}{Influences of conservative and non-conservative
  {L}orentz forces on energy conservation properties for incompressible
  magnetohydrodynamic flows}.
\newblock \bibinfo{journal}{J. Comput. Phys.} \bibinfo{volume}{491},
  \bibinfo{pages}{112372 (36 pages)}.
\newblock \DOIprefix\doi{https://doi.org/10.1016/j.jcp.2023.112372}.
\bibitem[{Yanaoka and Inafune(2023)}]{Yanaoka&Inafune_2023}
\bibinfo{author}{Yanaoka, H.}, \bibinfo{author}{Inafune, R.},
  \bibinfo{year}{2023}.
\newblock \bibinfo{title}{Frequency response of three-dimensional natural
  convection of nanofluids under microgravity environments with gravity
  modulation}.
\newblock \bibinfo{journal}{Numer. Heat Tr. A-Appl.} \bibinfo{volume}{83},
  \bibinfo{pages}{745--769}.
\newblock \DOIprefix\doi{https://doi.org/10.1080/10407782.2022.2161437}.
\bibitem[{Yanaoka et~al.(2007)Yanaoka, Inamura and Suzuki}]{Yanaoka_et_al_2007}
\bibinfo{author}{Yanaoka, H.}, \bibinfo{author}{Inamura, T.},
  \bibinfo{author}{Suzuki, K.}, \bibinfo{year}{2007}.
\newblock \bibinfo{title}{Numerical analysis of bioconvection generated by
  chemotactic bacteria}.
\newblock \bibinfo{journal}{JSME, Ser. B} \bibinfo{volume}{73},
  \bibinfo{pages}{575--580}.
\newblock \DOIprefix\doi{https://doi.org/10.1299/kikaib.73.575}.
\bibitem[{Yanaoka et~al.(2008)Yanaoka, Inamura and Suzuki}]{Yanaoka_et_al_2008}
\bibinfo{author}{Yanaoka, H.}, \bibinfo{author}{Inamura, T.},
  \bibinfo{author}{Suzuki, K.}, \bibinfo{year}{2008}.
\newblock \bibinfo{title}{Three-dimensional numerical analysis of bioconvection
  generated by chemotactic bacteria}.
\newblock \bibinfo{journal}{JSME, Ser. B} \bibinfo{volume}{74},
  \bibinfo{pages}{135--141}.
\newblock \DOIprefix\doi{https://doi.org/10.1299/kikaib.74.135}.
\bibitem[{Yanaoka and Nishimura(2022)}]{Yanaoka&Nishimura_2022}
\bibinfo{author}{Yanaoka, H.}, \bibinfo{author}{Nishimura, T.},
  \bibinfo{year}{2022}.
\newblock \bibinfo{title}{Pattern wavelengths and transport characteristics in
  three-dimensional bioconvection generated by chemotactic bacteria}.
\newblock \bibinfo{journal}{J. Fluid Mech.} \bibinfo{volume}{952},
  \bibinfo{pages}{A13--1--31 (31 pages)}.
\newblock \DOIprefix\doi{https://doi.org/10.1017/jfm.2022.898}.
\bibitem[{Zadeha et~al.(2020)Zadeha, Mehryanb, Sheremetc, Izadid and
  Ghodrate}]{Zadeha_et_al_2020}
\bibinfo{author}{Zadeha, S.M.H.}, \bibinfo{author}{Mehryanb, S.},
  \bibinfo{author}{Sheremetc, M.A.}, \bibinfo{author}{Izadid, M.},
  \bibinfo{author}{Ghodrate, M.}, \bibinfo{year}{2020}.
\newblock \bibinfo{title}{Numerical study of mixed bio-convection associated
  with a micropolar fluid}.
\newblock \bibinfo{journal}{Therm. Sci. Eng. Prog.} \bibinfo{volume}{18},
  \bibinfo{pages}{100539}.
\newblock \DOIprefix\doi{https://doi.org/10.1016/j.tsep.2020.100539}.

\end{thebibliography}

\end{document}